\documentclass{aa}  

\usepackage{graphicx}
\usepackage{natbib}
\bibpunct{(}{)}{;}{a}{}{,} 
\usepackage{txfonts}
\begin{document}

   \title{The Fornax Deep Survey with VST.}

   \subtitle{V. Exploring the faintest regions of the bright early-type galaxies inside the virial radius}

   \author{E. Iodice
          \inst{1}
          \and
          M. Spavone\inst{1}
          \and
          M. Capaccioli\inst{2}
          \and
          R.F. Peletier\inst{3}          
          \and
          G. van de Ven\inst{4}          
          \and
          N.R. Napolitano\inst{1}          
          \and
          M. Hilker\inst{4}          
          \and 
          S. Mieske\inst{5}          
          \and
          R. Smith\inst{6}          
          \and
          A. Pasquali\inst{7}          
          \and
          L. Limatola\inst{1}          
          \and
          A. Grado\inst{1}          
          \and
          A. Venhola\inst{3,8}          
          \and
          M. Cantiello\inst{9}          
          \and
          M. Paolillo\inst{2}          
          \and
          J. Falcon-Barroso\inst{10,11}          
          \and
          R. D'Abrusco\inst{12}          
          \and
          P. Schipani\inst{1}          
          }

   \institute{INAF-Astronomical Observatory of Capodimonte, via Moiariello 16, Naples, I-80131, Italy\\
              \email{enrichetta.iodice@inaf.it}
         \and
             University of Naples ``Federico II'', C.U. Monte Sant'Angelo, Via Cinthia, 80126, Naples, Italy\\
         \and
             Kapteyn Astronomical Institute, University of Groningen, PO Box 72, 9700 AV Groningen, The Netherlands\\
         \and
               European Southern Observatory, Karl-Schwarzschild-Strasse 2, D-85748 Garching bei Munchen, Germany\\
          \and           
            European Southern Observatory, Alonso de Cordova 3107, Vitacura, Santiago, Chile\\ 
            \and
            Department of Astronomy and Institute of Earth-Atmosphere-Astronomy, Yonsei University, Seoul 03722, Korea\\
            \and
            Astronomisches Rechen-Institut Zentrum f\''{u}r Astronomie der Universit\''{a}t Heidelberg, Monchhofstrasse 12-14, D-69120 Heidelberg, Germany\\
            \and
             Division of Astronomy, Department of Physics, University of Oulu, Oulu, Finland\\
            \and
             INAF-Astronomical Abruzzo Observatory, Via Maggini, 64100, Teramo, Italy\\
             \and
            Instituto de Astrof\'isica de Canarias, C/ Via L\'actea s/n, 38200 La Laguna, Canary Islands, Spain\\
             \and
	    Departamento de Astrof\'isica, Universidad de La Laguna, E-38200 La Laguna, Tenerife, Spain\\             
	    \and
             Smithsonian Astrophysical Observatory/Chandra X-ray centre, 02138 Cambridge (MA), US\\
             }

   \date{Received ....; accepted ......}

 
  \abstract
   {This paper is based on the multi-band ($ugri$) Fornax Deep Survey (FDS) with the VLT Survey Telescope (VST). 
   We study  bright early-type galaxies ($m_B \leq 15$~mag) in the $9$~square degrees around the core of the Fornax cluster, 
   which covers the virial radius (R$_{vir}$ $\sim 0.7$~Mpc).}
   {  The main goal of the present work is to provide the analysis of the light distribution for all galaxies out to unprecedented limits
   (in radius and surface brightness) and to release the main products resulting from this analysis in all FDS bands.   
    We give a first comprehensive view of the galaxy structure and evolution as function of the cluster environment.}
   {From the isophote fit,  we derive  the azimuthally averaged surface brightness profiles, the position angle and ellipticity profiles
    as a function of the semi-major axis. In each band, we derive the total magnitudes, effective radii, integrated colors and stellar mass-to-light ratios. }
   {  The long integration times, the arcsec-level  angular resolution of OmegaCam@VST and the large covered area of FDS allow us to 
   map the light and color distributions out to large galactocentric distances (up to about $10-15 R_e$) and surface brightness levels 
   beyond $\mu_r = 27$~mag/arcsec$^2$ ($\mu_B \geq 28$~mag/arcsec$^2$). Therefore, the new FDS data allow us to explore in  
   great detail the morphology and structure of cluster galaxies out to the region of the stellar halo.
   The analysis presented in this paper allows us to study how the structure of galaxies and the stellar population content vary 
   with the distance from the cluster center. In addition to the intra-cluster features 
   detected in previous FDS works, we found a new faint filament between FCC~143 and FCC~147, suggesting an ongoing interaction.}
   {  The observations suggest that the Fornax cluster is not completely relaxed inside the virial radius.
 The bulk of the gravitational interactions between galaxies happens in the W-NW core region of the 
   cluster, where most of the bright early-type galaxies are located and where the intra-cluster baryons (diffuse light and GCs) are found. 
 We suggest that the W-NW sub-clump of galaxies results from an infalling group onto the cluster, which  has modified 
   the structure of the galaxy outskirts (making asymmetric stellar halos) and has produced the intra-cluster 
   baryons (ICL and GCs), concentrated in this region of the cluster.}


   \maketitle
%

\section{Introduction}\label{intro}


Clusters of galaxies are  excellent sites to study the effects of the environment on the galaxy structure and on the star formation rate
\citep[e.g.][]{Trujillo2001,Lewis2002,Aguerri2004,Gutierrez2004,Mendez-Abreu2012,Aguerri2016,Merluzzi2016,Lisker2018}. 
{  There are several types of environmental mechanisms that can act, also simultaneously, on the cluster members, which induce strong 
changes in the stellar distribution and gas content. 
The process of repeated high-speed tidal encounters of the galaxies inside the cluster, which induce  galaxy-galaxy gravitational interactions and/or with the intra-cluster 
medium, is known as {\it galaxy harassment}  \citep{Moore1998}. It generates a tidal mass loss from the galaxies, which depends on the orbit of the galaxy
 inside the cluster  \citep{Mastropietro2005,Smith2015}.
The tidal forces can strip dark matter, stars, and gas from galaxies, which generates faint streams detectable along the orbit of the galaxy through the cluster. This mechanism is more efficient on low-mass galaxies \citep{Smith2010}.
The interaction of  gas-rich galaxies with the hot interstellar medium, known as {\it ram-pressure stripping}, can sweep off their atomic gas and therefore halt the star 
formation \citep{Gunn1972}. 
The ram-pressure stripping can also remove the hot, ionized gas within the halo of a galaxy, so that the supply of cold gas is lost. 
This is the so-called "strangulation"  that is also able to shut down the star formation in a galaxy  \citep{Larson1980,Dekel2006,Peng2015}.
{ The strangulation also occurs when the inflow of new gas onto the galaxy hot gas halo is stopped \citep{Dekel2006}.}
The mechanisms listed above, which induce morphological transformation and/or the quenching of the star formation are
thought to be partly responsible for the {\it morphology-density relation} \citep{Dressler1997,vdWel2010,Fasano2015}, where early-type galaxies (ETGs) 
dominate the central regions of the clusters, while late-type galaxies (LTGs), spirals and irregulars, populate the outskirts. 
Recently, to investigate the many processes acting in clusters,  the ``phase-space diagram'' has been used, where the cluster-centric velocities and radii are combined in 
one plot \citep[see][and references therein]{Smith2015,Jaffe2015,Rhee2017,Pasquali2018}. 
}

In the deep potential well at the cluster center, the galaxies continue to undergo active mass assembly and, in this process, gravitational interactions 
and merging between systems 
of comparable mass and/or smaller objects play a fundamental role in defining the galaxies' morphology and the build-up of the stellar halos.
During the infall of groups of galaxies to form the cluster, the material stripped from the galaxy outskirts builds up the intracluster light, 
ICL \citep{Delucia2007,Puchwein2010,Cui2014}. 
This is a diffuse faint component that grows over time with the mass assembly of the cluster, to which the relics of the interactions 
between galaxies (stellar 
streams and tidal tails) also contribute \citep[see][as reviews]{Arnaboldi2010,Tutukov2011,Mihos2015}. 


The evolution of galaxies in a cluster is mainly driven by the processes listed above. Their effects strictly depend on some parameters 
related to the internal structure of the galaxy, like the total mass, to the orbits, relative velocities and to the location of the galaxy in the cluster.

In this framework, the study of  ETGs, elliptical and lenticular (S0) galaxies,  has a special role. 
Since these galaxies are mainly located close to the cluster center, they are good tracers of  past galaxy interactions and mass assembly that shaped their 
structure, from the inner, bright regions to their faint outskirts. 
Important hints for the formation and evolution of ETGs reside in the morphology, age, metallicity, mass-to-light ratio and in the kinematics \citep[see][as review]{Buta2011,deZeeuw2002}.
The most recent studies on the photometry of galaxies in nearby clusters are the extensive analysis by \citet{Kormendy2009} for the Virgo cluster and by 
\citet{Gutierrez2004} for the Coma cluster. Recent integral-field spectroscopic surveys offer the largest kinematic database to study the stellar population in ETGs 
\citep[e.g. SAURON, ATLAS$^{3D}$, CALIFA, SAMI,][]{deZeeuw2002,Cappellari2011,Sanchez2012,Bryant2015}.

In this context, the {\it VST Early-type GAlaxy Survey} (VEGAS\footnote{see http://www.na.astro.it/vegas/VEGAS/Welcome.html})
aims to provide deep multi-band imaging and quantitative photometric analysis for a large sample of ETGs in different environments, 
including giant cD galaxies in the core of clusters \citep{Capaccioli2015,Spavone2017}. The main goal of 
VEGAS is to study the galaxy structure down to the faintest levels of surface brightness of $\mu_g\sim27-30$~mag/arcsec$^2$, in order to map also the 
regions of the stellar halos and the ICL. VEGAS is producing competitive results to the several pioneering deep imaging surveys \citep{Mihos2005,Jan2010,MarDel2010,Roediger2011,Ferrarese2012,Duc2015,Dokkum2014,Munoz2015,Trujillo2016,Mihos2017,Merritt2016,Crnojevic2016}. 

The {\it Fornax Deep Survey} at VST (FDS) is  one part of the deep-surveys campaign. It is a joint project based on VEGAS (P.I. E. Iodice)  and the 
OmegaCam guaranteed time (P.I. R. Peletier), which aims to cover the whole Fornax cluster out to the virial radius \citep[$\sim0.7$~Mpc,][]{Drinkwater2001},  
with an area of about $26$~square degrees around the central galaxy NGC~1399 in the cluster core and including the SW subgroup centred on NGC~1316. 
To date, together with the Hubble Space Telescope data \citep{Jordan2007} and DECAM data for the Next Generation Fornax Cluster Survey \citep{Munoz2015}, 
FDS is the deepest and widest dataset mapping the Fornax cluster in the optical wavelength range. These surveys are part of the multi-wavelengths observations available   for this cluster, as the Hershel survey \citep{Davies2013}, Chandra X-ray imaging of the cluster core \citep{Scharf2005}, UV GALEX imaging \citep{Martin2005}.  
Upcoming ALMA data (P.I. T. Davis) and neutral hydrogen HI data from the MeerKat survey \citep{Serra2016} will provide a complete census of the cool interstellar medium in Fornax. In addition, the integral-field spectroscopy is now available for dwarf and giant galaxies in Fornax with several instruments, including SAMI   \citep{Scott2014} and MUSE \citep{Sarzi2018}.

The Fornax cluster is the second most massive galaxy concentration within 20 Mpc, after the Virgo cluster, {  with a virial mass of M~$= 7 \times 10^{13}$~M$_\odot$} \citep{Drinkwater2001}. 
It is among the richest nearby sites to study galaxy evolution and dynamical harassment in a dense environment.
Previous works indicate that it has a complex structure and the mass assembly is still ongoing \citep{Drinkwater2001,Scharf2005}.
The core is in a quite evolved phase \citep{Grillmair1994,Jordan2007}, since most of the bright ($m_B<15$~mag) cluster members are ETGs  
\citep{Ferguson1989}, more than in the Virgo cluster.
It hosts a vast population of dwarf galaxies and ultra compact galaxies \citep{Munoz2015,Hilker2015a,Schulz2016,Venhola2017,Venhola2018,Eigenthaler2018}, 
an intra-cluster population of globular clusters (GCs) \citep{Schuberth2008,Schuberth2010,Dabrusco2016,Cantiello2018,Pota2018} and of planetary nebulae \citep{Napolitano2003,McNeil2012,Spiniello2018}.

{  FDS provided the mosaic of $3\times2$ square degrees around the central galaxy NGC~1399\footnote{See the ESO photo release at https://
www.eso.org/public/news/eso1612/} in the core and around the brightest cluster member in the SW subgroup, NGC~1316\footnote{See the ESO photo release at https://www.eso.org/public/news/eso1734/}.
With FDS we {\it i)} mapped the surface brightness around NGC~1399 and NGC~1316 out to an unprecedented distance of about $\sim 200$~kpc 
($R\sim6R_e$) and down to $\mu_g \simeq 29-31$ mag/arcsec$^2$  \citep{Iodice2016,Iodice2017}; 
{\it ii)} traced the spatial distribution of candidate globular clusters (GCs) inside $\sim0.5$~deg$^2$ of the cluster core \citep{Dabrusco2016,Cantiello2018};
{\it iii)} detected new and faint ($\mu_g \sim 29-30$~mag~arcsec$^{-2}$) features in the intra-cluster region between NGC~1399 and NGC~1387  \citep{Iodice2016}
 and in the outskirts of NGC~1316 \citep{Iodice2017}; 
 {\it iv)} detected a previously unknown region of ICL in the core of the cluster, on the West side of NGC~1399 \citep{Iodice2017b};
 {\it v)} provided a census of the low-surface brightness and dwarf galaxies in the whole area covered by FDS \citep{Venhola2017,Venhola2018}.

The FDS observations were completed at the end of 2017. A full description of the data and survey plans will be presented soon in a forthcoming paper 
(Peletier et al. in preparation). 
Taking advantage of the multi-band deep observations, the large field-of-view of OmegaCam@VST and the optimised observing strategy, the ``novelty'' 
of FDS consist the systematic characterisation of many basic, global properties of the galaxies in the Fornax cluster over a
wide baseline of radius, as luminosity profiles, isophote shapes, galaxy outskirts, color gradients and  inventories of satellite galaxies and GCs. 
The published papers on FDS proved the ability to study the galaxy outskirts, mapping the regions of the stellar halos and intra-cluster space to detect  
any kind of low-surface brightness structures  and their connection with the environment.

In this work we focus on the inner $9$~square degrees around the core of the Fornax cluster, which corresponds to about the viral radius ($\sim 0.7$~Mpc). 
We study the ETGs brighter than $m_B=15$~mag ($M_B =-16.4$~mag)) inside this area (19 objects), as a complete sample, in magnitude and morphological type, of galaxies 
inside the virial radius of the Fornax cluster. 
The main goal of the present work is to provide the analysis of the light and color distribution out to unprecedented limits (in radius and surface
brightness) and to release the main  products resulting from this analysis (i.e. total magnitudes, effective radii, integrated colors and M/L ratios).
We also give a global comprehensive view of the galaxy structure as function of the cluster environment from observations,
focusing on the results mainly based on the new FDS data. 
Therefore, with the present paper, we aim at providing the observables to be used in forthcoming/ongoing works on specific science topics. }

%


\section{FDS data}\label{data}

\begin{figure*}
   \centering
   \includegraphics[width=\hsize]{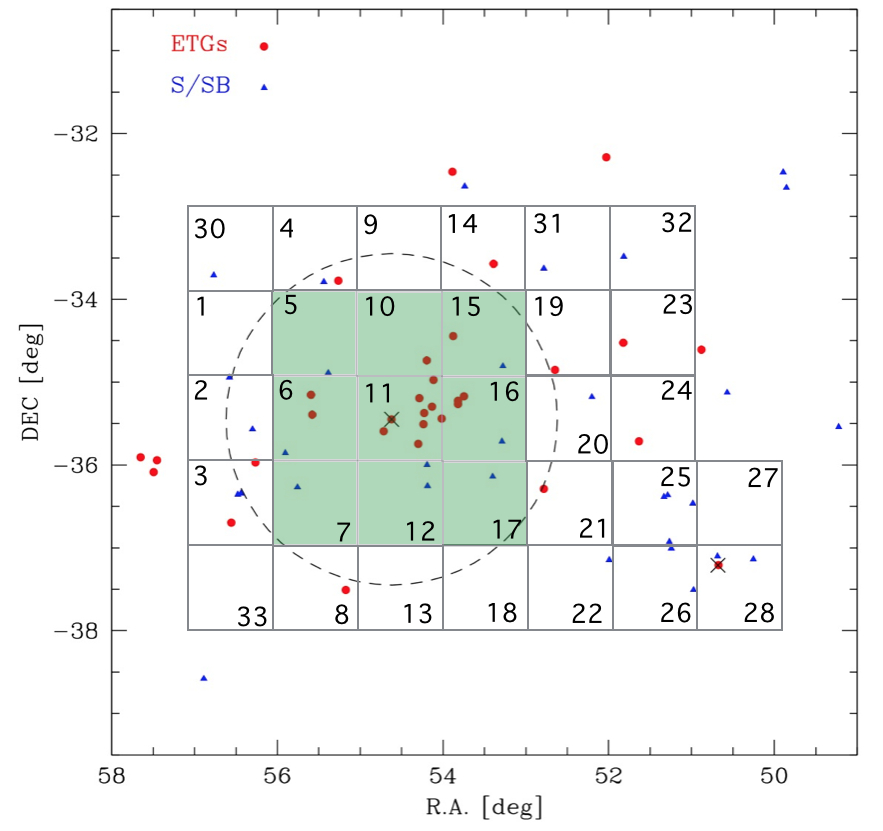}
      \caption{Distribution of early-type (Elliptical and S0s, red circles) and late-type galaxies (blue triangles) in the Fornax cluster and in the sub-group Fornax~A, brighter than $m_b<15$~mag (see also Tab.~\ref{galaxies}). The two crosses indicate the location of NGC~1399 at the centre of the cluster (R.A.=54.62 degree, DEC=-35.45 degree) and of NGC~1316 at the centre of the SW subgroup (R.A.=50.67 degree, DEC=-37.21 degree). The dashed circle corresponds to the virial radius $R_{vir}\sim0.7$~Mpc. Squares are the FDS fields of $1 \times 1$ degree. The green boxes are the FDS fields presented in this work.}
         \label{FCC}
   \end{figure*}

This work is based on the Fornax Deep Survey (FDS) data in the $u$,$g$,$r$ and $i$ bands,  inside an area of $9$~square degrees around the core of the Fornax cluster, ($\alpha=03h38m29.024s$, $\delta=-35d 27' 03.18''$, see Fig.~\ref{FCC}). 
The area covers the virial radius  of the cluster, which is $R_{vir}\sim0.7$~Mpc \citep{Drinkwater2001}.

FDS observations are part of the Guaranteed Time Observation surveys, {\it FOCUS}
(P.I. R. Peletier) and {\it VEGAS} \citep[P.I. E. Iodice,][]{Cap2015} obtained with the ESO VLT Survey Telescope (VST). 
VST is a 2.6-m wide field optical survey telescope, located at Cerro Paranal in Chile \citep{Schipani2012}, 
equipped with the wide field camera ($1 \times1$~degree$^2$) OmegaCam, having a pixel scale of 0.21~arcsec/pixel.

The central FDS fields (F11, F16) were presented by \citet{Iodice2016,Iodice2017} and \citet{Venhola2017}, which include a detailed description of the observing strategy and the data reduction pipelines used for FDS. 
The data used in this work (see Fig.~\ref{FCC}), which includes also the already published FDS fields cited above, were collected during several visitor mode runs (from 2013 to 2017, see Tab.~\ref{FDS_fields}) in dark time for the $u$, $g$ and $r$ bands and in grey time for the $i$ band and were reduced with the VST-Tube pipeline  \citep{Grado2012,Cap2015}. The image quality of all FDS fields is presented by \citet{Venhola2018}.

As pointed out by \citet{Iodice2016,Iodice2017} and \citet{Venhola2017},  in order to track the background variation exposure by exposure and, therefore, have a very accurate estimate of the sky background around bright and extended galaxies,
the FDS fields were obtained by using a {\it step-dither} observing strategy. It consists of a cycle of short exposures (150~sec) of adjacent fields, close in time and space ($\leq 1$~deg away). The "empty" fields, i.e. with no bright stars or galaxies, were adopted to derive an average sky frame for each night and each band. This is scaled and subtracted off from each science frame. 
The average sky frame derived for each observing night takes into account the small contribution to the sky brightness by the smooth components (i.e galactic cirrus, zodiacal light and of the terrestrial airglow)  plus the extragalactic background light.
In the sky-subtracted science frame {  could remain some ``residual fluctuations'', of a few percent in flux, which are due to the flux variation of the background during the night. These are estimated on the final stacked image (see Sec.~\ref{phot}) and sets the accuracy of the sky-subtraction step \citep[see also][]{Iodice2016,Venhola2017}.}

For each field we  obtained 76 exposures (of 150~sec each) in the $u$ band, 54 in the $g$ and $r$ bands, and 35 in the $i$ band,
giving a total exposure time of 3.17 hrs in the $u$ band, 2.25 hrs in the $g$ and $r$ bands and of 1.46 hrs in the $i$ bands.
Only images with seeing $FWHM\le1.5$~arcsec were co-added. The resulting total exposure times for each filter are listed in Tab.~\ref{FDS_fields}.

\begin{figure*}
   \centering
   \includegraphics[width=\hsize]{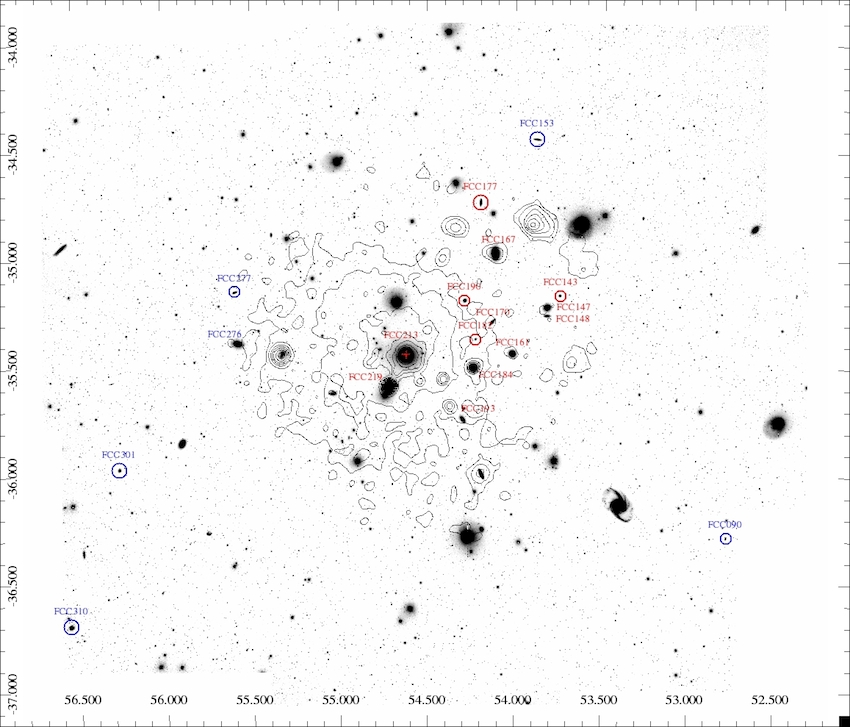}
      \caption{VST mosaic in the $g$ band of the Fornax cluster inside the virial radius (see Fig.~\ref{FCC}). The image size is about 9 square 
      degrees. Black contours are the X-Ray emission from ROSAT \citep{Paolillo2002}. {  The X-ray contours are spaced by a factor of 1.3, with 
      the lowest level at $3.0x10^{-3}$ counts/arcmin$^2$/s}. 
      All ETGs studied in the present work (see Tab.~\ref{galaxies}) are labelled on the image. 
      Galaxies marked with blue symbols have $g-i \leq0.8$~mag and $g-r \leq 0.4$~mag (see Sec.~\ref{results} for details). North is up, East is on 
      the left.}
         \label{mosaic_g}
   \end{figure*}

The full $g$-band mosaic of the $9$~square degrees around the core of the Fornax cluster is shown in Fig.~\ref{mosaic_g}.
The dataset studied in this work includes all ETGs brighter than $m_B\leq15$~mag selected from \citet{Ferguson1989} (see Tab.~\ref{galaxies} and Fig.~\ref{FCC}).
Fainter objects are classified as dwarf galaxies.
The $r$-band image, in surface brightness levels, for each galaxy of the sample is shown in Appendix~A.

\begin{sidewaystable*}
\caption{\label{FDS_fields} }
\begin{tabular}{lcccccccccccc}
\hline\hline
 Field&& $u$ band &  && $g$ band & && $r$ band &&& $i$ band \\
 &  &&  && & &&  &&&  \\
 & date & Exp. T. & FWHM & date & Exp. T. & FWHM & date & Exp. T. & FWHM & date & Exp. T. & FWHM\\
         &          &  [hrs]        & [arcsec] &        &   [hrs]        & [arcsec] &       &   [hrs]        & [arcsec] &        &   [hrs]        & [arcsec] \\
(1) & (2) & (3) & (4) & (5)& (6)& (7)& (8)& (9)& (10)& (11)& (12) & (13)\\
 &  &&  && & &&  &&&  \\
\hline
 &  &&  && & &&  &&& & \\
F5 & Nov 2015 & 2.9 & 1.33 & Nov 2014/2015 & 2.5 & 1.15 & Nov 2014/2015 & 2.7 & 1.39 & Nov 2014/2015 & 2.2 & 1.08\\
F6 & Nov 2013 & 2.12 & 1.11  & Nov 2013 & 2.3 & 0.84 & Nov 2013 & 2.2 & 1.08 &Nov 2014 & 1.0 & 1.21\\
F7 & Nov 2013 & 2.12 & 1.04  & Nov 2013 & 2.3 & 0.83 & Nov 2013 & 2.2 & 0.95 &Nov 2014 & 1.0 & 1.42\\
F10 & Nov 2015 & 2.9 & 1.34 & Nov 2014/2015 & 2.5 & 1.15 & Nov 2014/2015 & 2.7 & 1.02 & Nov 2014/2015 & 2.2 & 1.09\\
F11 & Nov 2014 & 3.7 &  1.27 & Nov 2014 & 2.2 & 1.06 & Nov 2014& 2.1 & 1.09 &Nov 2014& 1.3 & 1.15\\
F12 & Nov 2013 & 2.12 & 1.15  & Nov 2013 & 2.3 & 0.83 & Nov 2013 & 2.2 & 1.04 &Nov 2014 & 1.0 & 1.17\\
F15 & Nov 2013/2014 & 3.7 & 1.30  & Nov 2013/2014 & 2.2 & 1.13 & Nov 2013/2014& 2.1 & 0.9 &Nov 2013/2014& 1.3 & 0.97\\
F16 & Nov 2014 & 3.7 & 1.31  & Nov 2014 & 2.2 & 1.26 & Nov 2014& 2.1 & 0.94 &Nov 2014& 1.3 & 1.08\\
F17 & Nov 2014 & 3.7 &  1.27 & Nov 2014 & 2.2 & 1.11 & Nov 2014& 2.1 & 0.87 &Nov 2014& 1.3 & 1.01\\
%
\hline
\hline
\end{tabular}
\tablefoot{{\it Col.1 -} FDS field, see Fig.\ref{FCC}.  {\it Col.2 - Col.3 - Col.4} Date of the observations,  total exposure time in hours and average seeing in arcsec for the $u$ filter. {\it Col.5 to Col.13} same as Col.2 to Col.4 for the $g$, $r$ and $i$ bands.
The FWHM is computed for all FDS fields by \citet{Venhola2018}.}
\end{sidewaystable*}

\begin{table*}
\caption{\label{galaxies} Early-type galaxies inside the virial radius ($R_{vir}\sim0.7$~Mpc) of the Fornax cluster, brighter than $m_B<15$~mag.}
\centering
\begin{tabular}{lccccccc}
\hline\hline
object & $\alpha$ & $\delta$ & Morph Type & Distance & $m_B$ &FDS field& Names\\
           & h:m:s     & d:m:s     &                     & Mpc & mag & & \\
(1) & (2) & (3) & (4) & (5) & (6) & (7) & (8)\\
\hline
\hline
%
FCC090 & 03 31 08.1 & -36 17 19 & E4 pec &  $19.5\pm1.7$ & 15.0 &F16-F17 &\\
FCC143 & 03 34 59.1 & -35 10 10 & E3 &  $19.3\pm0.8$ & 14.3 &F16 & NGC1373, ESO358-G21\\
FCC147 & 03 35 16.8 & -35 13 34 & E0 &  $19.6\pm0.6$ & 11.9 &F16 & NGC1374, ESO358-G23\\
FCC148 & 03 35 16.8 & -35 15 56 & S0 &  $19.9\pm0.7$ & 13.6 & F16 & NGC1375, ESO358-G24\\
FCC153 & 03 35 30.9 & -34 26 45 & S0 &  $20.8\pm0.7$ & 13.0 &F10-F15 & IC1963, ESO358-G26\\
FCC161 & 03 36 04.0 & -35 26 30 & E0 &  $19.9\pm0.4$ & 11.7 &F11-F16 & NGC1379, ESO358-G37\\
FCC167 & 03 36 27.5 & -34 58 31 & S0/a &  $21.2\pm0.7$ & 11.3 &F10-F11 & NGC1380, ESO358-G28\\
FCC170 & 03 36 31.6 & -35 17 43 & S0 &  $21.9\pm0.8$ & 13.0 &F11 & NGC1381, ESO358-G29\\
FCC177 & 03 36 47.4 & -34 44 17 & S0 &  $20.0\pm0.6$ & 13.2 &F10 & NGC1380A, ESO358-G33\\
FCC182 & 03 36 54.3 & -35 22 23 & SB0 pec &  $19.6\pm0.8$ & 14.9 & F11 & \\
FCC184 & 03 36 56.9 & -35 30 24 & SB0  &  $19.3\pm0.8$ & 12.3 & F11 & NGC1387, ESO358-G36\\
FCC190 & 03 37 08.9 & -35 11 37 & SB0  &  $20.3\pm0.7$ & 13.5 &F11 & NGC1380B, ESO358-G37\\
FCC193 & 03 37 11.7 & -35 44 40 & SB0  &  $21.2\pm0.7$ & 12.8 &F11 & NGC1389, ESO358-G38\\
FCC213 & 03 38 29.2 & -35 27 02 & E0  &  $20.9\pm0.9$ & 10.6 &F11 & NGC1399, ESO358-G45\\
FCC219 & 03 38 52.1 & -35 35 38 & E2  &  $20.2\pm0.7$ & 10.9 &F11 & NGC1404, ESO358-G46\\
%
%
%
FCC276 & 03 42 19.2 & -35 23 36 & E4  &  $19.6\pm0.6$ & 11.8 &F6 & NGC1427, ESO358-G52\\
FCC277 & 03 42 22.6 & -35 09 10 & E5  &  $20.7\pm0.7$ & 13.8 &F6 & NGC1428, ESO358-G53\\
FCC301 & 03 45 03.5 & -35 58 17 & E4  &  $19.7\pm0.7$ & 14.2 &F7 & ESO358-G59\\
FCC310 & 03 46 13.7 & -36 41 43 & SB0  &  $19.9\pm0.6$ & 13.5 &F3-F7 & NGC1460, ESO358-G62\\
\hline
\end{tabular}
\tablefoot{{\it Col.1 -} Fornax cluster members from \citet{Ferguson1989}. {\it Col.2 and Col.3 -} Right ascension and declination.  {\it Col.4, Col.5 and Col.6 -} Morphological type, distance and total magnitude in the B band given by \citet{Ferguson1989} and \citet{Blakeslee2009}. For FCC~161 the distance is from \citet{Tonry2001}. {\it Col.7 -} Location in the FDS field (see Fig.\ref{FCC}). {\it Col.8 -} Other catalogue name.}
\end{table*}

\section{Surface photometry of the ETGs in the Fornax cluster}\label{phot}

\begin{figure*}
   \centering
   \includegraphics[width=\hsize]{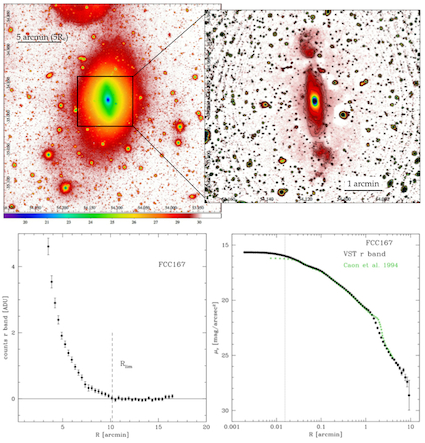}
      \caption{ {\it Top panels -} Extracted region of the VST mosaic of  $0.34\times0.34$~square degrees around FCC167 in $r$ band (left), plotted in surface brightness levels (shown in the colorbar). High-frequency residual image obtained from the $r$-band VST  image of FCC~167 (right). 
          {\it Lower panels -} Intensity profile (left), in the $r$ band, in the outer regions of FCC167. For $R \ge 10.2$~arcmin, the galaxy's light blends into the average sky level, which has a residual scatter of about 0.04 counts around the zero (continuous black line).
          Azimuthally-averaged surface brightness profiles of FCC167 in the $r$ band (right) derived from VST data (black  points),   compared with literature data from \citet{Caon1994} in the B band (green asterisks), transformed to $r$-band.  {  The vertical dotted line (right panel) delimits the region where the light profiles are "flattened" due to  the convolution effect with the point-spread function (PSF). The FWHM of the PSF is about 1 arcsec, which is the average seeing for FDS observations (see Tab.~\ref{FDS_fields}). }}
   \label{FCC167_image}
   \end{figure*}

In this section we describe the method adopted to study VST images and the main steps to provide the parameters characterising the galaxy structure (i.e., total magnitudes, colors and effective radii). 
We have chosen one galaxy of the sample, FCC167 (see Fig.~\ref{FCC167_image}), to describe each step of the analysis in detail. 
FCC167 (also known as NGC~1380) is a lenticular galaxy in the NW side of the Fornax cluster, located at $\sim0.6$~degree ($\sim219$~kpc) from NGC~1399  (see Fig.~\ref{mosaic_g}) and at a distance of $\sim21$~Mpc from us \citep{Blakeslee2010}. This object was chosen as show case since it is close to the core of the cluster but outside the extended stellar halo of NGC~1399, which is mapped out to 33~arcmin \citep{Iodice2016}.

Results for all galaxies of the sample are described in Appendix~\ref{note} and related images and profiles are shown in Appendix~\ref{VST_image} and Appendix~\ref{colormap}.

{\it Method: isophote fitting -}
From  the sky-subtracted FDS field, we extracted the azimuthally-averaged intensity profile for each object of the sample in each band, by using the IRAF\footnote{The Image Reduction and Analysis Facility is distributed
  by the National Optical Astronomy Observatory (NOAO), which is
  operated by the Association of Universities for Research in
  Astronomy (AURA), Inc. under cooperative agreement with the National
  Science Foundation.} task ELLIPSE. 
The method is described in detail by \citet{Iodice2016}, where the galaxy properties of the central cluster galaxy NGC 1399 were presented.
The main steps, performed for each images in each filter, are:

\begin{enumerate}
\item mask  all the bright sources (galaxies and stars) and background objects close to the galaxy under study; 
\item fit the isophotes in elliptical annuli centred on the galaxy  out to the edge of the FDS field {  (out to about 0.5 deg)}, where all fitting parameters (center, position angle and ellipticity) were left free. {   For those galaxies that are in the overlapping regions between two FDS fields (see Fig.~\ref{FCC}), we performed this analysis on the mosaic of the two fields;}
\item from the intensity profiles in each band, estimate the outermost radius $R_{lim}$ from the centre of the galaxy where the galaxy's light blends into the average background level\footnote{The average background level is the residual after subtracting the sky frame, and therefore very close to zero \citep[see][]{Iodice2016}.}. 

{ We estimated the average sky fluctuation levels ($\Delta$ sky) and the ''amplitudes'' of the distribution, 
i.e. the RMS of the mean value at one sigma. 
We assume as the latest valuable science point in the surface brightness profile the point where flux $\geq \Delta$ sky.
The average RMS on the sky subtraction  increases from the $u$ to the $i$ band, with the following average values: $\sim 0.01 - 0.08$~counts in the $u$ band; $\sim 0.02 - 0.13$~counts in the $g$ band; $\sim 0.03 - 0.2$~counts in the $r$ band and $\sim 0.08 - 0.3$~counts in the $i$ band. }For FCC167, this step is shown in  Fig.~\ref{FCC167_image} (lower-left panel), where $R_{lim}=10.2$~arcmin in the $r$ band.
\end{enumerate}

{\it Products:  total magnitudes, effective radii, colors, mass-to-light ratio}

\begin{enumerate}

\item  From the isophote fit,  we derive  the azimuthally averaged surface brightness (SB) profiles (corrected for the residual background level estimated at $R \ge R_{lim}$), the position angle (P.A.) and ellipticity ($\epsilon$) profiles (where $\epsilon = 1 - b/a$, and $b/a$ is the axial ratio of the ellipses)  as a function of the semi-major axis, in each band. The error estimates on the magnitudes take the uncertainties  on the photometric calibration ($\sim 0.001 - 0.006$~mag) and sky subtraction into account \citep{Cap2015,Iodice2016}. 
For FCC167 the SB profiles in the $r$ band are shown in Fig.~\ref{FCC167_image}.
For all the galaxies of the sample, results from the isophote fit are shown in Appendix~\ref{VST_image}.
\item From the azimuthally averaged SB profiles we derive the $g-i$ color profiles and from the corresponding images we make the $g-i$ color maps. 
For FCC167 they are shown in Fig.~\ref{FCC167_col}.
For all the galaxies of the sample, the $g-i$ color maps and color profiles are shown in Appendix~\ref{colormap}. 
\item Using a growth curve analysis, derived from the isophote fit in the elliptical apertures, we compute the total magnitudes and effective radii ($R_e$) inside the outermost radius $R_{lim}$ in each band. They are listed in Tab.~\ref{mag}.
\item From the total magnitudes (see Tab.~\ref{mag}) we computed the $g-i$ and $g-r$ average colors.
We also derived the average $g-i$ colors in the inner regions of the galaxy at $R\leq0.5R_e$ and in the outskirts at $R\geq3R_e$.
 They are listed in Tab.~\ref{Mtot_tab}, which also includes the absolute magnitudes and the projected distance of each galaxy from NGC~1399 (assumed as the "center" of the cluster).
\item From the $g-i$ average color we estimate the stellar mass $M_{*}$ by using the empirical relation $\log_{10} \frac{M_{*}}{M_{\odot}} = 1.15 + 0.70(g-i) - 0.4M_{i}$ from \citet{Taylor2011}, where $M_{i}$ is the absolute magnitude in the $i$ band{  \footnote{The empirical relation proposed by \citet{Taylor2011} assumed a Chabrier IMF.}}. 
According to \citet{Taylor2011}, this relation provides an estimate of the stellar mass-to-light ratio ($M_{*} / L_i$) to a $1 \sigma$ accuracy of $\sim0.1$ dex.
The $M_{*} / L_i$ value for each galaxy of the sample is given in Tab.~\ref{Mtot_tab}.

\end{enumerate}

 \begin{figure*}
 \centering
    \includegraphics[width=\hsize]{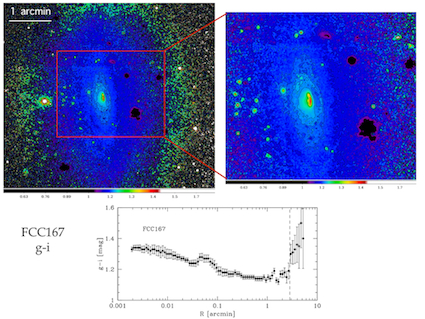}
    \caption{ {\it Top panels -} $g-i$ colormap of FCC~167 inside  $4.5 \times 4.3$~arcmin from the center (left panel) and a zoom toward the nuclear regions (right panel), for $2.6 \times 1.8$~arcmin. The $g-i$ levels are shown in the colorbar. {\it Bottom panel -} Azimuthally-averaged $g-i$ color profile  as function of the  semi-major axis by the fit of the ellipses. The vertical long-dashed line indicates the transition radius at $R \sim 3$~arcmin, between the disk and the outer stellar exponential halo (see Sec.~\ref{sec:fcc167}).}
         \label{FCC167_col}
   \end{figure*}


\begin{table*}
\caption{\label{mag} Total magnitudes and effective radii for the early-type galaxies inside the virial radius ($R_{vir}\sim0.7$~Mpc) of the Fornax cluster, brighter than $m_B<15$~mag, in the $ugri$ bands.}
\centering
\begin{tabular}{lcccccccc}
\hline\hline
object & $m_u$ & $m_g$ & $m_r$ & $m_i$ & $Re_u$ & $Re_g$ & $Re_r$ & $Re_i$\\
           & mag     & mag     & mag    & mag & arcsec & arcsec & arcsec & arcsec \\
(1) & (2) & (3) & (4) & (5) & (6) & (7) & (8) & (9)\\
\hline
\hline
%
FCC090 & $15.18 \pm 0.04$ & $14.16 \pm 0.02$ & $13.62 \pm 0.03$ & $13.38 \pm 0.02$ & $10.34 \pm 0.10$& $10.31 \pm 0.06$& $12.11 \pm 0.10$ & $11.31 \pm 0.07$ \\
FCC143 & $14.88 \pm 0.06$ & $13.36 \pm 0.02$ & $12.66 \pm 0.03$ & $12.39 \pm 0.07$ & $13.51 \pm 0.09$& $11.86 \pm 0.01$& $11.00 \pm 0.05$ & $10.46 \pm 0.09$ \\
FCC147 & $12.67 \pm 0.03$ & $11.14 \pm 0.01$ & $10.50 \pm 0.01$ & $10.15 \pm 0.02$ & $38.34 \pm 0.12$& $28.53 \pm 0.01$& $24.8 \pm 0.2$ & $24.1 \pm 0.2$ \\
FCC148 & $13.77 \pm 0.02$ & $12.33 \pm 0.01$ & $11.70 \pm 0.01$ & $11.47 \pm 0.02$ & $30.4 \pm 0.2$& $27.2 \pm 0.2$& $28.3 \pm 0.2$ & $25.04 \pm 0.14$ \\
FCC153 & $13.88 \pm 0.02$ & $11.94 \pm 0.02$ & $11.70 \pm 0.03$ & $11.17 \pm 0.01$ & $24.5 \pm 0.3$ & $26.3 \pm 0.3$ & $19.8 \pm 0.7$ & $27.4 \pm 0.5$ \\
FCC161 & $12.85 \pm 0.03$ & $11.18 \pm 0.01$ & $10.47 \pm 0.01$ & $10.14 \pm 0.03$ & $28.2 \pm 0.2$& $28.6 \pm 0.2$& $28.6 \pm 0.2$ & $26.96 \pm 0.03$ \\
FCC167 & $11.70 \pm 0.02$ & $9.86 \pm 0.01$ & $9.27 \pm 0.02$ & $8.83 \pm 0.01$ & $57.9 \pm 0.3$& $72.7 \pm 0.5$& $56.4 \pm 0.5$ & $60.0 \pm 0.4$ \\
FCC170 & $13.30 \pm 0.01$ & $11.64 \pm 0.01$ & $10.99 \pm 0.01$ & $10.57 \pm 0.01$ & $20.9 \pm 0.5$& $17.97 \pm 0.15$& $15.89 \pm 0.13$ & $16.34 \pm 0.12$ \\
FCC177 & $13.88 \pm 0.04$ & $12.52 \pm 0.01$ & $11.8 \pm 0.01$ & $11.44 \pm 0.02$ & $31 \pm 1$& $33.9 \pm 0.3$& $35.9 \pm 0.2$ & $35.6 \pm 0.2$ \\
FCC182 & $15.66 \pm 0.04$ & $14.24 \pm 0.01$ & $13.58 \pm 0.01$ & $13.21 \pm 0.01$ & $12.66 \pm 0.05$& $10.27 \pm 0.05$& $9.90 \pm 0.05$ & $9.78 \pm 0.06$ \\
FCC184 & $12.64 \pm 0.01$ & $10.76 \pm 0.01$ & $10.00 \pm 0.04$ & $9.62 \pm 0.03$ & $32.26 \pm 0.03$& $36.8 \pm 0.01$& $34.45 \pm 0.01$ & $31.17 \pm 0.11$ \\
FCC190 & $14.20 \pm 0.06$ & $12.92 \pm 0.01$ & $12.26 \pm 0.01$ & $11.88 \pm 0.01$ & $26.32 \pm 0.13$& $18.98 \pm 0.08$& $18.34 \pm 0.08$ & $18.2 \pm 0.08$ \\
FCC193 & $13.09 \pm 0.06$ & $11.42 \pm 0.01$ & $10.69 \pm 0.02$ & $10.25 \pm 0.02$ & $28.2 \pm 0.3$& $27.6 \pm 0.3$& $28.2 \pm 0.3$ & $32.8 \pm 0.4$ \\
FCC213 & $10.55 \pm 0.12$ & $8.57 \pm 0.09$ & $7.89 \pm 0.09$ & $7.59 \pm 0.03$ & $232 \pm 2$& $346 \pm 3$& $308 \pm 2$ & $248 \pm 2$ \\
FCC219 & $11.33 \pm 0.04$ & $9.28 \pm 0.08$ & $8.57 \pm 0.04$ & $8.32 \pm 0.01$ & $97.47 \pm 0.06$& $201.5 \pm 0.1$& $161 \pm 0.2$ & $116.6 \pm 0.9$ \\
%
%
%
FCC276 & $12.52 \pm 0.05$ & $10.45 \pm 0.06$ & $10.15 \pm 0.02$ & $9.81 \pm 0.04$ & $56.1 \pm 0.5$& $56.1 \pm 0.5$& $44.67 \pm 0.07$ & $42.4 \pm 0.3$ \\
FCC277 & $14.64 \pm 0.02$ & $12.74 \pm 0.01$ & $12.34 \pm 0.02$ & $11.97 \pm 0.04$ & $14.45 \pm 0.10$& $12.61 \pm 0.05$& $12.78 \pm 0.04$ & $12.61 \pm 0.03$ \\
FCC301 & $15.0 \pm 0.1$ & $12.99 \pm 0.05$ & $12.65 \pm 0.01$ & $12.29 \pm 0.03$ & $14.9 \pm 0.4$& $13.7 \pm 0.4$& $11.7 \pm 0.3$ & $11.9 \pm 0.3$ \\
FCC310 & $13.0 \pm 0.01$ & $12.19 \pm 0.05$ & $11.81 \pm 0.04$ & $11.41 \pm 0.01$ & $63 \pm 1$& $36.3 \pm 0.14$& $35.57 \pm 0.02$ & $36.41 \pm 0.06$ \\
\hline
\end{tabular}
\tablefoot{{\it Col.1 -} Fornax cluster members from \citet{Ferguson1989}. {\it Col.2 to Col.5 -} Total magnitudes in the $u$, $g$, $r$ and $i$ band respectively, derived from the isophote fit.  Values were corrected for the Galactic extinction by using the absorption coefficient by \citet{Schlegel98}. {\it Col.6 to Col.9 -} Effective radius in the $u$, $g$, $r$ and $i$ band respectively, derived from the isophote fit.}
\end{table*}

\begin{table*}
\caption{\label{Mtot_tab}  Stellar mass estimates of the ETGs in the $i$-band. }
\centering
\begin{tabular}{lcccccccccc}
\hline\hline
object & D$_{core}$ & M$_r$ & M$_i$ & $R_e$ & $g-r$ & $g-i$  & $g-i$ ($R\leq0.5R_e$)  & $g-i$ ($R\geq3R_e$) & M$_*$   &   M/L  \\
           & [degree]     & [mag] & [mag] & [kpc]      &  [mag] & [mag]& [mag]&[mag]&$10^{10}$ M$_{\odot}$&      \\
&&&&&&&\\
(1) & (2) & (3) & (4) & (5) & (6) & (7) & (8) & (9) & (10) & (11)\\
\hline
\hline
%
FCC090 & 1.7 & -18.83 & -18.06 & 1.07 &  $0.54\pm0.05$ & $0.78\pm0.04$ & $0.62\pm0.02$ & $0.8\pm0.1$ & 0.082  & 0.72\\
FCC143 & 0.76 & -18.77 & -19.04 & 0.98 &  $0.07\pm0.05$ & $0.97\pm0.09$ & $1.08\pm0.02$& $0.8\pm0.2$ & 0.28 &  0.99\\
FCC147 & 0.67 & -20.96 & -21.31 & 2.29 & $0.64\pm0.02$ &$0.99\pm0.03$ & $1.13\pm0.01$ & $0.98\pm0.07$ & 2.4 & 1.04\\
FCC148 & 0.67 & -19.79 & -20.02 & 2.42 &  $0.63\pm0.02$ & $0.86\pm0.03$ & $0.92\pm0.02$ & $0.7\pm0.3$ & 0.58 &  0.83\\
FCC153 & 1.17 &  -19.89 & -20.42 & 2.84 & $0.24\pm0.07$ & $0.77\pm0.07$ & $0.78\pm0.06$ & $0.77\pm0.13$ & 0.76 &  0.74 \\
FCC161 & 0.49 & -21.02 & -21.35 & 2.60 & $0.71\pm0.02$ & $1.04\pm0.04$ & $1.04\pm0.04$ & $1.04\pm0.04$ & 2.63 &  1.11\\
FCC167 &  0.62 & -22.36 & -22.8 & 6.17 & $0.59\pm0.03$ & $1.03\pm0.02$ & $1.19\pm0.02$ & $1.31\pm0.11$ & 9.85 &  1.10\\
FCC170 &  0.42 & -20.71 & -21.13 & 1.73 & $0.65\pm0.02$ & $1.07\pm0.02$ & $1.13\pm0.01$ & $1.07\pm0.07$ & 2.25 &  1.17\\
FCC177 &  0.79 & -19.71 & -20.07 & 3.45 & $0.72\pm0.02$ & $1.08\pm0.03$ & $1.00\pm0.02$ & $1.1\pm0.6$ & 0.85 &  1.19\\
FCC182 &  0.32 & -17.88 & -18.25 & 0.93 & $0.66\pm0.02$ & $1.03\pm0.02$ & $1.09\pm0.01$ & $1.03\pm0.15$ & 0.15 &  1.1\\
FCC184 & 0.31 & -21.43 & -21.81 & 2.91 & $0.76\pm0.05$ & $1.14\pm0.04$ & $1.30\pm0.01$ & $1.34\pm0.14$ & 4.70 &  1.31\\
FCC190 & 0.37 & -19.28 & -19.66 & 1.79 & $0.66\pm0.02$ & $1.04\pm0.02$ & $1.09\pm0.01$ & $0.89\pm0.14$ &0.54 &  1.10\\
FCC193 & 0.39 & -20.93 & -21.38 & 3.37 & $0.73\pm0.03$ & $1.17\pm0.03$ & $1.12\pm0.01$ & $1.14\pm0.06$ &3.32 &  1.37\\
FCC213 & 0.0 & -23.71 & -24.01 & 25.1 & $0.7\pm0.2$ & $0.98\pm0.12$ & $1.19\pm0.02$ & $0.8\pm0.5$ &27.5 &  1.0\\
FCC219 & 0.17 & -22.95 & -23.21 & 11.4 & $0.71\pm0.12$ & $0.96\pm0.09$ & $1.18\pm0.01$ & $1.07\pm0.04$ & 12.7 &  0.98\\
FCC276 & 0.79 & -21.31 & -21.65 & 4.03 & $0.3\pm0.08$ & $0.64\pm0.10$ & $0.81\pm0.01$ & $0.61\pm0.15$ & 1.81 &  0.58\\
FCC277 & 0.86 & -19.24 & -19.61 & 1.26 & $0.4\pm0.03$ & $0.77\pm0.05$ & $0.75\pm0.01$ & $0.76\pm0.09$ &0.34 &  0.72\\
FCC301 & 1.44 &  -18.82 & -19.17 & 1.14 & $0.34\pm0.06$ & $0.70\pm0.08$ & $0.70\pm0.05$ & $0.65\pm0.15$ & 0.20 &  0.63\\
FCC310 & 2.0 & -19.70 & -20.10 & 3.51 &$0.37\pm0.09$  &  $0.77\pm0.06$ & $0.76\pm0.01$ & $0.66\pm0.07$ & 0.54 &   0.72  \\
\hline
\end{tabular}
\tablefoot{{\it Col.1 -} Fornax cluster members from \citet{Ferguson1989}. {\it Col.2 -} Projected distance from the galaxy center (in degree), i.e. from NGC~1399 (FCC~213). {\it Col.3 and Col.4 -} Absolute magnitudes in the $r$ and $i$ bands.  {\it Col.5 -}  Effective radius (in kpc) in the $i$ band. {\it Col.6 and Col.7 -} Average $g-r$ and $g-i$ colors. {\it Col.8 and Col.9 -} Average $g-i$ colors for $R\leq0.5R_e$ and for $R\geq3R_e$. 
{\it Col.10 and Col.11 -} Stellar mass and Mass-to-Light (M/L) in the $i$ band.}
\end{table*}



\section{Mapping the galaxy light distribution beyond $\mu_B= 28$~mag/arcsec$^2$ in FDS}\label{results}

The bright galaxies inside the Fornax cluster are well studied in a wide wavelength range and at several resolution levels (see Sec.~\ref{intro}). 
{  In this section we show that 
the main and new contribution from the FDS is to map the light and color distribution in the optical bands up to unprecedented limits. 
As shown in Sec.~\ref{overview}, this allows to systematically characterise, for the first time, the galaxy structure and color profiles out to the regions of the stellar halos (where $\mu_g \geq 26$~mag/arcsec$^2$).}

\begin{figure}
   \centering
   \includegraphics[width=\hsize]{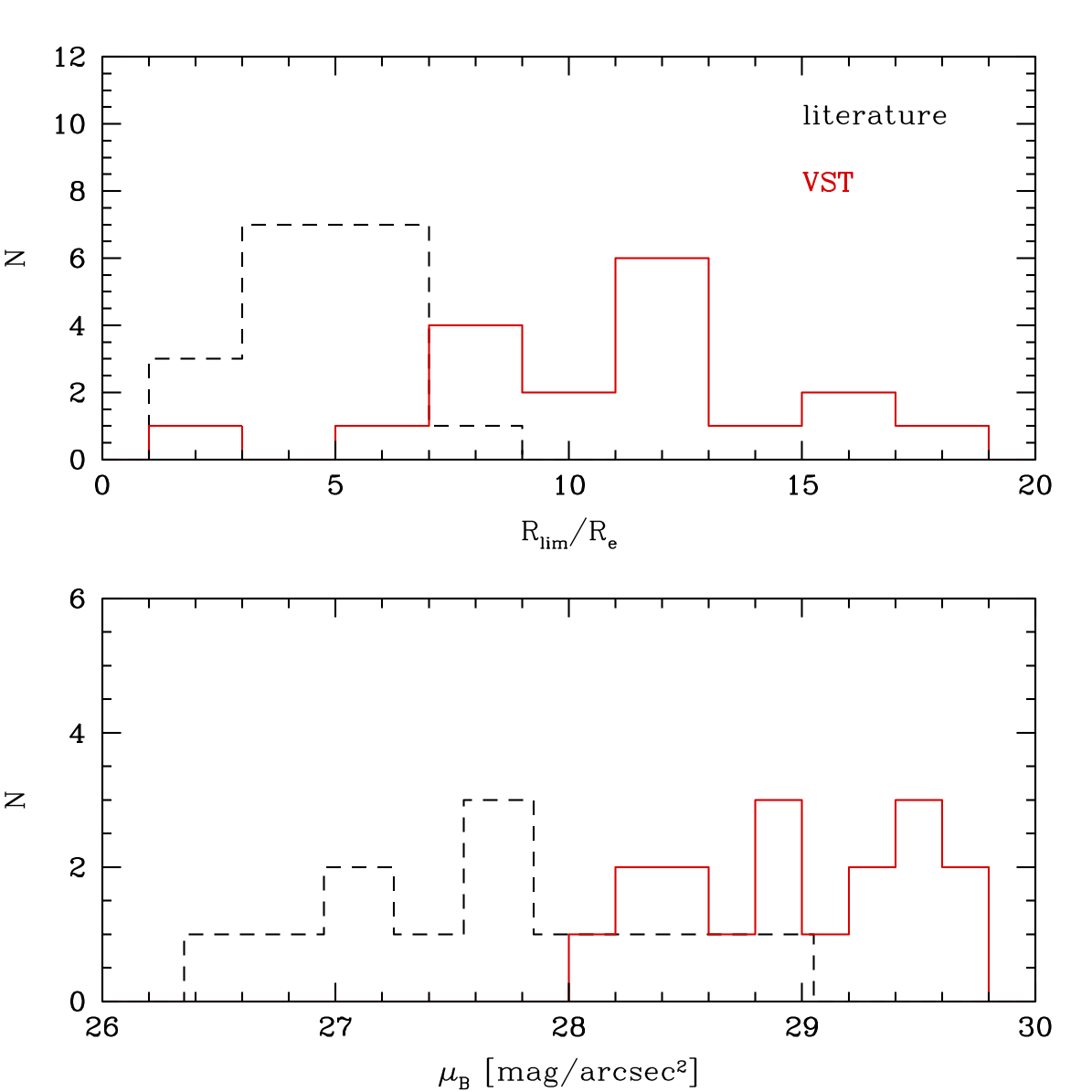}
      \caption{ Comparison of the limiting magnitudes, derived from the surface brightness profiles shown in Appendix~\ref{VST_image}, (bottom panel) and effective radii (top panel) obtained by the new FDS images (red histograms) and the estimates from previous works (dashed histograms), mainly from \citet{Caon1994}. }
         \label{conf}
   \end{figure}

In Fig.~\ref{conf} we compare the limiting magnitudes and effective radii obtained by FDS  with the estimates from previous works, mainly from 
\citet{Caon1994}, which contains the most extended analysis of the light distribution in the B band for the majority of the ETGs in the FDS sample: FDS is providing 
the deepest images of  ETGs in the Fornax cluster. For all ETGs in  FDS the faintest levels of the azimuthally averaged surface brightness  are about two 
magnitudes deeper than the previous photometry and about twice as extended (see Fig.~\ref{conf}).

Looking at FCC167, adopted as illustrative example, the $r$-band SB profile extends out to 10~arcmin ($\sim52$~kpc) from the center, which is about $10 R_e$, and down to $\mu_r = 29 \pm 1$~mag/arcsec$^2$ (see lower-right panel of Fig.~\ref{FCC167_image}). 
The $r$-band SB profile was compared with  the light profile from \citet{Caon1994} in the 
B band\footnote{Following the colour transformation by \citet{Fukugita1996}, where the average $g-r=0.59$ color for FCC~167 (see Tab.~\ref{Mtot_tab}) is taken into account to estimate 
the color term, we derived $B-r=1.33$ and the B band profile is shifted to the $r$ band, accordingly.}, 
which is not azimuthally averaged but  was derived along the major axis of the galaxy. This causes the small difference at $1 \le R \le 3$~arcmin, where the B-band profile has a more pronounced "bump". This feature at $ R \simeq 2$~arcmin is reasonably due to the bright knots detected along the major axis of the galaxy from the high-frequency residual image\footnote{A Gaussian filter of the $r$-band VST image of FCC~167 given in
Fig.~\ref{FCC167_image} was performed by using the IRAF task FMEDIAN. The ratio between the original
and filtered image gave the high-frequency residual image shown in the top-right panel of Fig.~\ref{FCC167_image}.} 
(see right panel of Fig.~\ref{FCC167_image}) and which are smoothed out in the azimuthally averaged profiles.
Except for this, the agreement is satisfactory out to 4~arcmin, the outermost measured SB value from previous photometry.
The $r$-band SB profile from the VST data is  two times more extended and about four magnitudes deeper then that from \citet{Caon1994} in the B band.

\begin{figure*}
   \centering
   \includegraphics[width=7.7cm]{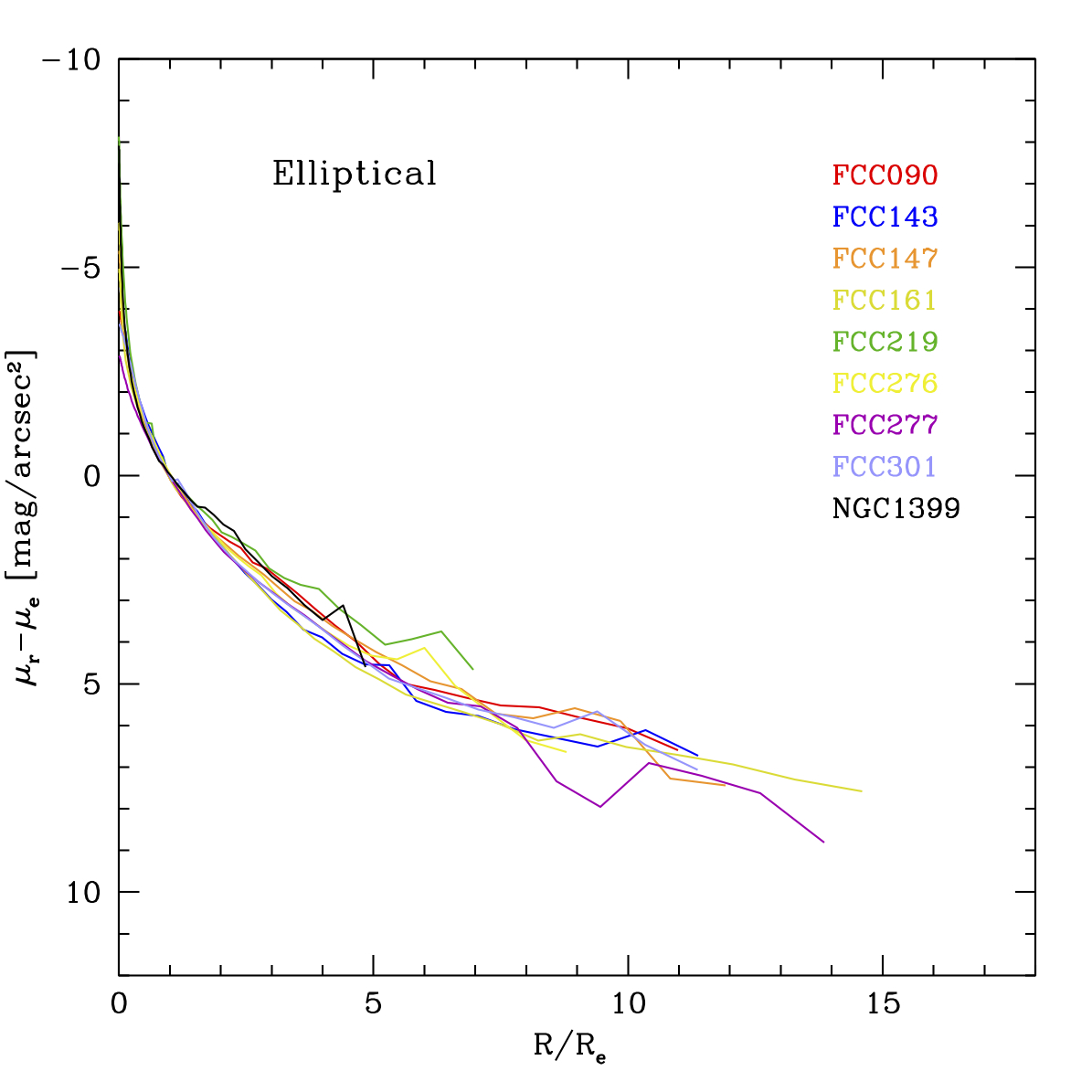}
   \includegraphics[width=7.7cm]{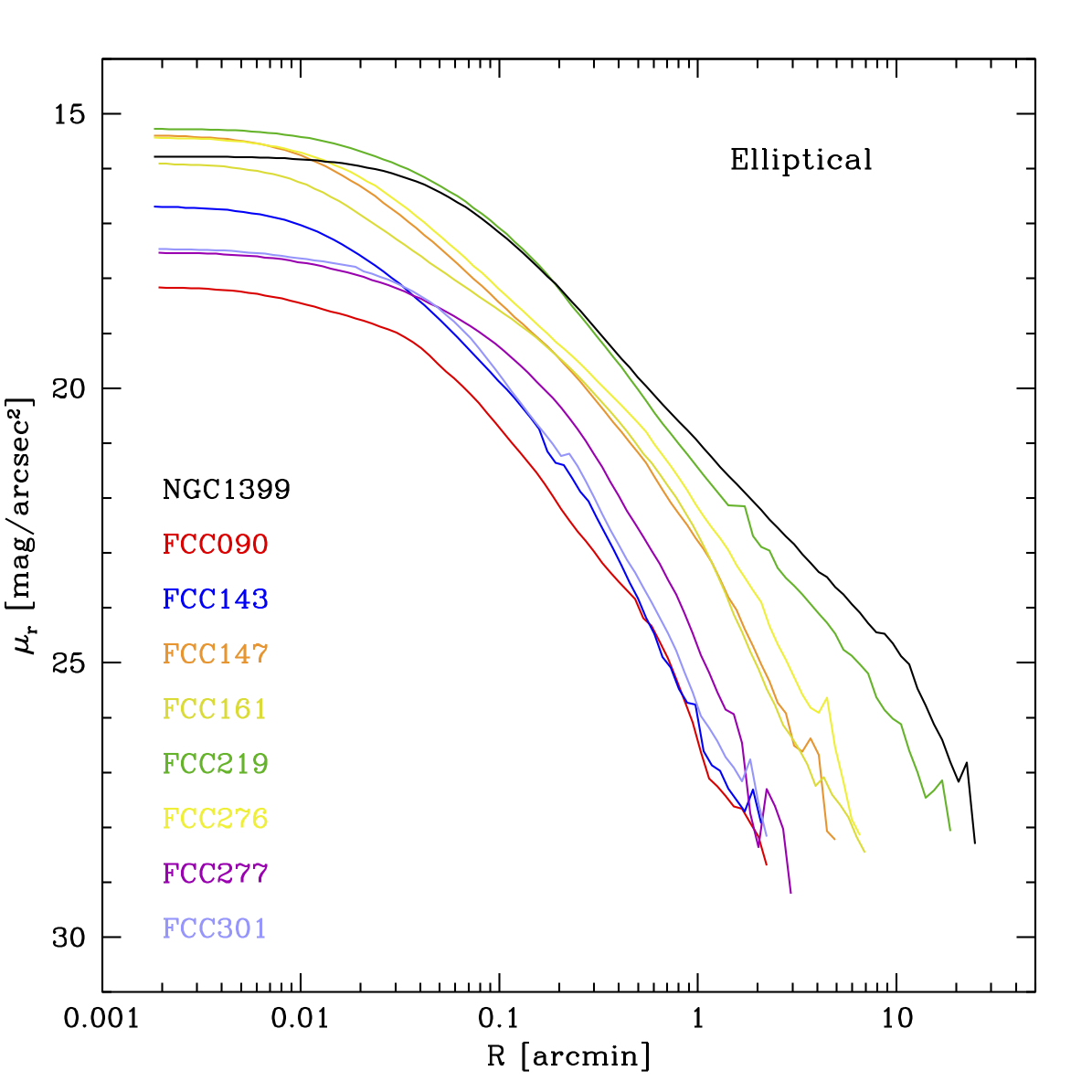}
   \includegraphics[width=7.7cm]{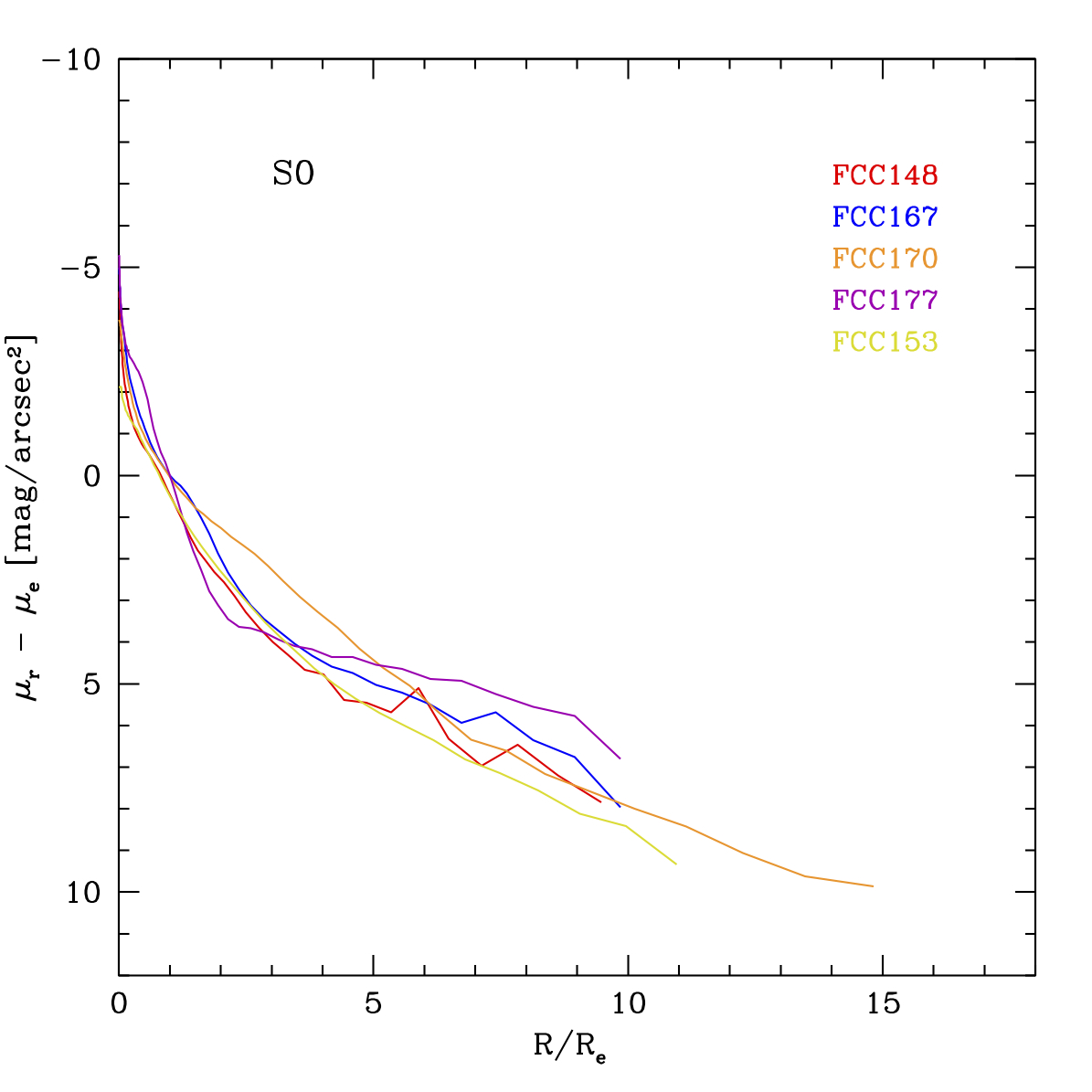}
   \includegraphics[width=7.7cm]{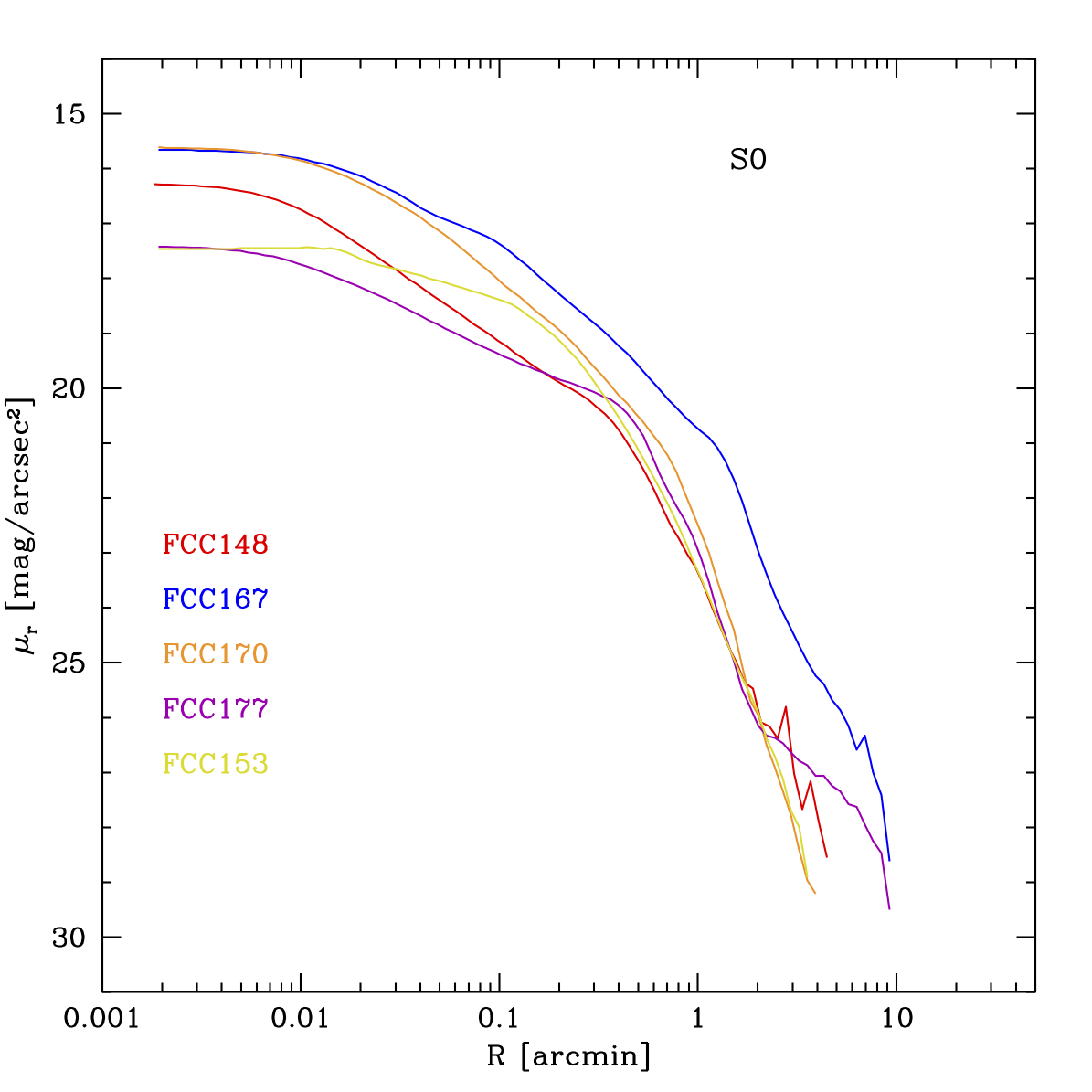}
   \includegraphics[width=7.7cm]{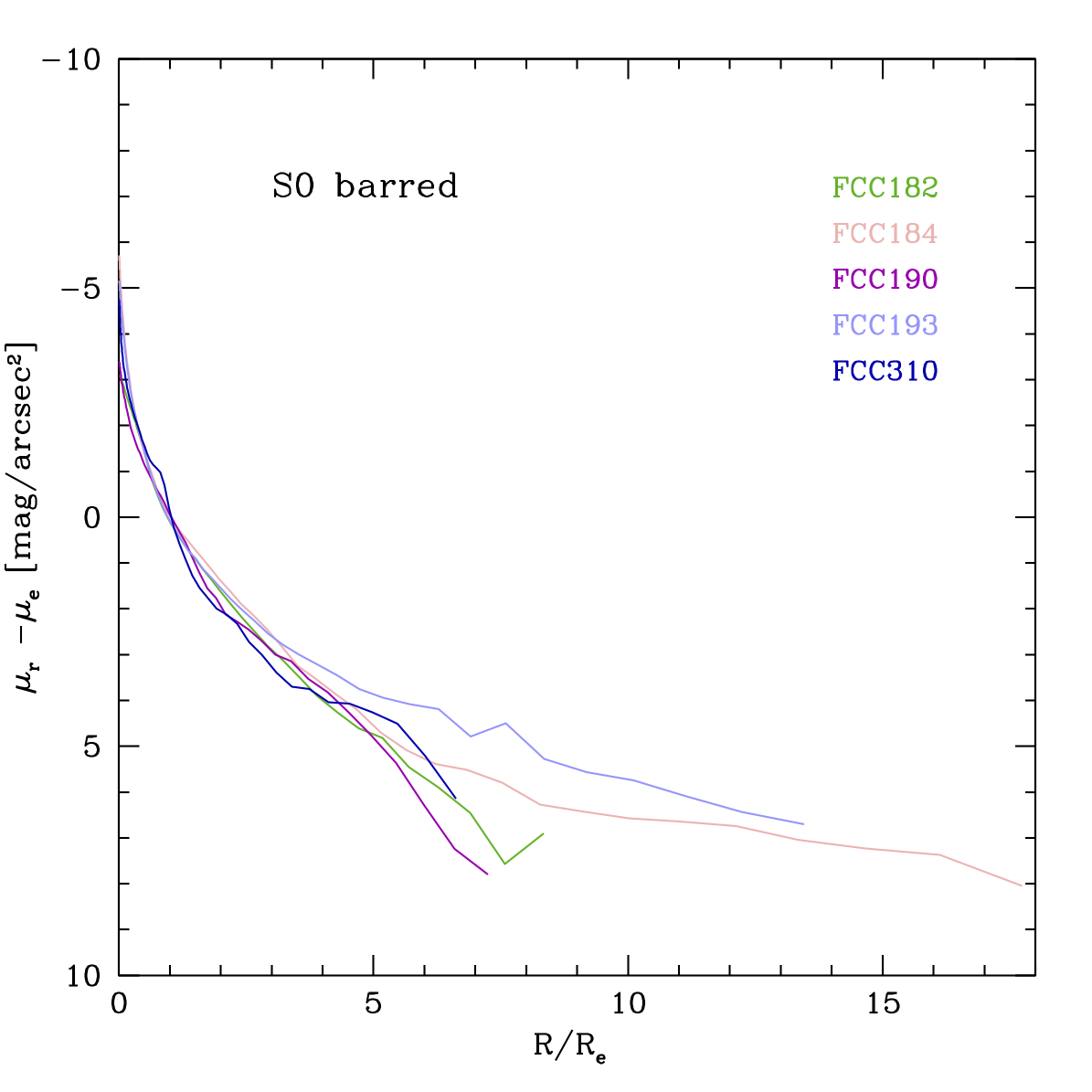}
     \includegraphics[width=7.7cm]{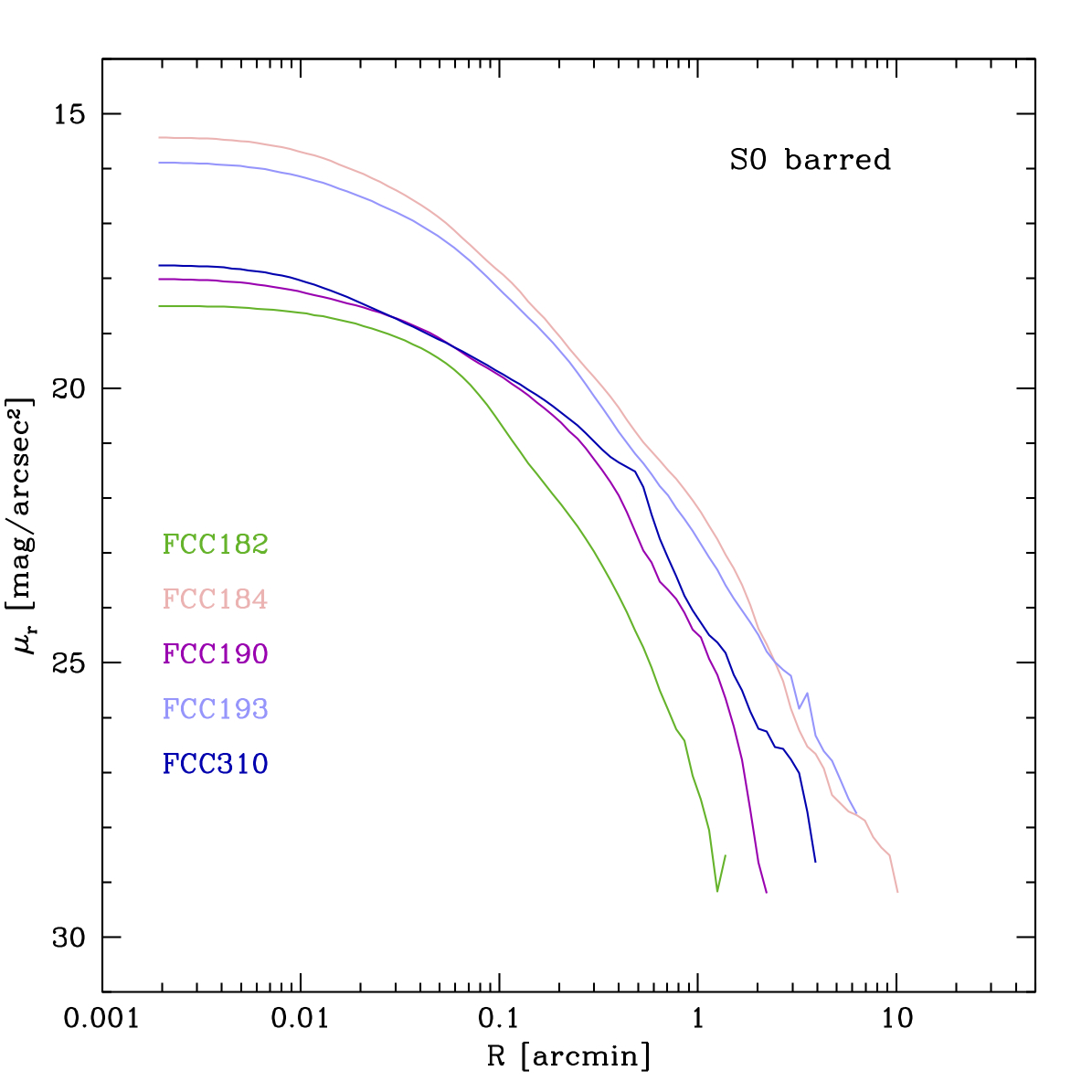}
    \caption{ Azimuthally averaged surface brightness profiles in the $r$ band for elliptical galaxies (top panels), S0s (middle panels) and barred S0s (bottom panels). In the left panels they are shown in linear scale as function of $R/R_e$ where $R_e$ is the effective radius given in Tab.~\ref{mag}. In the right panels they are shown in logarithmic scale.}
         \label{prof_all}
   \end{figure*}

\begin{figure*}
   \includegraphics[width=9cm]{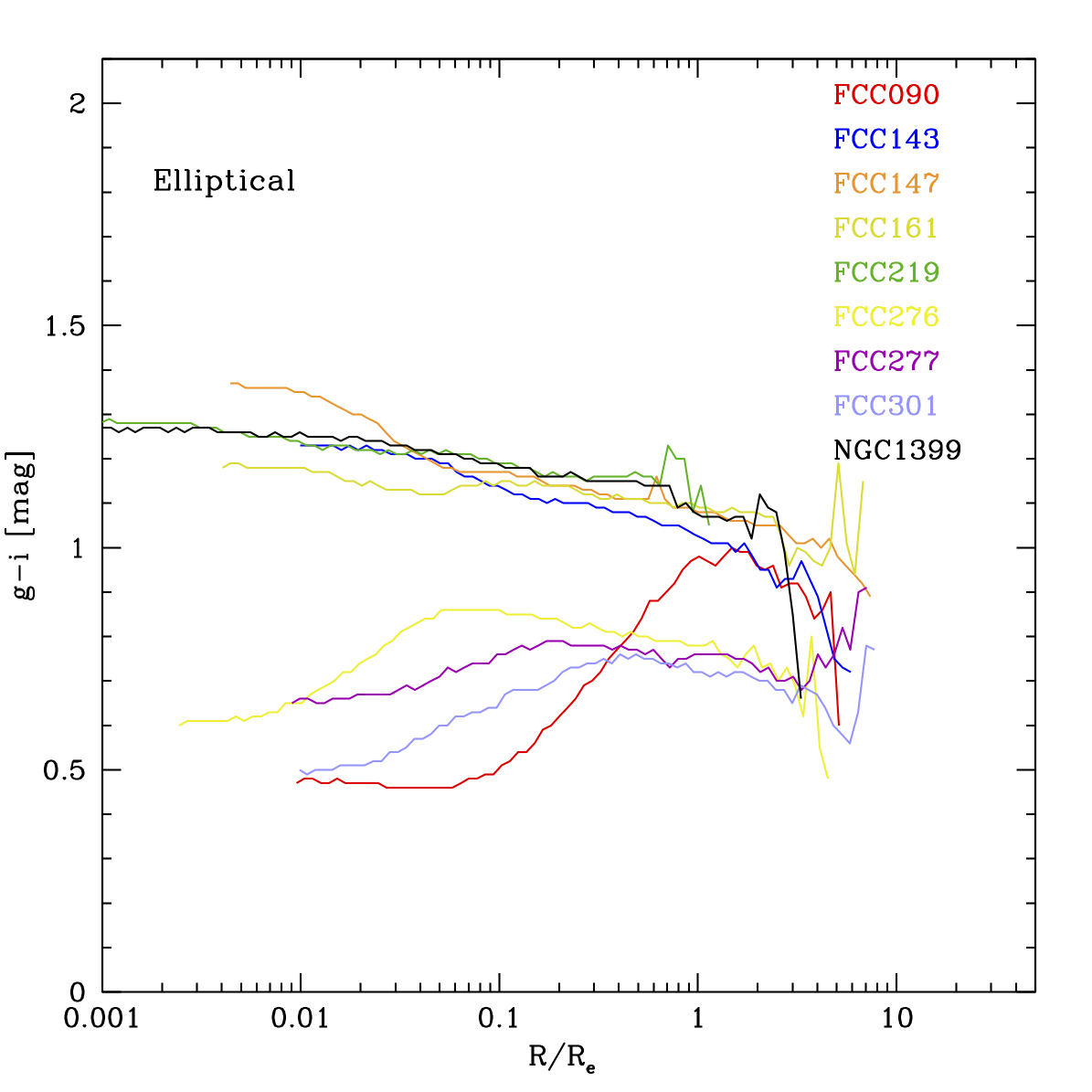}
   \includegraphics[width=9cm]{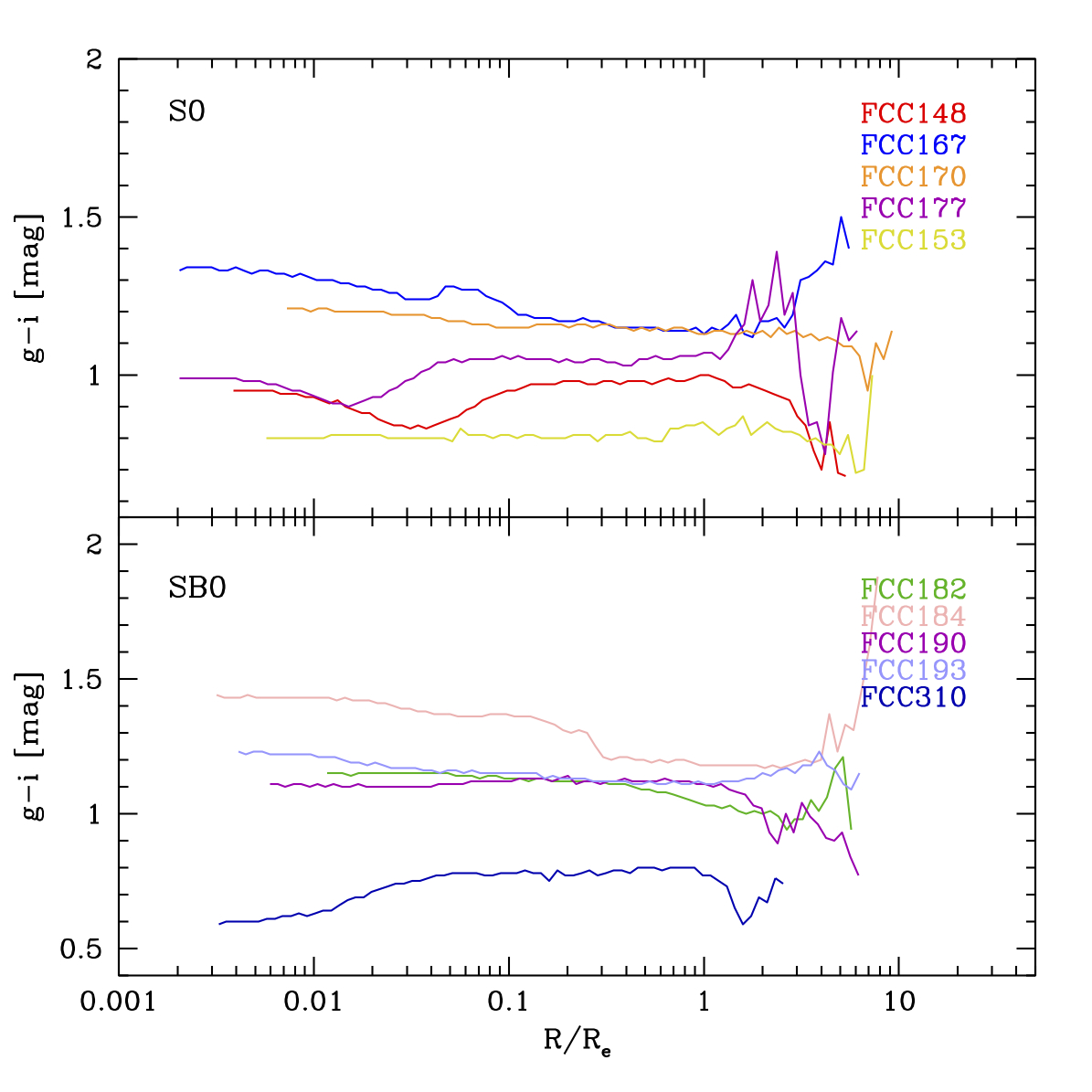}
      \caption{ $g-i$ color profiles for elliptical galaxies (left panel) and normal S0s (top right panel) and barred S0s (bottom right panel) inside the virial radius of the Fornax cluster.}
         \label{profcol_all}
   \end{figure*}

{  The SB profiles for all ETGs of the sample in the $r$ band are shown in Fig.~\ref{prof_all}. As for FCC167, all of them extend out to $10- 15 R_e$ and down to $\sim 8 \mu_e$. 
The shape of the SB profiles appears similar for all the elliptical galaxies (see top panels of Fig.~\ref{prof_all}), including a shallower decline at larger radii ($R\geq5R_e$), which resembles the typical behaviour of the outer stellar envelope as observed in other ETGs \citep{Seigar2007,Spavone2017,Iodice2016}. 
Most of the S0 galaxies have a Type~II profile or a composite Type~II + Type~III profile (see middle and lower panels of Fig.~\ref{prof_all}), where one or two {\it breaks} are observed in the SB profiles \citep{Erwin2008}.
According to \citet{Erwin2008}, the inner break in the SB profile is due to the presence of a bar. 
The outer break could be due to a truncation of the disk or indicates the presence of an outer additional component, like the stellar halo.
A quantitative analysis of the SB profiles for all galaxies of the sample, by using a multi-component fit to estimate the scale lengths and total luminosity of each 
galaxy component (as bulge, bar, disk and stellar halo) will be presented in a forthcoming paper (Spavone et al. in preparation).

Fig.~\ref{profcol_all} shows the azimuthally averaged $g-i$ color profiles: they are mapped out to $5 - 10 R_e$ with an error less than $10\%$. 
A color gradient is evident for five elliptical galaxies of the sample (FCC~143, FCC~147, FCC~161, FCC~219, FCC~213), all of them are located in the core of the 
cluster (see Fig.~\ref{mosaic_g}). For the remaining four elliptical galaxies (FCC~090, FCC~276, FCC~277, FCC~301), which are at larger cluster-centric radius (R$
\geq 0.8-1.0$~deg),  the color profiles show a dip in  their inner regions, where colors are bluer ($g-i \sim0.5$~mag), and they remain almost constant at larger radii. 
For all these galaxies, the dip starts at a larger radii ($\sim 1.8 - 6$ arcsec) than the regions affected by the seeing ($\sim 1$ arcsec).    
There are no other comparable estimates of the color distribution available in the literature for the ETGs in the Fornax cluster. 


}

\section{Results}\label{overview}

\begin{figure*}
   \centering
   \includegraphics[width=8.5cm]{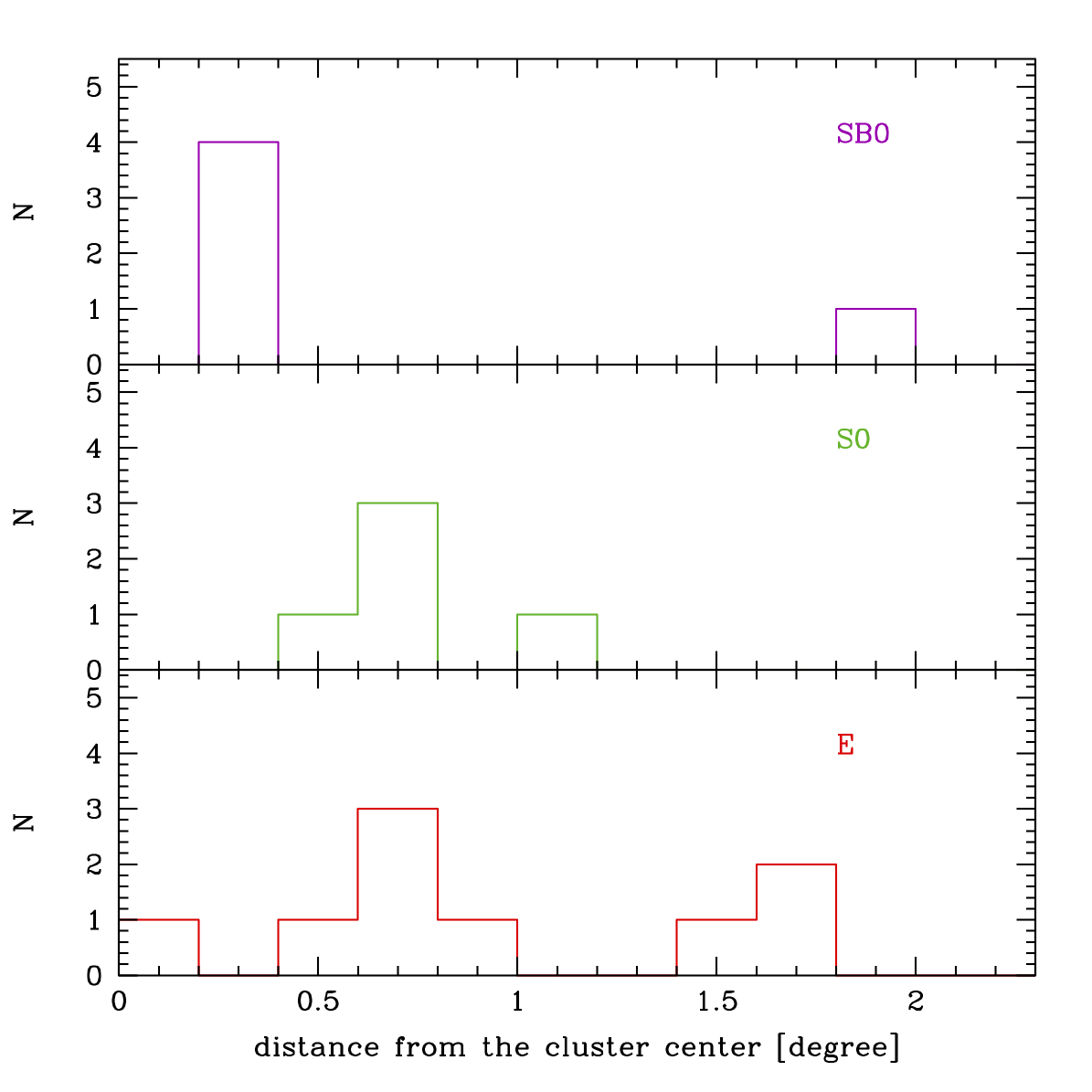}
   \includegraphics[width=8.5cm]{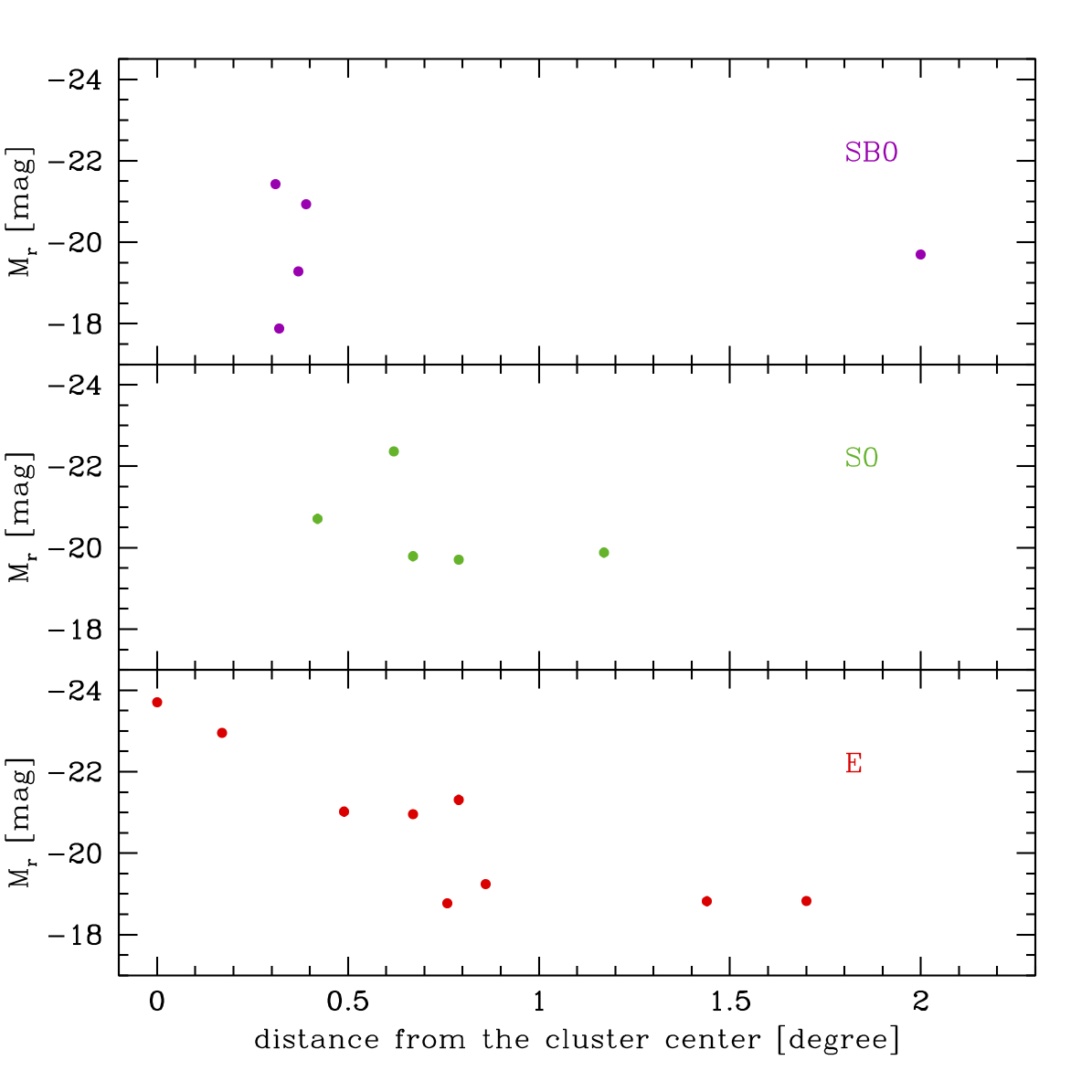}
      \caption{ Distribution (left panel) and absolute magnitude in the $r$ band (right panel) of ETGs  (barred S0 in the top panel, normal S0 and elliptical galaxies in the middle and bottom panels respectively) inside the virial radius as function of the projected distance from the cluster center. }
         \label{type}
   \end{figure*}

\begin{figure}
   \centering
   \includegraphics[width=9cm]{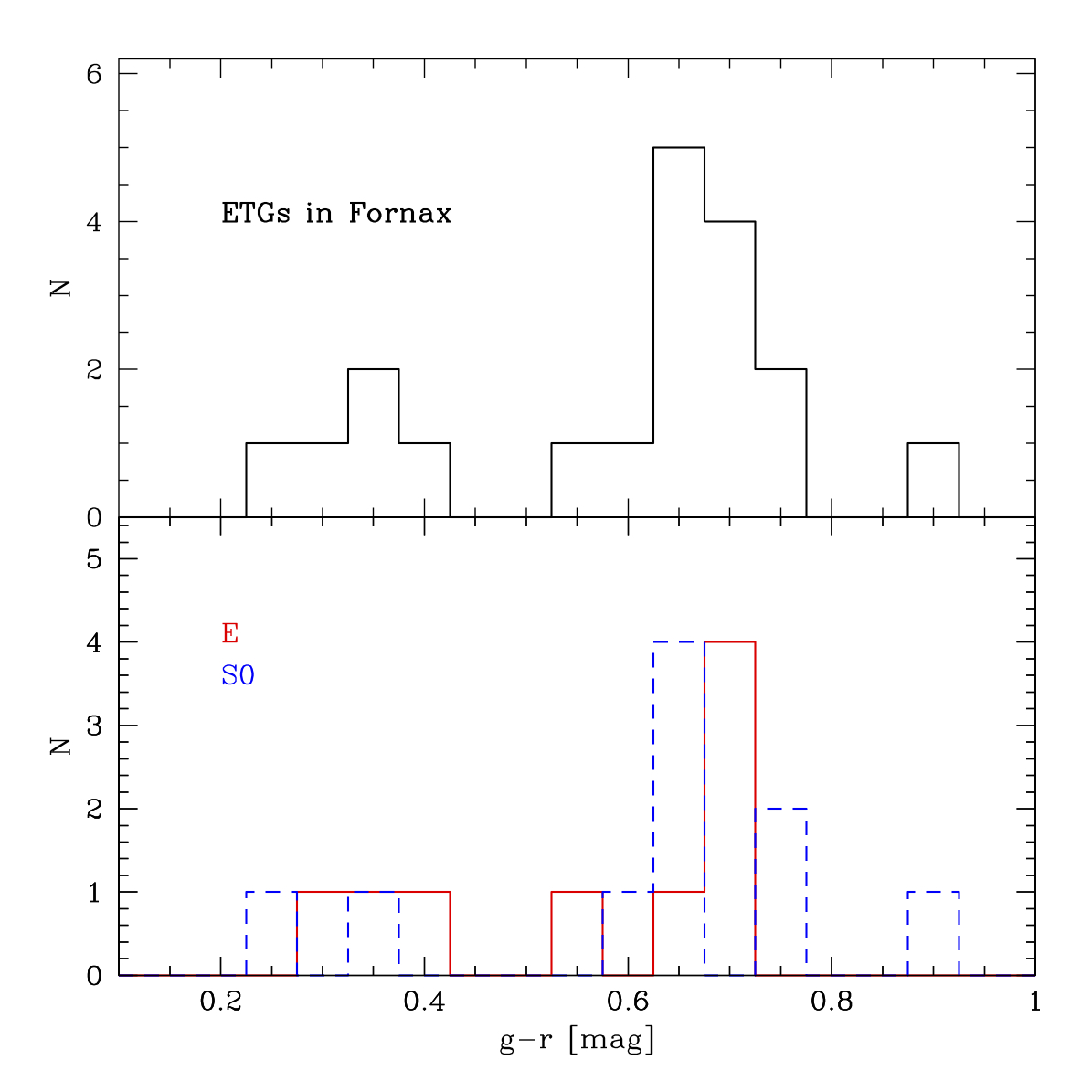}
      \caption{  $g-r$ average color distribution of ETGs inside the virial radius of the Fornax cluster. In the bottom panel is plotted the $g-r$ average color for elliptical and S0 galaxies.}
         \label{col_ETG}
   \end{figure}

\begin{figure*}
   \centering
   \includegraphics[width=9cm]{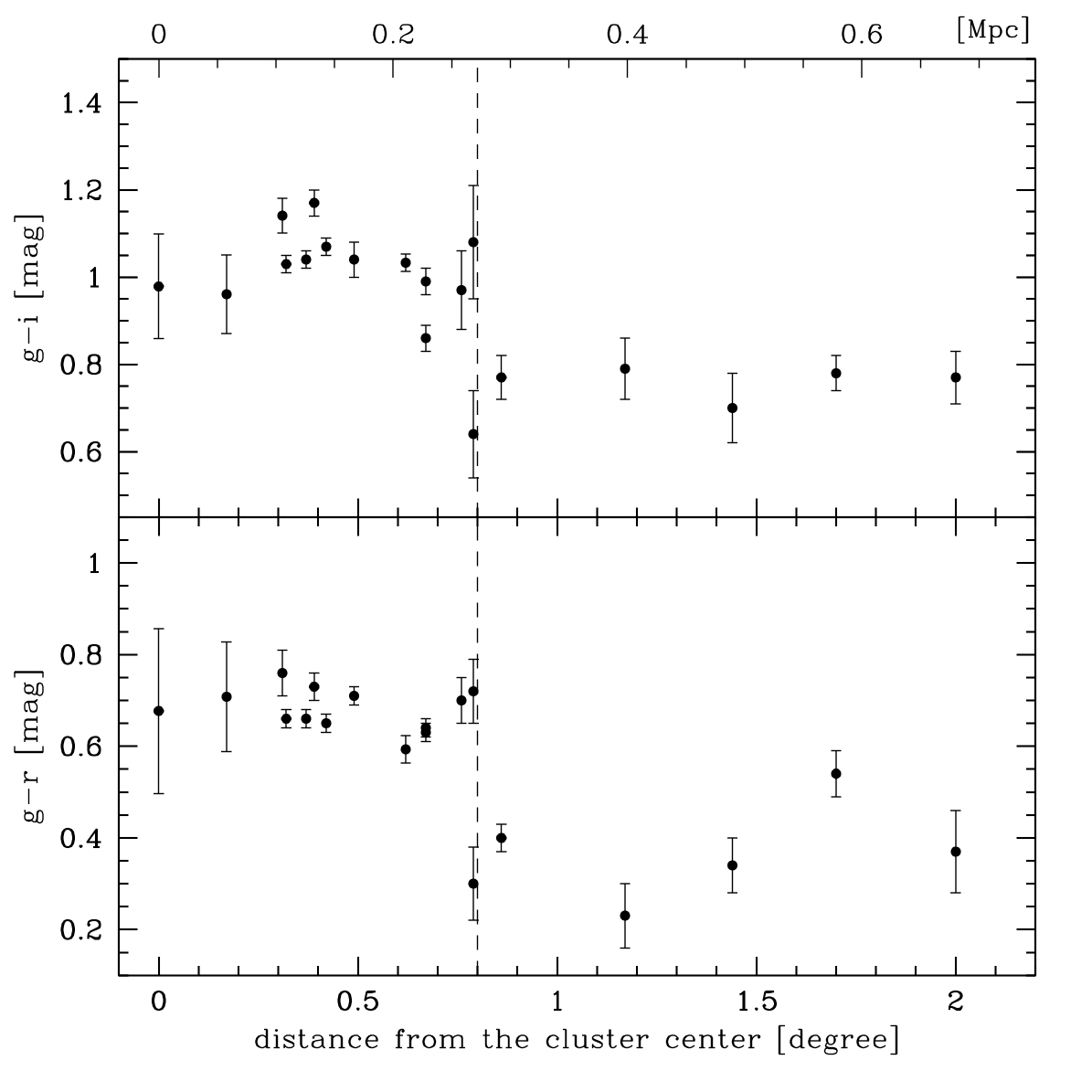}
   \includegraphics[width=9cm]{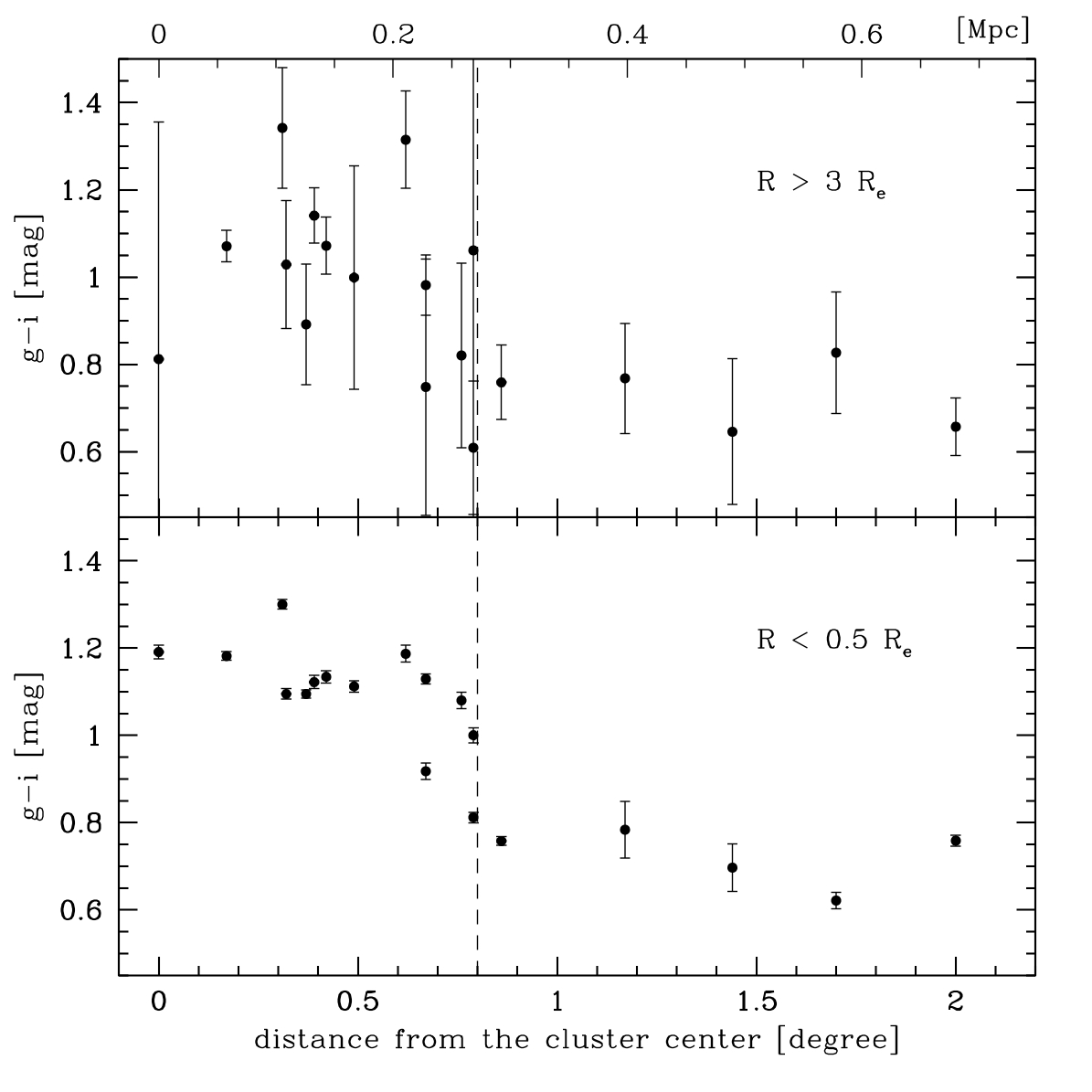}
      \caption{ {\it Left panel -} $g-r$ (bottom panel) and $g-i$ (top panel) average colors of ETGs inside the virial radius of the Fornax cluster as function of the projected distance from the cluster center. {\it Right panel -} $g-i$ average colors derived in the inner parts, for $R\leq0.5R_e$ (lower panel), and in the galaxy outskirts at $R\geq3R_e$ (top panel). The dashed line in each panel indicates the transition radius from the high-density to low-density regions of the cluster, at R=0.8 degree = 0.27~Mpc ($\sim0.4 R_{vir}$). }
         \label{col_radius}
   \end{figure*}

\begin{figure}
   \centering
   \includegraphics[width=9cm]{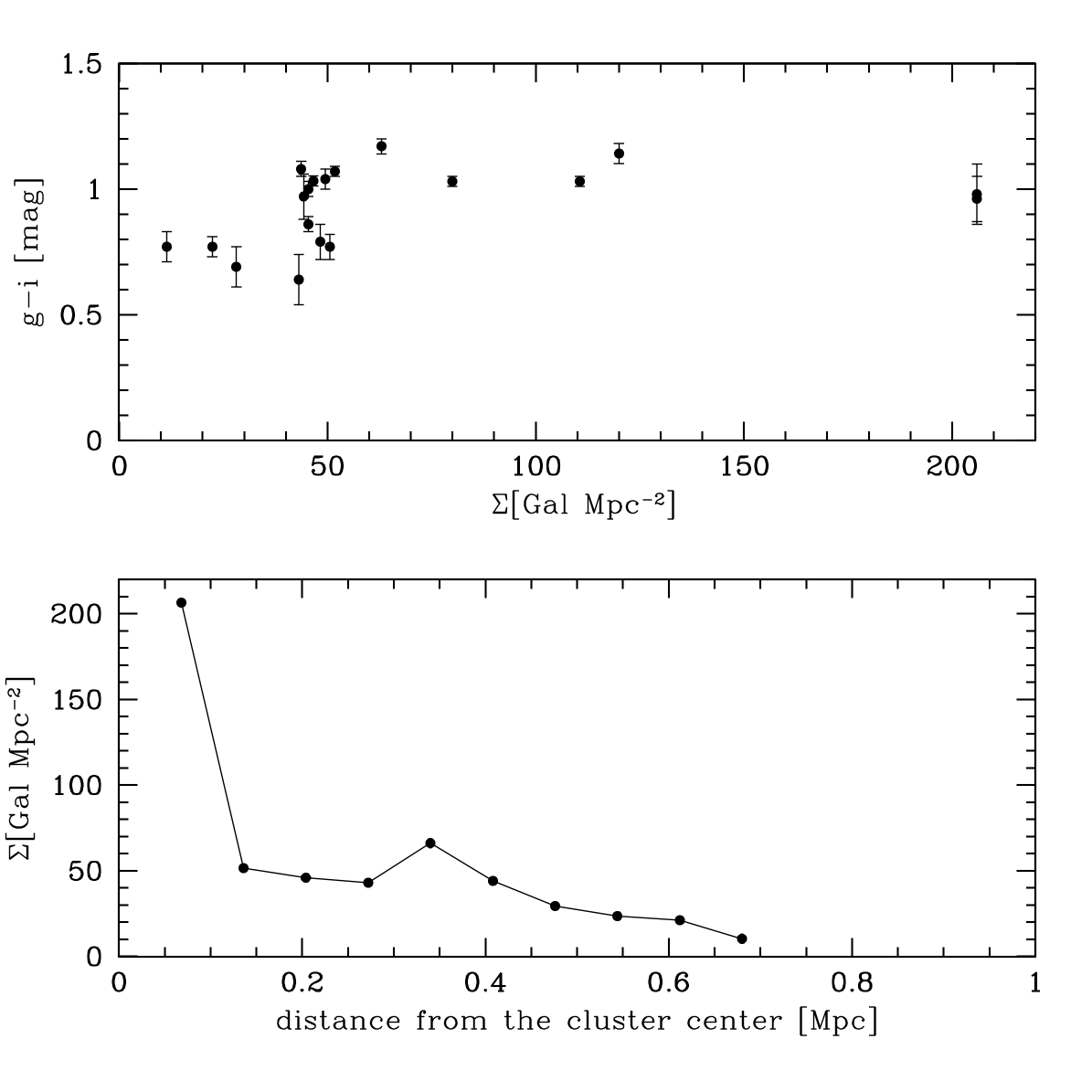}
      \caption{ Color-density diagram (top) of the  ETGs inside the virial radius of the Fornax cluster as function of the cluster 
density $\Sigma$. The cluster density as as function of the projected distance from the cluster center in Mpc is shown in the lower panel. This is derived in circular annuli with bin of 0.2 degrees (out to 2 degrees), accounting for all cluster members given by \citet[][]{Ferguson1989}. }
         \label{density}
   \end{figure}

\begin{figure}
   \centering
   \includegraphics[width=10cm]{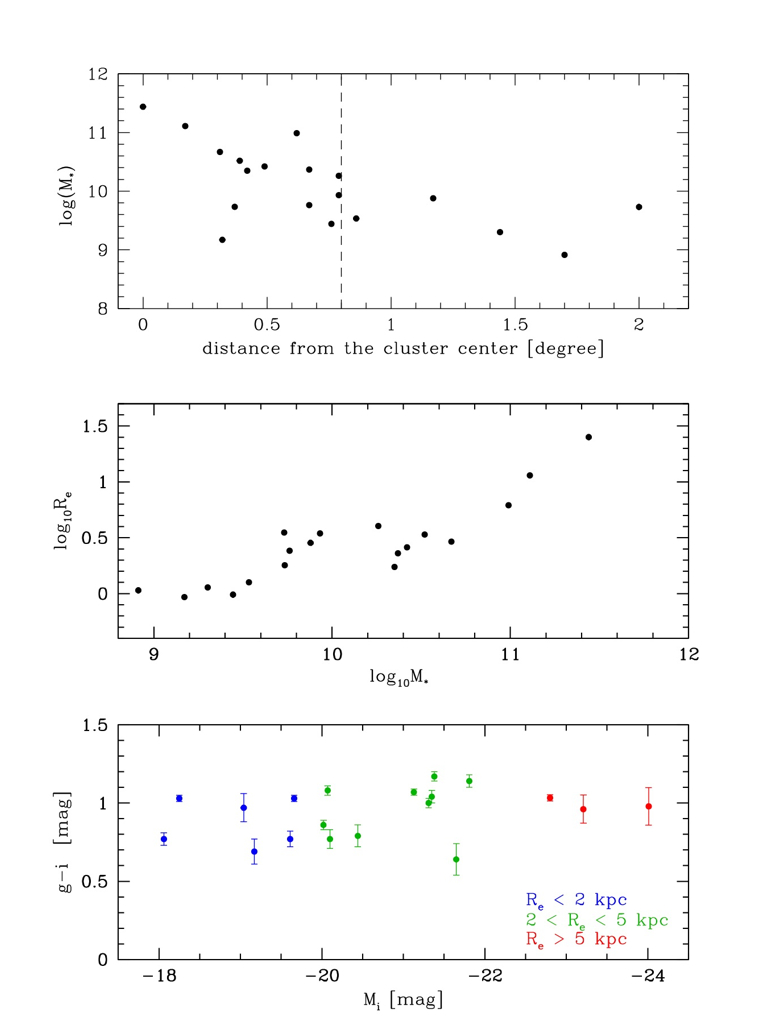}
      \caption{  Mass-size (middle panel) and color-magnitude (lower panel) relations for the ETGs inside the virial radius of the Fornax cluster. Different colors are for galaxies in different ranges of the effective radius (values are listed in Tab.~\ref{Mtot_tab}). In the top panel is shown the stellar mass as function of the cluster-centric radius. The dashed vertical line corresponds to the transition radius ~0.8 degree from the high-density to the low-density regions of the cluster.
 }
         \label{mass_radius}
   \end{figure}

\begin{figure*}
   \centering
   \includegraphics[width=\hsize]{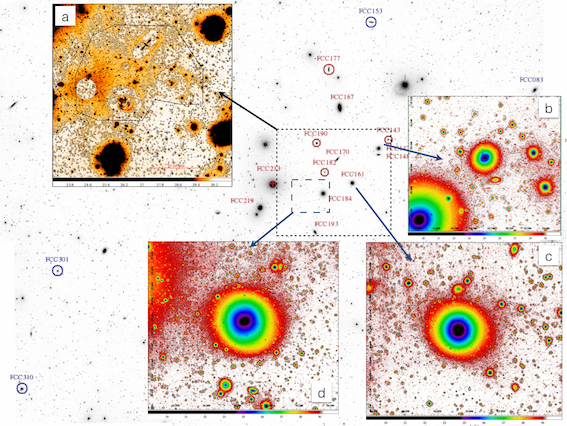}
      \caption{   Diffuse intra-cluster light in the Fornax cluster. The intra-cluster light detected in the core of the cluster, by \citet{Iodice2017b}, is shown 
      in the panel {\it a} (on the top-left). The asymmetric stellar halos in FCC~184 and FCC~161 are shown in panel  {\it c} (lower-right) and {\it d} (lower-left). 
      The panel {\it d} also shows the stellar bridge between NGC~1399 and FCC~184 detected by \citet{Iodice2016}. The panel {\it b} (top-right) shows the diffuse light detected between FCC~147 and FCC~143, on the North-West side of the cluster.}
         \label{ICL}
   \end{figure*}

{  
In this section, we give a global comprehensive view of the main properties for all bright ETGs members inside the virial radius of the cluster, 
as function of the environment. 
We focus, in particular, on  the analysis of the deep photometry and large covered area of FDS.
Results are discussed in Sec.~\ref{concl}.

According to the pioneering studies of \citet{Ferguson1988}, half of the bright ETGs inside the virial radius of the  Fornax cluster ($\sim 0.7$~Mpc) are elliptical galaxies 
and half of the lenticular (S0) galaxies are barred (see Tab.~\ref{galaxies}). 
Looking at the FDS mosaic in this area (shown in Fig.~\ref{mosaic_g}), the ETGs  are not homogeneously distributed within the 
core of the cluster:  most of the ETGs are located on the West side of the core, concentrated along a "stripe" in the North-South direction. 
All S0 galaxies are inside 1.2~degree ($\sim 0.4$~Mpc) from the 
core of the cluster (see left panel of Fig.~\ref{type}) and only on the West side (see Fig.~\ref{mosaic_g}). 
Except for FCC310, which is at $R\sim R_{vir}$, all barred galaxies are  very close to the core of the cluster, projected inside the stellar halo of NGC~1399 (i.e. $\sim0.5$~degree),  on the West side. 
As the total magnitudes derived from FDS confirm, the most luminous elliptical galaxies are close to the cluster 
center (see the lower-right panel of Fig.~\ref{type}). Such a trend is not obvious for S0 galaxies (see the middle-right panel of Fig.~\ref{type}) and it seems completely 
absent for the barred S0 galaxies (see the top-right panel of Fig.~\ref{type}).

The $g-r$ average colors of the ETGs are in the range $0.5-0.8$~mag, with a peak around $g-r\sim0.7$~mag, 
and there is no significant difference in colors between ellipticals and S0s (see Fig.~\ref{col_ETG}). 
Both $g-r$ and $g-i$ colors are almost constant ($g-r\sim0.7$~mag and $g-i\sim1$~mag) out to $\sim1$~degree from the core. 
At larger cluster-centric distances ETGs tend to be bluer, with $g-r\sim0.2-0.4$~mag and $g-i\sim0.6-0.8$~mag (see the left panel of Fig.~\ref{col_radius}). 
The color-density diagram shows that the bluest galaxies are in the lower density regions of the cluster (see top panel of Fig.~\ref{density}), 
where $\Sigma \leq 40$~Gal/Mpc$^2$ at a distance larger than 0.8 degree, i.e. $\sim0.3$ Mpc ($\sim0.4 R_{vir}$) from the cluster center. 
The bluest galaxies are the less luminous ($M_i\leq -20$~mag) and less massive objects ($M_*\leq 6.3 \times 10^9$~M$_{\odot}$) of the sample 
(see the mass-size and color-magnitude relations in  Fig.~\ref{mass_radius}). The massive and reddest galaxies are located in the high-density 
region of the cluster (see the top panel in Fig.~\ref{mass_radius}), at  $R\leq0.3$ Mpc ($\sim0.4 R_{vir}$) from the cluster center. 
By comparing the X-ray emission with the color distribution of ETGs inside the cluster (see Fig.~\ref{mosaic_g}), it clearly appears that     
the transition from red to blue colors (at about 0.3 Mpc) corresponds  to a decrease in the hot gas density.

\subsection{A deep look at the Fornax cluster from FDS}\label{halos}

{\it Signs of interaction in the diffuse light -} 
The deep observations, the large covered area and the optimised observing strategy of FDS allow, for the first time, 
to detect and analyse the faintest structures in the galaxy outskirts and in the intra-cluster area.
Fig.~\ref{ICL} shows which are the faint features detected inside the virial radius of the cluster and where they are located.

In previous studies on FDS we detected a faint ($\mu_g \sim 29-30$~mag~arcsec$^{-2}$) stellar bridge, about 5~arcmin long ($\sim 29$~kpc), 
in the intracluster region, on the west side of NGC~1399 and towards FCC~184 \citep{Iodice2016}. 
An over density of blue GCs, also based on FDS data, was found in the same region \citep{Dabrusco2016}, 
which confirms the  ongoing interaction between the two galaxies, where the outer envelope of NGC1387 on its east side is stripped away \citep[see also][]{Bassino2006}.

On the west side of the cluster core, we detected several patches of diffuse light in the intra-cluster region \citep{Iodice2017b}.
These features are very faint ($\mu_r \sim 28 - 29$~mag/arcsec$^2$ in the $r$ band) and extend out to 250 kpc within the core.
The most prominent  is concentrated between the three bright galaxies in the core, 
FCC~184, FCC~161 and FCC~170. The spatial distribution of the blue GCs coincides with the intra-cluster patches of light, 
suggesting that these regions in the core of the cluster are populated by intra-cluster material.

In this work, we have explored a wider region of the cluster at the faintest magnitude levels and we detected a new, quite faint, bridge of light 
$\mu_r \simeq 29.5$~mag/arcsec$^2$ between the compact elliptical galaxy FCC~143 and the luminous and larger elliptical  FCC~147 
(see Fig.~\ref{ICL} and also the top-left panel of Fig.~\ref{FCC143}), located on West side of the core. 
The two galaxies are quite close in space \citep[distance differs by 0.3 Mpc][]{Blakeslee2009},  therefore they are likely going through an interaction.

By analysing the outskirts of each galaxy of the sample (see Appendix.~\ref{note}), i.e. at $\mu_g \geq 26$~mag/arcsec$^2$, 
we found that many of the bright ETGs in the high-density region on the West 
side of the cluster show an asymmetric stellar envelope, more elongated and twisted in one direction (like FCC~161, FCC~167, FCC~184 shown in Fig.~\ref{ICL}).
Far away from the cluster center, there are no other detectable faint structures in the intracluster space and the galaxy outskirts appear  more regular than those of 
galaxies in the high-density region. 

In summary, signs of interaction in the diffuse light are found on the West side of the cluster, 
where the galaxy density is highest.

\smallskip

{\it Colors in the galaxy outskirts -} 
Since the FDS data allow to map the color profiles out to $5 -10 R_e$, we derived the average $g-i$ colors in the inner and brightest regions of the galaxies, 
at $R\leq0.5R_e$ (excluding the seeing disk), and in the outskirts, at $R\geq3R_e$. 
We found that the color distribution shows the same trend observed for the average global colors, i.e. at larger cluster-centric distances the inner regions of ETGs, 
as well as the outskirts,  tend to be bluer (see the right panel of Fig.~\ref{col_radius}).  This is quite evident for colors inside $0.5R_e$, 
while  larger scatter and errors affect the  colors in the galaxy outskirts (at $R\geq3R_e$). 
In the high-density regions of the cluster, the galaxy outskirts are about 0.12 mag bluer than the inner and brightest parts, i.e. 
 $g-i=1.0\pm0.2$~mag at $R\geq3R_e$ and $g-i=1.12\pm0.09$~mag at $R\leq0.5R_e$ (see the right panel of Fig.~\ref{col_radius}).
In the low-density regions,  the difference is about 0.03 mag compared to the colors of the inner parts, even if it is 
still within the errors ($g-i=0.74\pm0.07$~mag at $R\leq0.5R_e$ and $g-i=0.71\pm0.08$~mag at $R\geq3R_e$).
The galaxy outskirt colors reflect differences in stellar populations as function of the environment, 
therefore a different process of mass assembly. This point is addressed in Sec.~\ref{concl}. 

\smallskip

{\it Flaring of the disks in the edge-on S0 galaxies -}
Mapping the light distribution beyond $\mu_r = 27$~mag/arcsec$^2$, FDS data have revealed the flaring of the disk 
in the outer and fainter regions of the three edge-on S0 galaxies, FCC~153, FCC~170  and FCC~177, 
all of them located on the North-West side of the cluster.
For the three galaxies, we derived  the two-dimensional model from the fit of the isophotes (see Sec.\ref{phot})  by 
using the IRAF task BMODEL. The residuals are obtained by subtracting the model from the galaxy image. 
For all  three galaxies, the results are shown in Fig.~\ref{fig:flares}.

The most extended and thickest flaring is  detected in FCC~170 (see the middle panels of Fig.~\ref{fig:flares}), since it reaches a distance of 
1.7 arcmin from the galaxy center and an height of 0.5 arcmin at $\mu_r\sim29$~mag/arcsec$^2$. 
The flaring in FCC~153 has a similar morphology and spans a comparable surface brightness range as that in FCC~170. 
It is less extended ($R\sim1.3$~arcmin) but it is as thick as that in  FCC~170, since the height is $\sim0.5$ arcmin at $\mu_r\sim29$~mag/arcsec$^2$ 
(see the top panels of Fig.~\ref{fig:flares}).
With a quite different shape and luminosity from the previous ones, the flaring in FCC~177 appears thin and more luminous close to the galaxy center, with $\mu_r
\sim23$~mag/arcsec$^2$ at $\sim0.5$ arcmin. At larger radii ($\sim1$ arcmin) at $\mu_r\sim27$~mag/arcsec$^2$, the maximum thickness is only $\sim0.2$ arcmin (see the lower panel of Fig.~\ref{fig:flares}).

The $g-i$ color maps, which are model-independent,  also show a flaring structure for the disk in each object (see Fig.~\ref{col_2}). 
For FCC~153 and FCC~177 they are quite evident from the color map, even if the flaring has different color distributions: 
FCC~153 shows bluer colors 
($g-i\sim0.8$~mag, see also Sec.~\ref{sec:fcc153}) at large radii, while FCC~177 shows a dip in the color distribution  ($g-i\sim0.9$~mag) in the regions of the 
light peak at $\sim0.5$ arcmin, and red colors ($g-i\sim1.1$~mag) at larger radii (see also Sec.~\ref{sec:fcc177}). 
The flaring in the disk of FCC~170 is less pronounced in the color map, which is dominated by a central red ($g-i\sim1.2$~mag) and thin disk. In the outer region it appears thicker and bluer  ($g-i\sim1$~mag, see also Sec.~\ref{sec:fcc170}). 

The two-dimensional stellar kinematics and stellar population analysis of FCC~170, from the Fornax3D (F3D) project with MUSE \citep{Sarzi2018}, 
reveal the presence of a fast-rotating and flaring thin disk and a slower rotating thick disk in this galaxy \citep{Pinna2018}. Preliminary results from the F3D survey for the other two galaxies, FCC~153 and FCC~177, also confirm the existence of a flaring disk (Pinna et al., in preparation).
Different morphology, luminosities and colors  suggest a different origin for the flaring of the disk in each galaxy (see discussion in Sec.~\ref{concl}). 

}

%
%

  \begin{figure*}
   \centering
   \includegraphics[width=17cm]{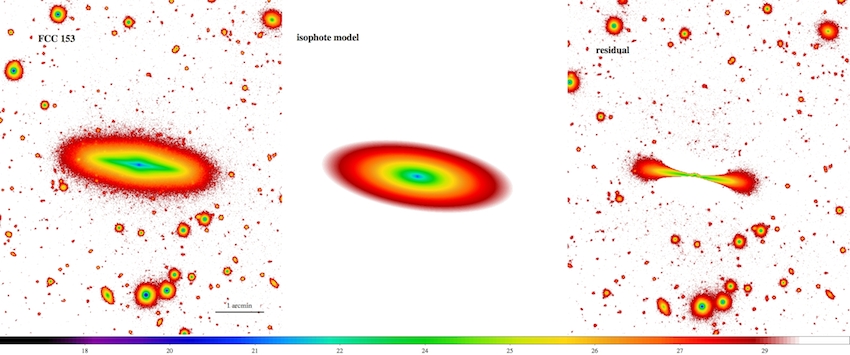}
   \includegraphics[width=17cm]{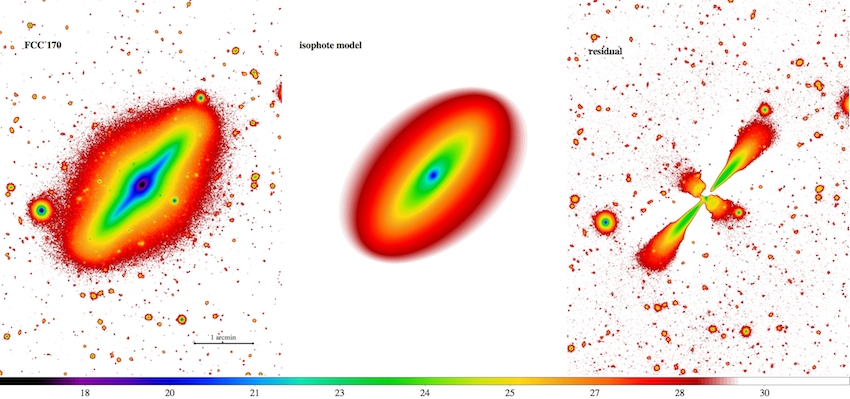}
    \includegraphics[width=17cm]{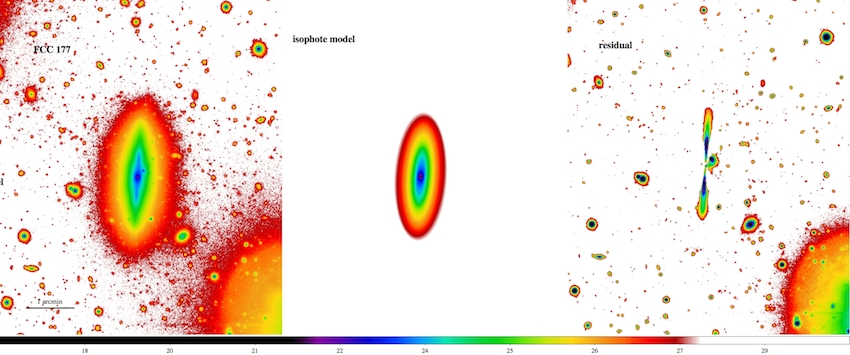}
     \caption{ Results from the isophote fits for the three S0 edge-on galaxies of the sample: FCC~153 (top panels), FCC~170 (middle panels) and FCC~177 (lower panels). For each galaxy, the $r$-band VST image is shown in the left panels, the two-dimensional models and residuals are shown in the middle and right panels, respectively. Images are in surface brightness levels with values reported in the color bar.}
         \label{fig:flares}
   \end{figure*}


\section{Discussion}\label{concl}

This paper is based on FDS data and it focuses on the bright ETGs galaxies ($m_B \leq 15$~mag)  inside the virial radius of the Fornax cluster 
($R_{vir} \sim 0.7$~Mpc). 
For all galaxies of the sample (given in Tab.~\ref{galaxies}), we analysed the light and color distribution in all OmegaCam filters, $ugri$, 
(see Sec.~\ref{phot}).  A detailed description of the results for each galaxy is provided in Appendix~\ref{note}. 
 The deep images and the color maps are given in  Appendix~\ref{VST_image} and Appendix~\ref{colormap}, respectively.
{  In this work we release the main data products resulting from the surface photometry\footnote{  The tables and the profiles 
resulting from the isophote fit (i.e. surface brightness, ellipticity and P.A. profiles) given in the present paper will be available in the CDS 
database. The reduced FDS images for the whole FDS area will be released by the team later in 2019.}  
  in all OmegaCam filters, $ugri$  
(i.e. total magnitudes, effective radii, integrated colors and M/L ratios) for all the bright ETGs galaxies in the sample (see Tab.~\ref{mag} and Tab.~\ref{Mtot_tab}).

 Taking advantage of the multi-band deep observations, the large field-of-view of OmegaCam@VST and the optimised observing strategy, 
the main contribution of this work to the previous science on the Fornax cluster is tracing with 
great detail the structure of galaxies and color gradients, out to the largest galactocentric radii  reached compared to  previous studies 
(up to about $10-15 R_e$) and beyond $\mu_r = 27$~mag/arcsec$^2$ ($\mu_B \geq 28$~mag/arcsec$^2$, see Fig.~\ref{conf}). 
Therefore, in addition to the previous study by \citet{Ferguson1988}, which presented the morphologies and luminosity function of the 
galaxies in the Fornax cluster inside a radius of 2.4~degrees, from FDS we are able to map the faint outskirts of galaxies and the intra-cluster 
space over a comparable area ($\sim  2$~degrees). 

The main results are summarised below.

\begin{enumerate}

\item All signs of interactions between galaxies, in the form of faint ($\mu_r\sim 29$~mag/arcsec$^2$) bridges of stars connecting galaxies, diffuse patches of ICL and disturbed galaxy envelopes, are located in the high-density region of the cluster ($\leq0.3$~Mpc) on the West side. 
In particular, we found that many of the bright ETGs in the high-density region on the west side of the cluster show 
an asymetric and elongated stellar envelope, that is twisted in one direction (like FCC~161, FCC~167, FCC~184, FCC~190).
In the intra-cluster region, the new feature detected from the FDS data is a very faint bridge of light $\mu_r \simeq 29.5$~mag/arcsec$^2$ between 
the compact elliptical galaxy FCC~143 and the luminous and larger elliptical  FCC~147.
This new intra-cluster patch of light is added to the other two detected in the core of the cluster in previous FDS studies. 
Fig.~\ref{ICL} shows most of these features.

\item The outskirts (at $R\geq 3R_e$) of galaxies located in the high density region of the cluster have redder colors ($g-i =1.0 \pm 0.2$~mag) than the outskirts of cluster members in the low density region ($g-i =0.71 \pm 0.08$~mag).

\item For the three edge-on S0 galaxies of the sample (FCC~153, FCC~170 and FCC~177) the disk is flaring in the outer and fainter regions, down to $\mu_r\sim29$~mag/arcsec$^2$.

\end{enumerate}

Below, we give a global comprehensive view of the galaxy structure and cluster properties based on the main results obtained by exploiting the FDS depth.


\subsection{The building-up of the Fornax cluster}

Inside the virial radius, the ETGs of the Fornax cluster are not uniformly distributed in space around the core: 
most of the ETGs are located on the West side of NGC~1399, concentrated along a "stripe" in  North-South direction (see Fig.~\ref{mosaic_g}). 
All S0 galaxies are inside 1.2~degree ($\sim 0.4$~Mpc) from the center of the cluster (see left panel of Fig.~\ref{type}) and only on the West side 
(see Fig.~\ref{mosaic_g}). 
Few elliptical galaxies are  found on the East side at larger projected distances ($\geq 1.5$ degree $\sim0.5$~Mpc), which are fainter and bluer than the elliptical 
galaxies in the core, and show nuclear substructure. 
This difference might suggest a different formation and/or evolution process for the galaxies in the  two regions of the cluster.

The W-NW sub-clump may result from the accretion of a group of galaxies during the gradual build-up of the cluster,  that induced this evident asymmetry in the 
spatial distribution of bright galaxies. Since groups of galaxies host different types of galaxies, 
the accretion of a group onto a cluster contributes to a large range of luminosities and morphologies \citep{Tully2008}. 
Galaxy interactions and merging are more frequent in groups than in clusters, due to a lower velocity dispersion in the former  \citep{Tremaine1981},
therefore the galaxy cluster  members accreted as part of groups can undergo a pre-processing  that have modified their structure and internal dynamics. 
\citet{Vija2013} found that a sudden burst of enhanced ram pressure occurs when the group passes the pericentre of the cluster.
Simulations show that the tidal interactions induce a decrease in rotational support and a thickening of the stellar structure \citep{Toomre1972,Bialas2015,Smith2015}.
This would explain the high fraction of early-type systems in W-NW sub-clump of the Fornax cluster. 

The different morphology and colors of the flaring observed in the three edge-on S0s of the sample (FCC~153, FCC~170 and FCC~177, see Sec.~\ref{halos})
and  the presence of a thick disk in FCC~170  \citep[see also][]{Pinna2018}, 
which is in the core of the sub-clump and close to the core of the cluster, could be reconciled with different kinds of interactions.
During the infalling of a group in a cluster, the field late-type galaxies could be swept up and interact for a short time with the galaxies in group before 
becoming cluster members \citep{Vija2013}. In this process, these galaxies experience ram pressure stripping and appear as S0 galaxies. 
{  \citet{Boselli2006a} also addressed the origin of lenticular galaxies induced by the gravitational interactions in infalling groups.  
Due to the interaction with the hot and dense intra-cluster medium,  the late-type galaxies, after several crossing times, were depleted of the 
gas supply, loose their angular momentum and a possible heating of the stellar disk. Therefore, they will result 
in disk dominated and quiescent galaxies, similar to S0s.
With the help of simulations, \citet{Boselli2006b} showed that the gas stripping in the disk galaxies, caused by the interaction with the 
intra-cluster medium, reflects on the color gradients. By studying the color profiles of the spiral galaxy NGC~4569 in the Virgo cluster,
they found an inversion in the color profiles that appear redder in outer disk and bluer toward the center.}
FCC~177, on the North in the Fornax cluster, could be one example of this mechanism, since it shows a different structure with respect to the other edge-on S0s in 
Fornax (see Sec.~\ref{halos} and Appendix~\ref{note}). 
This galaxy shows an extended thin disk ($R\leq 2$~arcmin and $\mu_r\sim 20-27$~mag/arcsec$^2$), bluer in the inner regions, thicker and redder in the outer parts 
(see Fig.~\ref{col_2}). FCC~177 could be the result of a early-type spiral galaxy that has lost the cold gas and is evolving into an S0 galaxy.

\subsection{The growth of bars}

We have reported that all barred galaxies are very close to the core of the cluster (i.e. $\sim0.5$~degree),  on the West side
(see Sec.~\ref{overview}). This is consistent with predictions from N-body simulations, which found that tidal interactions trigger bar  
formation in a cluster core, rather than in the outskirts or in isolation, causing therefore a larger fraction of barred galaxies toward the cluster center \citep{Lokas2016,Martinez2017}.
In a Virgo-like cluster, simulations show that a bar forms at the first pericenter passage of a disky galaxy and survive many giga years after the bar formation.
Moreover, a pronounced boxy/peanut shape could appear in those galaxies that undergo  stronger tidal forces \citep{Lokas2016,Martinez2017}. 
This would be the case of FCC~148 (see Fig.~\ref{sec:fcc148}).

\subsection{The region of tidal interactions}

From the observational side, the deep FDS data  further confirm that 
the bulk of the gravitational interactions between galaxies  happened on the W-NW sub-clump of the cluster. In fact, this is the only
region of the cluster, inside the virial radius, where the intra-cluster baryons (diffuse light and GCs) are found, i.e. the bridge between NGC~1399 
and FCC~184 \citep{Iodice2016}, the intra-cluster light and GCs between FCC~184, FCC~170 and FCC~161 \citep{Iodice2017b} and the new 
faint filaments between FCC~143 and FCC~147 (see Sec.~\ref{halos} and Fig.~\ref{ICL}). 
The  gravitational interactions could have also modified the structure of the galaxy outskirts and produced the intra-cluster baryons. 
Compared with simulations by \citet{Rudick2009}, the diffuse form observed for the ICL is consistent with the scenario 
where this component formed by stripped material from the outskirts of a galaxy in a close passage with the cD \citep{Iodice2017b}.
In this area of the cluster, the stellar envelope of some ETGs is asymmetric, appearing more elongated and twisted in one direction, while the outskirts 
of galaxies at larger distances from the cluster center have a more regular shape. 
\citet{Mastropietro2005} show how harassment can induce twists in the outer isophotes of dwarf galaxies. More massive and luminous galaxies, like the
elliptical galaxies in this region of the Fornax cluster, have a deeper potential well, therefore the stripping of stars by  harassment  implies even stronger 
tidal forces. 
A disturbed morphology in galaxy outskirts could also results from the ongoing accretion of smaller satellites.
Simulations on the mass accretion and stellar halo formation for different stellar masses ($10^{10} - 10^{13}$~M$_\odot$), 
show that the outskirts of galaxies (for $\mu_r \sim 27 - 31$~mag/arcsec$^2$) appear with a quite disturbed morphology and with an overall elongated shape \citep{Michel2010,Cooper2015,Monachesi2018}. The structure of the stellar envelope, as well as the shape of the SB profile, 
depends on the mass and numbers of the accreted progenitors. 
The ETGs showing asymmetric and diffuse envelope (FCC~161, FCC~167, FCC~184, see Sec.~\ref{halos} and Fig.~\ref{ICL}), 
are in range of stellar mass ($\sim 3 - 10$ $10^{10}$~M$_\odot$) comparable with simulations and they could still build up their envelope. 
As noticed by \citet{Iodice2017b}, a fraction of the ICL population in this region of the cluster could also come from lower-mass dwarf galaxies 
that are tidally disrupted in the potential well of the massive galaxies, which are therefore contributing to the mass assembly in their halo. 
This is further supported by a recent study from \citet{Venhola2017}, based on FDS data, that found a drop in the number density of LSB galaxies 
at cluster-centric distances smaller than $\sim180$~kpc.

\subsection{The color segregation inside the cluster}

We have found that the galaxies in the W-NW sub-clump of the cluster have redder colors than galaxies at larger cluster-centric radii.
This sub-clump of reddest galaxies is the major contribution to the galaxy density ($\Sigma \geq 40$~Gal Mpc$^{-2}$, see Fig.~\ref{density}), 
inside 0.3~Mpc ($\sim0.5$~R$_{vir}$), 
and they are the most luminous (M$_i \leq -20$~mag) and massive M$_* \geq 10^{10}$~M$_{\odot}$ 
objects of the sample (see Fig.~\ref{mass_radius}).
Similar color segregation was found by \citet{Kodama2001} for the Abell~851 cluster at $z\sim0.4$ and also by \citet{Aguerri2018} for the nearby ($z\sim0.05$) cluster Abell~85, which is considered  a strong indication of environmental influences on the galaxy evolution.

The transition from redder to bluer ETGs, at a projected distance of $\sim0.8$~degree from the cluster center (see 
Fig.~\ref{col_radius}), coincides with the decrease of the X-ray emission (see Sec.~\ref{overview} and Fig.~\ref{mosaic_g}).
This  might confirm that the reddest and massive galaxies in the high-density region of the cluster, where the X-ray emission is still detected, 
have been depleted of their gas content by processes such as harassment or suffocation induced by ram-pressure stripping, 
so the star formation stopped earlier than for the galaxies in the low-density regions of the cluster. 
In fact the ROSAT data have shown that the Fornax X-ray halo is composed of several components whose centroids are offset with respect to the optical galaxies. 
This is likely due to the sloshing movement of NGC~1399 and other bright ellipticals within the central dark matter halo traced by the X-ray gas 
\citep{Paolillo2002,Sheardown2018} combined with the infall of the NGC~1316 group into the cluster, 
where the collisional component (e.g. the hot gas) is lagging behind the non-collisional ones (e.g. stars, GCs and galaxies) 
due to ram pressure effects.  These  results  seem confirmed  by  recent  XMM  observations presented in \citet{Su2017}.

The dependence of the colors with the local density suggests a decrease of the star formation rate in galaxies from low to high density regions of the 
cluster. According to the observational evidences given above, there is a strong indication that the star formation in galaxies in the W-NW sub-clump, which are the 
reddest objects, quenched due to external interactions (environmental quenching). The few galaxies in the low density regions of the cluster, uniformly distributed 
inside the virial radius, have bluer colors, this could mean that they contain, on average, a younger stellar population, i.e. they were able to continue forming stars.
Anyway, since the bluest galaxies in Fornax are also the least massive, the color segregation could also 
result from the mass-metallicity relation, where low-mass galaxies
 in the outer part of the cluster would have a blue color due to their lower metallicity \citep{Peng2010,Peng2015}.
This issue needs more investigations, in particular,  with the help of spectroscopic data, the metallicity and the age of the stellar populations can be derived for 
each galaxy of the sample in oder to study how they change as function of the cluster-centric radius.

\subsection{The mass assembly in the galaxy outskirts}

The deep FDS data allow us to map the colors out to the galaxy outskirts ($R\leq 5-10 R_e$) and we have found that they are red in the galaxies located 
in the high-density region of the cluster compared to those measured in the outskirts of the galaxies in the low-density regions. 
The differences in colors in the outskirts of galaxies can reflect differences in stellar populations as function of the environment, 
therefore a different process of mass assembly. 
Simulations show that galaxies with a low fraction of accreted mass in the stellar halo have a steeper metallicity gradient, while
a high-fraction of accreted mass tends to flatten the metallicity gradient \citep{Cook2016}. 
Therefore, the blue colors in the outskirts of the galaxies in the low-density regions of the cluster may suggest a lower fraction of the 
accreted mass with respect to the galaxies in the high-density regions. Also different progenitors can be responsible for differences in colors and metallicities:
in the dense regions of the Fornax cluster, the build-up of the stellar halos could result from the accretion of rather massive, 
metal-rich satellite galaxies, whereas blue colors in the outskirts only allow for contribution of minor mergers. 
The estimate of the stellar halo mass fraction in the bright ETGs of the Fornax cluster  
will be addressed and discussed in a forthcoming paper (Spavone et al. in preparation).
A direct measure of the age and metallicity can be done by using spectroscopic data available for all galaxies of the 
sample \citep{Sarzi2018} out to the regions of the galaxy outskirts. Results can be directly compared with theoretical predictions.

\subsection{How does Fornax compare with  nearby clusters?}

Compared to the other nearby ($< 20$~Mpc) massive  Virgo cluster, Fornax has generally been considered as classical example of a virialized cluster for its more  
regular galaxy distribution, high fraction of early-type systems and lower velocity dispersion  \citep{Grillmair1994,Jordan2007}.
The new deep observations of the Virgo cluster, which detected many other new patches of ICL and tidal streams between galaxies \citep{Mihos2017}, 
have confirmed that it is a dynamically young  and unrelaxed cluster.
The deep FDS data brought to light the previously unknown intra-cluster baryons (ICL and GCs), including streams of stars between galaxies.
These are all detected in the West side of the cluster, between the sub-clump and the cluster core. Outside this region, but 
inside the virial radius, the deep FDS data do not show any other 
significant sign of gravitational interactions. We suggest that the build-up of the cluster is still in process, with the W-NW group entering into the cluster, 
and how this mechanism is important in generating ICL.
Compared to Virgo, Fornax hosts $\sim 300$ galaxies with $B_T\sim18$ \citep{Ferguson1989}, which is an order of magnitude lower than the galaxy 
population in Virgo, mostly concentrated inside 0.3~Mpc from the core. 
Differently, in Virgo there are at least three main sub-groups dominated by the giant ETGs M87,  M84 and the south sub-clump around M49 \citep{Binggeli1987}. 
Therefore,  in Fornax we expect to find the peak of the gravitational interactions in the core, as detected in the FDS data.
In Virgo, the bulk of the projected ICL is found in the core, i.e. around M87 and in the adjacent region including M86/M84, and there is evidence 
for an extensive ICL component in the field around M49  \citep{Mihos2017}.
Like Fornax, also the Coma cluster was considered a virialized and evolved large-scale structure, given its spherical symmetry and compactness \citep{Zwicky1957,Kent1982,Mellier1988}. To date, Coma also appears in a more dynamical active phase.
As observed in Fornax, the ICL distribution is more elongated in one direction (toward East) and over a common envelope between the two giant galaxies in the core 
\citep{Jimenez2018}. It is dominated by younger/low-metallicity stars, suggesting that the this component could be formed by the stripping of stars from the galaxy 
outskirts and/or the disruption of dwarf galaxies during merging \citep[see][and references therein]{Jimenez2018}. 

The hot gas in the Fornax cluster is four times less massive than in Virgo \citep{Schindler1999,Paolillo2002,Scharf2005}. 
This suggests that the role of ram pressure stripping is lower in Fornax than in Virgo, as also found from predictions by \citet{Davies2013}, who estimated 
that the ram pressure stripping is about 16 times less important in Fornax than in Virgo.
FDS data have shown that the galaxies in the low density regions do not show detectable sign of tidal interactions as found in the cluster core, also the outskirts 
are quite regular in shape. Therefore,  in these regions, ram pressure stripping could have been the main process that stopped the star formation in the galaxy outskirts, 
while continued in the inner regions (see Sec~\ref{overview}).
The effect of ram pressure stripping on the Fornax cluster members is addressed in the analysis of the ALMA data \citep{Zabel2018} 
and in a forthcoming paper from the FDS data on the late-type galaxies (Raj et al., in preparation).


\section{Conclusions and future perspectives}

The outskirts of galaxies are the least bound, so they are most sensitive to being  tidally disturbed \citep{Smith2016}, as well as affected by the 
ram pressure, which acts on the gas from the outside inwards, first stripping the outer disk and later reaching the inner disk with increasing ram pressure.
Therefore, by studying the very faint outer parts of galaxies it is possible to map the regions where they are most sensitive to environmental effects. 
Inside the virial radius of the Fornax cluster, the deep FDS data allow the analysis of the light and color distribution of the ETGs down to 
the faintest magnitudes ever reached before, where the relics of the past gravitational interactions reside. 

Such a study has provided strong constraints for  the formation history of the Fornax cluster. 
It has provided strong constraints on the formation and evolution of the galaxies in the cluster and on the possible environmental mechanisms 
that have played a major role.
Observations suggest that the Fornax cluster is not completely relaxed inside the virial radius.
 The bulk of the gravitational interactions between galaxies happened in the W-NW core region of the 
   cluster, where most of the bright early-type galaxies are located and where the intra-cluster baryons (diffuse light and GCs) are found. 
 We suggest that W-NW sub-clump of galaxies results from the
accretion of a galaxy group which has modified 
   the structure of the galaxy outskirts (making asymmetric stellar halos) and have produced the intra-cluster 
   baryons (ICL and GCs), concentrated in this region of the cluster.

Future works on FDS data will extend the analysis presented in this paper to all galaxies in the FDS over $\sim 26$~degrees of Fornax, including late-type galaxies.
In particular, in a forthcoming paper, for all ETGs in our sample we will present the fit of the light distribution in order to estimate the structural 
parameters, deriving from them a quantitative morphological classification, and asses the stellar halo mass fraction to be compared 
with the theoretical predictions (Spavone et al. in preparation). 
The same kind of analysis is underway for the late-type galaxies (LTGs) inside the same area (Raj et al., in preparation).
Moreover, we aim at mapping the color gradients as function of the mass and cluster-centric radius out to the regions of the stellar halos
in order to constrain the age and metallicities of the stellar populations and the importance of the accreted satellites during the mass assembly 
process \citep[see e.g.][]{Labarbera2012}.

Auxiliary spectroscopic data, available for all ETGs inside the virial radius \citep{Sarzi2018}, 
will further constrain the star-formation rate, age and metallicity with of the galaxies and their stellar halos.
By taking advantage of kinematics and dynamical modelling, one could investigate the galaxy structure and, in particular, 
the nature of the disk flaring, since this 
could also naturally happen for galaxies in isolation as a result of weaker restoring force at larger distances \citep{Narayan2002} and thick disk 
might not be the result of a dynamical heating \citep{Brook2012}.

Therefore, FDS data represent a mine to study the galaxy structure, from the brightest inner regions to the faint outskirts, including the intra-cluster 
regions, and the evolution as function of the cluster-centric radius.
The deep observations can be directly compared with  predictions from galaxy formation models also in relation with environment. }

\begin{acknowledgements}
This work is based on visitor mode observations collected at the European Organisation for Astronomical Research in the Southern Hemisphere
under the following VST GTO programs: 094.B-0512(B), 094.B-0496(A), 096.B-0501(B), 096.B-0582(A).
{  Authors thank the anonymous referee for his/her suggestions that allowed us to much improve the paper.}
E.I. wishes to thank the ESO staff of the Paranal Observatory for their support during the observations at VST. 
E.I. and M.S. acknowledge financial support from the VST project (P.I. P. Schipani). 
E.I. is also very grateful to T. de Zeeuw and F. La Barbera for the discussions and suggestions on the present work.
J.~F-B. acknowledges support from grant AYA2016-77237-C3-1-P from the Spanish Ministry of Economy and Competitiveness (MINECO).
GvdV acknowledges funding from the European Research Council (ERC) under the European Union's Horizon 2020 research and innovation programme under grant agreement No 724857 (Consolidator Grant ArcheoDyn).
NRN and EI acknowledge financial support from the European Union Horizon 2020 research and innovation programme under the Marie Skodowska-Curie grant agreement n. 721463 to the SUNDIAL ITN network. 

\end{acknowledgements}

  \bibliographystyle{aa} 
   \bibliography{fornax} 
%


\begin{appendix} 

\section{Results on individual galaxies: a deep look at the outskirts}\label{note}

In this section we describe in detail the main properties of each galaxy of the sample from the FDS dataset presented  in this paper.
We focus on the faint features and structures in the outskirts of ETGs, which were unexplored in previous studies of the Fornax cluster.
We identify as  {\it stellar envelope} the regions at $\mu_r \geq 26$~mag/arcsec$^2$.
For each galaxy of the sample, we discuss the light and color distribution shown in Appendix~\ref{VST_image} and Appendix~\ref{colormap}, respectively. 

\subsection{FCC~090}
\begin{figure*}
   \centering
   \includegraphics[width=\hsize]{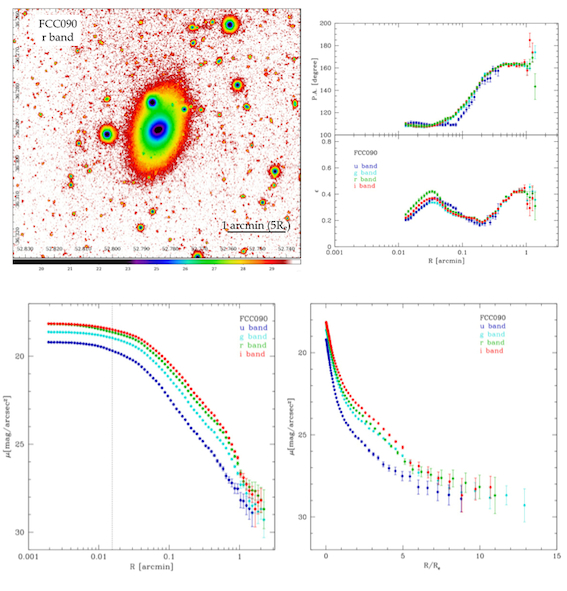}
      \caption{ Surface photometry for FCC~090. }
         \label{FCC090}
   \end{figure*}

FCC~090 is a small ($R_e = 1.5$ kpc) and faint ($M_i=-18.06$~mag)  elliptical galaxy (see Tab.~\ref{Mtot_tab}) in the SW region of the cluster at 1.7 degrees from NGC~1399 (see Fig.~\ref{mosaic_g}).
The VST image in the $r$ band is shown in Fig.~\ref{FCC090} (top-left panel). The surface brightness is mapped down to $\mu_r \sim29$~mag/arcsec$^2$ and out to about $11 R_e \sim 16.5$~kpc (see lower panels of Fig.~\ref{FCC090}). The isophotes show a large twisting, with P.A. varying by about 60 degrees from the center to large radii (see top-right panel of Fig.~\ref{FCC090}).
The stellar envelope, at $\mu_r \simeq 28-29$~mag/arcsec$^2$, is flatter ($\epsilon\sim0.4$) than the inner isophotes ($\epsilon\sim0.2$) and it has faint ($\mu_r \geq 29$~mag/arcsec$^2$) substructures on the SE, in the form of a tiny tail ($\sim0.6$~arcmin) toward the East, and on the NE side where an over-density of light makes the isophotes more elongated toward the North.  
The colormap brings to light a very irregular structure (see Fig.~\ref{col_1}), where the  isophotes show several ripples. The inner regions (for $R\leq0.1$~arcmin) are quite blue ($g-r\sim0.3-0.4$~mag) with respect to the outer and fainter envelope ($g-r\sim0.6-0.8$~mag). 
Taken together, this could suggest a recent minor merging event in this galaxy. 
{  The accretion of one or more smaller companion galaxies by an  early-type disk galaxy could form shells or ripples \citep[e.g.,][]{Quinn1984} like those
  observed in FCC090.
Preliminary results from the ALMA survey of the Fornax cluster (Zabel et al., in preparation) also suggest that past galaxy-galaxy
  interactions may be responsible for the irregular gas morphology and kinematics.}

\subsection{FCC~143 (NGC~1373)}\label{sec:fcc143}
\begin{figure*}
   \centering
   \includegraphics[width=\hsize]{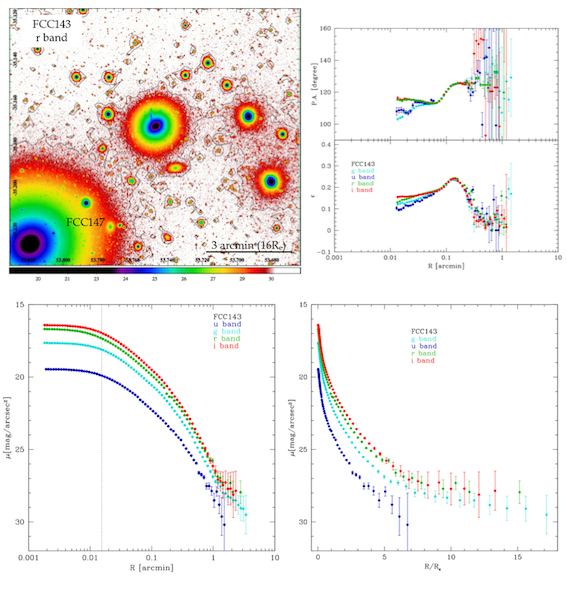}
      \caption{Surface photometry for FCC~143. }
         \label{FCC143}
   \end{figure*}

\begin{figure*}
  \centering
   \includegraphics[width=16cm]{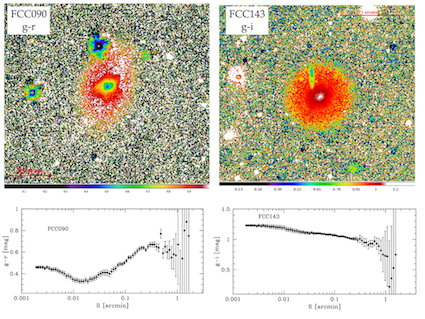}
  \includegraphics[width=16cm]{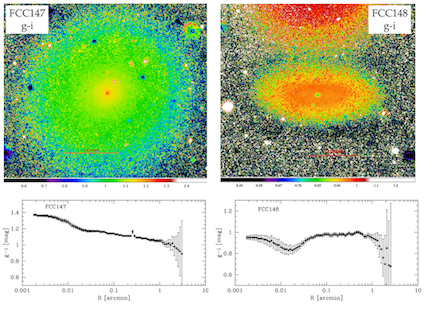}
  \caption{ $g-i$ color maps and color profiles for FCC~090 (top-left panels), FCC~143 (top-right panels),
  FCC~147 (lower-left panels) and FCC~148 (lower-right panels).}
       \label{col_1}
   \end{figure*}

FCC~143 is a compact ($R_e = 1.0$ kpc and $M_i=-19.04$~mag, see Tab.~\ref{Mtot_tab}) elliptical galaxy on the west side of the cluster, at  0.76 degrees from NGC~1399 (see Fig.~\ref{mosaic_g}). This galaxy is quite close in space to FCC~147, a more luminous and bigger elliptical galaxy  SE of FCC~143. 
According to \citet{Blakeslee2009}, the difference in distance is only  0.3 Mpc. The deep  VST image in the $r$ band shows a very faint bridge of light $\mu_r \simeq 
29.5$~mag/arcsec$^2$ between the two galaxies (see the top-left panel of Fig.~\ref{FCC143}), which suggests a possible ongoing interaction.
In the center, the bright and disk-like structure on the north is a superposed background edge-on disk galaxy. 
The outer and faint stellar envelope is rounder ($\epsilon\sim0.05$) than the inner isophotes ($\epsilon\sim0.2$). 
The major axis points towards and is aligned with the major axis of the companion galaxy  FCC~147  ($P.A.\sim 120$~degrees, see top-right panel of Fig.~\ref{FCC143} and of Fig.~\ref{FCC147}).
The $g-i$ color map and color profile show that the center is quite red ($g-i\sim1.3$~mag) and has a remarkable gradients towards bluer values with radius (see the top-left panels of Fig.~\ref{col_1}), with $g-i\sim0.7$~mag at $R\sim1$~arcmin. Therefore, the outer stellar envelope is  blue, with colors comparable to those observed in the 
outskirts of the companion galaxy FCC~147 (see the lower-left panels of Fig.~\ref{col_1}).

\subsection{FCC~147 (NGC~1374)}
\begin{figure*}
   \centering
   \includegraphics[width=\hsize]{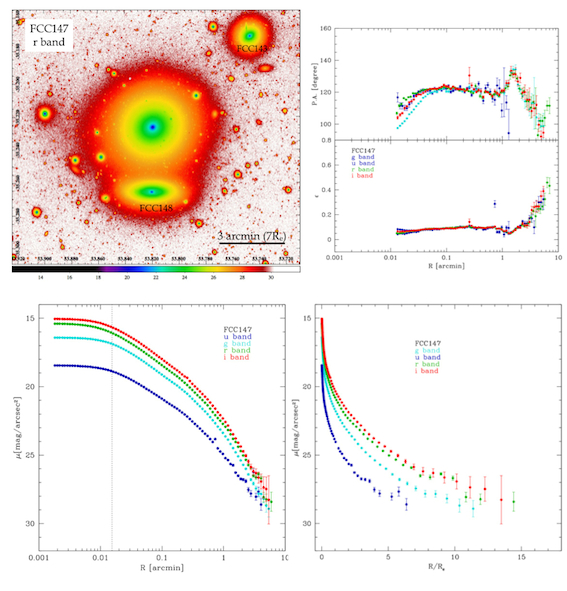}
      \caption{ Surface photometry for FCC~147.}
         \label{FCC147}
   \end{figure*}

FCC~147 is one of the brightest ($M_i\leq -21.31$~mag) elliptical galaxies in the core of the Fornax cluster, at 0.67 degrees from NGC~1399, on the west side (see Fig.~\ref{mosaic_g} and Tab.~\ref{Mtot_tab}).
This galaxy is close in space to FCC~143 (see also Sec.~\ref{sec:fcc143}), with which the deep VST image shows a possible ongoing interaction. 
It is also close to the S0 galaxy FCC~148,  the distances for the two galaxies are quite similar \citep{Blakeslee2009}.
The surface brightness is mapped down to $\mu_r \sim29$~mag/arcsec$^2$ and out to about $13-15 R_e \sim 34$~kpc (see lower panels of Fig.~
\ref{FCC147}). The stellar envelope  is quite extended ($1\leq R \leq10$~arcmin) and flatter ($\epsilon\sim0.4$) than the inner isophotes ($\epsilon\sim0.05$). 
The center is red ($g-i\sim1.4$~mag) and the color profile decreases with radius of about 0.7 magnitudes in the outskirts, where  $g-i\sim0.7$~mag at $R\sim2$~arcmin.

\subsection{FCC~148 (NGC~1375)}\label{sec:fcc148}
\begin{figure*}
   \centering
   \includegraphics[width=\hsize]{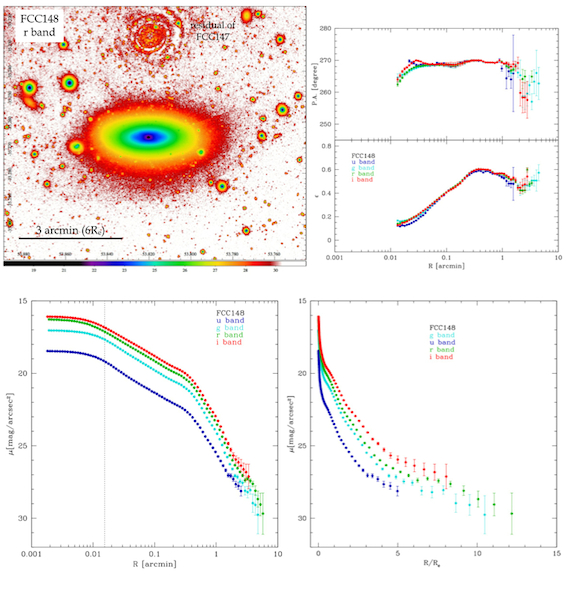}
      \caption{ Surface photometry for FCC~148. }
         \label{FCC148}
   \end{figure*}

This is an S0 galaxy with a marked boxy bulge (see Fig.~\ref{FCC148}, top-left panel), very close  to the elliptical galaxy FCC~147, located on the 
west side of the cluster at 0.67 degrees from NGC~1399.  
{  In order to account for the light from the closeby galaxy FCC~147, before performing the isophote fit on FCC~148, the 2D model of FCC147
was subtracted to the image. The 2D model for FCC~147 was obtained by the isophote fit for this galaxy, using the IRAF task BMODEL. 
In Fig.~\ref{FCC148} (top-left panel) is shown the resulting image for FCC~148.}
The SB profiles show two breaks in all bands (see lower-left panel of Fig.~\ref{FCC148}): the inner 
break is at $R\sim0.3$~arcmin ($\sim1R_e$ $\sim 1.3$~kpc), where the profile becomes steeper, and the second break occurs at about $4R_e$ ($\sim5$~kpc) where 
the profile shows an upturn. This is a typical example of a composite Type-II + Type-III profiles (see Sec.~\ref{overview}).
Since the ellipticity and P.A profiles also change at the second break (see top-right panels of Fig.~\ref{FCC148}), according to 
\citet{Erwin2008} this part of the surface brightness distribution is mapping the stellar envelope rather than being a truncated disk. 
In these regions, colors are bluer ($g-i\sim0.7-0.9$) than in the inner disk ($g-i\sim1$), even if a dip of about 0.2 magnitudes is observed in the center ($0.3\leq R
\leq3$~arcsec, see Fig.~\ref{col_1}, lower-right panels), suggesting the presence of bluer substructure. In the same range of radii, long-slit kinematics by \citet{Bedregal2006} pointed out the evidence of a kinematically decoupled component. 

\subsection{FCC~153 (IC~1963)}\label{sec:fcc153}
\begin{figure*}
   \centering
   \includegraphics[width=\hsize]{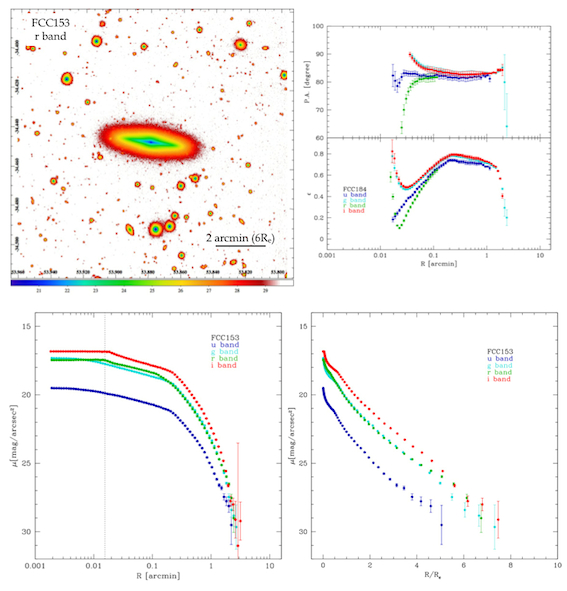}
      \caption{ Surface photometry for FCC~153.}
         \label{FCC153}
   \end{figure*}

FCC~153 is an edge-on S0 galaxy located in the low-density region on the North of the cluster, at a projected distance from NGC~1399 of 1.17 degrees (see 
Fig.~\ref{mosaic_g}). As observed for FCC~148, also in this case the SB profiles show two breaks. The inner one is at $R\sim0.2$~arcmin ($\sim1R_e$ $\sim 
3$~kpc), followed by a steep exponential-like decline out to about $5R_e$ (see lower panels Fig.~\ref{FCC153}). At larger radii, the SB profiles appear shallower out 
to $10R_e$ and down to $\mu_r \sim29$~mag/arcsec$^2$.
Together with the other two edge-on S0 galaxies of the sample (FCC~170 and FCC~177), FCC~153 shows a flaring in the disk, towards the outer radii, which is detected both in the light distribution (see the top panels of Fig.~\ref{fig:flares}) and in the color map (see Sec.~\ref{halos}).
The color map (see Fig.~\ref{col_2}, top-left panel) reveals a thin and red disk ($g-i\sim0.8$~mag) with a flaring in the outer regions ($R\geq1$~arcmin), where it appears a bit bluer ($g-i\sim0.7$~mag).

\subsection{FCC~161 (NGC~1379)}

\begin{figure*}
   \centering
   \includegraphics[width=\hsize]{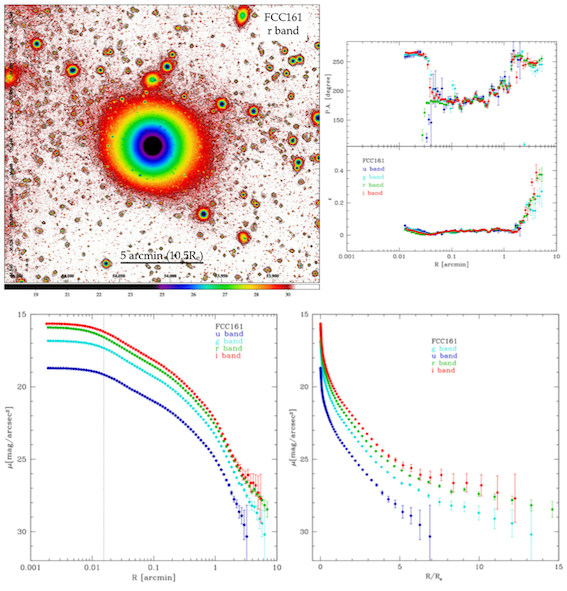}
      \caption{ Surface photometry for FCC~161.}
         \label{FCC161}
   \end{figure*}

\begin{figure*}
   \centering
   \includegraphics[width=16cm]{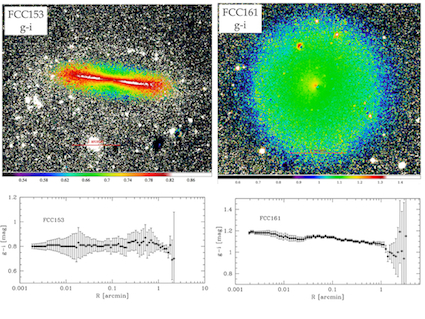}
  \includegraphics[width=16cm]{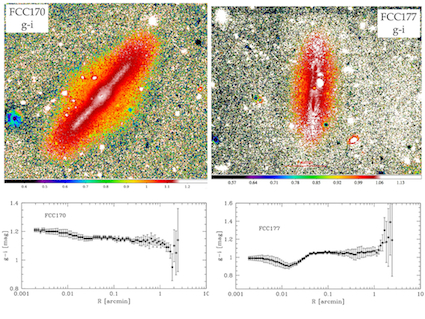}
      \caption{ $g-i$ color maps and color profiles for FCC~153 (top-left panels), FCC~161 (top-right panels),
  FCC~170 (lower-left panels) and FCC~177 (lower-right panels).}
         \label{col_2}
   \end{figure*}

This is the brightest  ($M_i=-21.35$~mag, see Tab.~\ref{Mtot_tab}) elliptical galaxy inside the stellar halo of NGC~1399.
In projection, it is located about 0.5 degree from NGC~1399, on the west side (see Fig.~\ref{mosaic_g}).
This galaxy is in the area where the intra-cluster light (ICL) was detected \citep{Iodice2017b}.
As also pointed out by  \citet{Iodice2017b}, the stellar halo of the galaxy is very asymmetric: it extends in the NE-SW direction and an excess of light is detected on the SW side  (see Fig.~\ref{FCC161}, top-left panel).
The surface brightness distribution is mapped down to $\mu_g \sim30$~mag/arcsec$^2$ and out to $\sim13R_e$ ($\sim 34$~kpc, see lower panels of Fig.~\ref{FCC161}).
A strong twisting ($\sim50$~degrees) and an increasing flattening ($0 \leq \epsilon \leq0.4$) is observed for $R\geq2$~arcmin, in the region of the stellar envelope, where the SB profiles show a shallower decline.
In the nuclear regions, for $R\leq2$~arcsec, where the P.A.  shows a twist of about 50 degrees and there is a small increase in the ellipticity ($\sim0.05$), the color map reveals the presence of  nuclear  dust (see Fig.~\ref{col_2}), where $g-i\sim1.2$~mag.
The outer stellar envelope, for $R\geq2$~arcmin, is bluer ($g-i\sim1.$~mag) than the inner and brightest region of the galaxy where the color profile shows a small gradient from $g-i\sim1.2$~mag to $g-i\sim1.1$~mag (see top right panels of Fig.~\ref{col_2}).

\subsection{FCC~167 (NGC~1380)}\label{sec:fcc167}
\begin{figure*}
   \centering
   \includegraphics[width=\hsize]{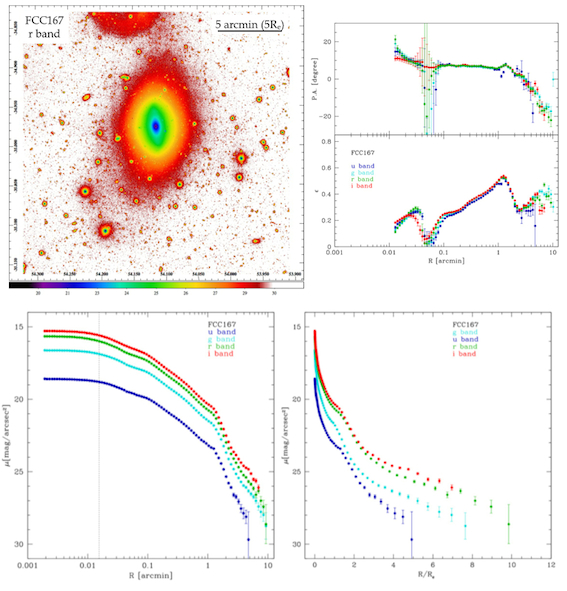}
      \caption{ Surface photometry for FCC~167.}
         \label{FCC167}
   \end{figure*}

FCC~167 was chosen as illustrative example to show each step of the analysis presented in this work.
Some details on its structure were already given in Sec.~\ref{results}.
It is the brightest S0 galaxy inside the virial radius of the cluster, with $M_i=-22.8$~mag (see Tab.~\ref{Mtot_tab}).
In the inner $\sim1$~arcmin, the high-frequency residual image reveals a bar-like boxy structure along the major axis of the galaxy, which ends  with a bright knot, on each side, at  $R\sim1.3$~arcmin (see right panel of Fig.~\ref{FCC167_image}). 
The outer and fainter isophotes ($R\geq 3$~arcmin and $\mu_r \sim27-30$~mag/arcsec$^2$) appear less flattened and twisted with respect to the inner and more disky ones (see top left panel of Fig.~\ref{FCC167_image} and Fig.~\ref{FCC167}).
The $r$-band SB profile extends out to 10~arcmin ($\sim52$~kpc) from the center, which is about $10 R_e$, and down to $\mu_r = 29 \pm 1$~mag/arcsec$^2$ (see lower-right panel of Fig.~\ref{FCC167_image}). 
The dust absorption affects the light distribution inside 0.1~arcmin, and perturbs  the P.A. and ellipticity profiles at the same distance (see top-right panel of Fig.~\ref{FCC167}). 
At larger radii, two {\it breaks} are observed. The first break marks the end of the bar-like boxy structure, it is between 1-1.5~arcmin, where the SB profile becomes steeper. The second break is at $R\sim3$~arcmin where the SB profile is shallower.
The ellipticity and P.A. profiles also show different behaviours at the break radii given above, in particular,  the isophotes are progressively rounder at larger radii. 
As suggested by \citet{Erwin2008}, this is an indication that light in this region is due to a spheroidal component rather than being part of the disk, therefore the SB profile is classified as {\it Type~III-s}. 
The $g-i$ colormap and the azimuthally-averaged $g-i$ color profile are shown in Fig.~\ref{FCC167_col}. The galaxy is quite red in the center, with 
$g-i \simeq 1.2-1.3$~mag for $R\leq 0.1$~arcmin. A peak is observed at $\sim5$~arcsec where $g-i \sim1.3$~mag. At larger radii, the galaxy is bluer, having  
$g-i \simeq 1.1-1.2$~mag for $0.1\leq R \leq 3$~arcmin (see bottom panel of Fig.~\ref{FCC167_col}).
Inside 1~arcmin, the spiral-like structure detected in the light distribution (see Fig.~\ref{FCC167_image}) appears also in the $g-i$ colormap (see the upper panels of Fig.~\ref{FCC167_col}). 
It seems more extended on the West side of the galaxy and has a color gradient toward bluer values ($g-i \leq 1.2$~mag). 
In the inner 10 arcsec, the ring of dust appears as a quite red ($g-i \sim 1.3$~mag) structure. 
At larger radii, for $R \geq 3$~arcmin, a strong gradient towards red colors is observed ($g-i  \sim 1.2-1.5$~mag),  which is significant even taking into account the larger errors (see lower panel of Fig.~\ref{FCC167}).

\subsection{FCC~170 (NGC~1381)}\label{sec:fcc170}
\begin{figure*}
   \centering
   \includegraphics[width=\hsize]{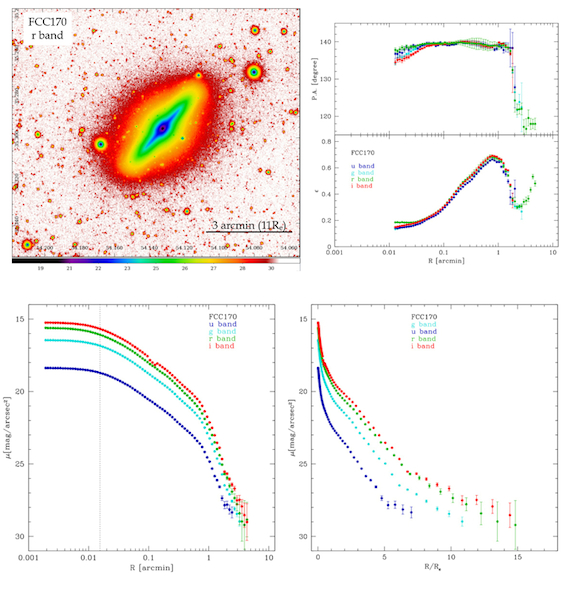}
      \caption{ Surface photometry for FCC~170.}
         \label{FCC170}
   \end{figure*}

This is an edge-on S0 galaxy close to the center of the cluster, located at a projected distance of $0.42$~degree from NGC~1399  (see Fig.~\ref{mosaic_g}).
Together with FCC~161 and FCC~184, FCC~170 is in the area were the diffuse intra-cluster light was detected by \citet{Iodice2017b}.
The inner and bright regions show a boxy bulge and a thin disk, while the faintest outskirts ($\mu_r \sim 27-30$~mag/arcsec$^2$) resemble the superposition 
of a thick disk and a rounder envelope (see Fig.~\ref{FCC170}).
This galaxy has a prominent flaring of the disk towards outer radii (see the middle panels of the Fig.~\ref{fig:flares}), which is more extended and thickest compared to the  similar features detected in FCC~153 and FCC~177 (see Sec.~\ref{halos}).
The SB profiles show two breaks (see lower panels of Fig.~\ref{FCC170}), the first in the inner regions at $R\sim1$~arcmin ($2R_e$ $\sim4$~kpc) and the 
second one at $R\sim3$~arcmin ($7R_e$ $\sim6$~kpc). To the second break radius corresponds a variation of the P.A. and ellipticity profiles, with a twisting of 
about 20 degrees and $\epsilon$ increases from $\sim 0.3$ to $\sim0.5$ (see top panel of Fig.~\ref{FCC170}). As discussed for the other S0 galaxies of the 
sample, this is a typical example of a composite Type-II + Type-III profile.
The inner boxy bulge and the thin disk are quite red, with $g-i\sim1.2$~mag (see lower-left panel of Fig.~\ref{col_2}). The thick disk has $g-i\sim 1-1.1$~mag 
and the stellar envelope is even bluer, with $g-i\leq0.9$~mag.

\subsection{FCC~177 (NGC~1380A)}\label{sec:fcc177}
\begin{figure*}
   \centering
   \includegraphics[width=\hsize]{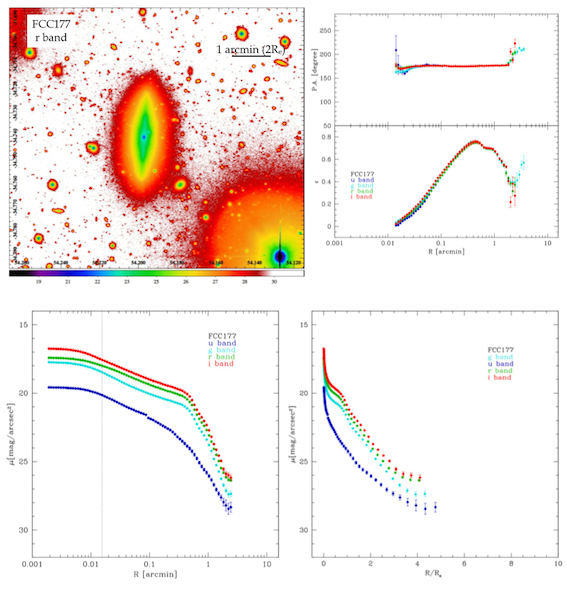}
      \caption{ Surface photometry for FCC~177.}
         \label{FCC177}
   \end{figure*}

This is an edge-on S0 galaxy in the North of the cluster, at a projected distance of $0.78$~degree from NGC~1399  (see Fig.~\ref{mosaic_g}).
It is characterised by a small and bright bulge ($R\leq30$~arcsec and $\mu_r\geq20$~mag/arcsec$^2$), an extended thin disk ($R\leq 2$~arcmin and $\mu_r\sim 20-27$~mag/arcsec$^2$) and a thick stellar halo ($R\geq2$~arcmin and $\mu_r\sim24$~mag/arcsec$^2$, see Fig.~\ref{FCC177}).
Therefore, the resulting SB profiles are of  Type-II + Type-III, as observed for many other S0 galaxies of the sample (see Sec.~\ref{overview}).
As for  FCC~153 and FCC~170, a flaring of the disk is observed in the light distribution (see Sec.~\ref{halos}), which is also detected in the  color map, in the outer regions of the disk ($R\geq 1$~arcmin), along the major axis. 
Differently from FCC~153, colors are red, $g-i\sim1.1 -1.3$~mag (see lower-right panel of Fig.~\ref{col_2}).
In the inner regions ($R\leq 0.5$~arcmin), the color map shows a thin and blue feature ($g-i\sim0.9$~mag). The thick stellar envelope, which can be mapped along the galaxy minor axis has a bluer color than the thin disk, with  $g-i\sim0.85 - 1$~mag.

\subsection{FCC~182}
\begin{figure*}
   \centering
   \includegraphics[width=\hsize]{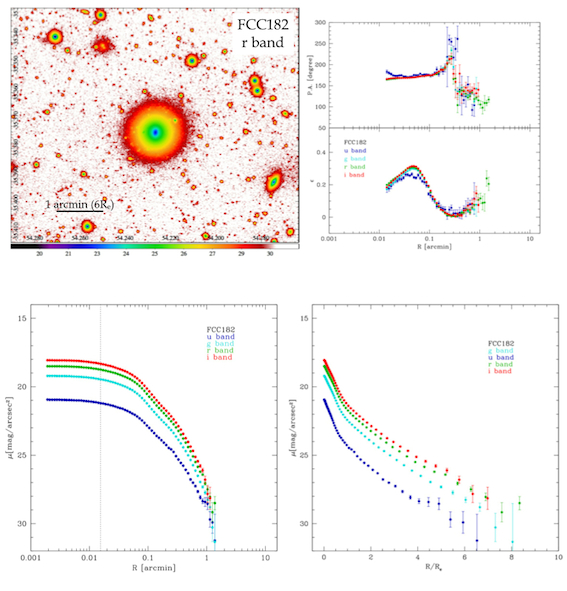}
      \caption{ Surface photometry for FCC~182.}
         \label{FCC182}
   \end{figure*}

FCC~182 is a small ($R_e=1.2$~kpc, see Tab.~\ref{Mtot_tab}) face-on S0 galaxy, very close to the cluster center, at a projected distance of 0.32~degrees from NGC~1399 (see Fig.~\ref{mosaic_g}). This galaxy is completely embedded in the ICL detected by \citet{Iodice2017b}.
Differently from the other galaxies in this area (FCC~161, FCC~170 and FCC~184), the outer isophotes of FCC~182 appear quite regular 
and symmetric (see the top-right panel of Fig.~\ref{FCC182}).  
The flattened isophotes ($\epsilon \sim 0.3$) and the constant P.A. ($\sim170$~degrees) inside 0.1~arcmin suggest the presence of a small bar.
For $0.1\leq R \leq 0.3$ arcmin, a moderate twisting of about 50 degrees is observed, where the isophotes are almost round.
At larger radii, the ellipticity increases again ($\epsilon\sim0.2$) and the P.A. varies by more than 100 degrees. In this regions, the SB profiles show an extended exponential-like decline down to $\mu_r \sim 29$~mag/arcsec$^2$ and out to $7R_e$ ($\sim8.4$~kpc).
The color map and color profile show that inside 0.1 arcmin the galaxy is very red ($g-i\sim 1.15$~mag), while outside, for $0.1\leq R \leq 0.3$ arcmin, a small gradient towards bluer colors is observed  ($g-i\sim 1.15-1$~mag). The stellar envelope, for $R\geq0.3$~arcmin, has redder colors, with $g-i$  increasing to $\sim 1.2$~mag (see the top-left panel of Fig.~\ref{col_3}).

\subsection{FCC~184 (NGC~1387)}
\begin{figure*}
   \centering
   \includegraphics[width=\hsize]{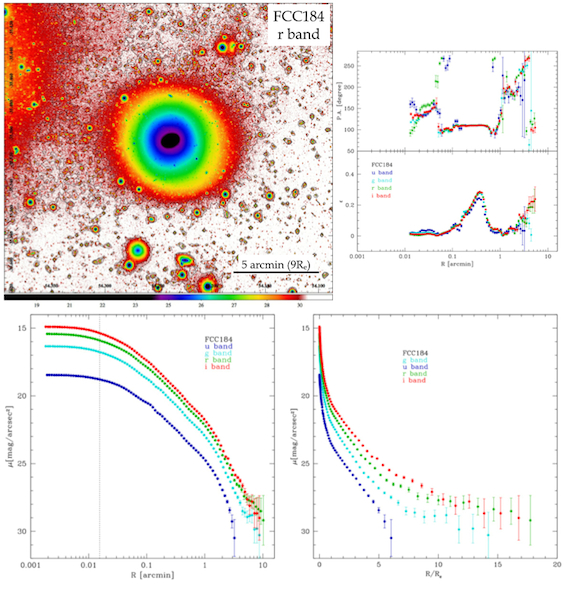}
      \caption{ Surface photometry for FCC~184.}
         \label{FCC184}
   \end{figure*}

\begin{figure*}
   \centering
   \includegraphics[width=16cm]{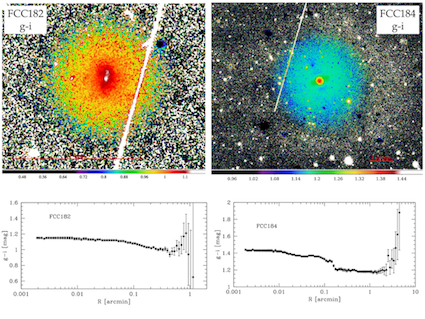}
   \includegraphics[width=16cm]{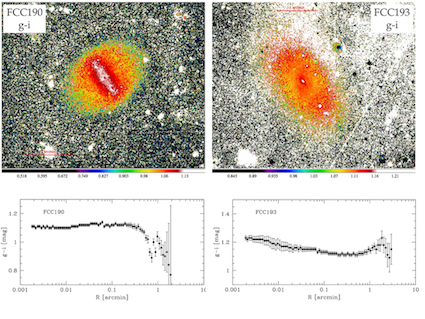}
      \caption{ $g-i$ color maps and color profiles for FCC~182 (top-left panels), FCC~184 (top-right panels),
  FCC~190 (lower-left panels) and FCC~193 (lower-right panels).}
         \label{col_3}
   \end{figure*}

This is the brightest ($M_i=-21.81$~mag, see Tab.~\ref{Mtot_tab}) elliptical galaxy closest in space to NGC~1399.
According to \citet{Blakeslee2009}, the distance differs by 1.6$\pm1.2$ Mpc.
In projection, it is located at about 0.3 degree from NGC~1399, on the west side (see Fig.~\ref{mosaic_g} and Tab.~\ref{Mtot_tab}).
In the intracluster region, on the west side of NGC~1399 and towards FCC~184, \citet{Iodice2016} detected a faint ($\mu_g \sim 30$~mag~arcsec$^{-2}$) 
stellar bridge, about 5~arcmin long ($\sim 29$~kpc). 
Based on the color analysis, it could result from the stripping of the outer envelope of FCC~184 on its east side. 
Such a feature was also detected in the spatial distribution of the blue GCs \citep{Bassino2006,Dabrusco2016}.
This galaxy is in the area of the intra-cluster light  \citep{Iodice2017b}.
As also pointed out by \citet{Iodice2016} and \citet{Iodice2017b}, the outer isophotes of the galaxy are very asymmetric: they extend on the W-SW side (see Fig.~\ref{FCC184}, top-left panel), opposite to the region where the bridge of light toward NGC~1399 is detected. 
The surface brightness distribution is mapped down to $\mu_r \sim29$~mag/arcsec$^2$ and out to $\sim15R_e$ ($\sim 43$~kpc, see lower panels of Fig.~\ref{FCC184}).
A strong twisting ($\sim100$~degrees) and an increasing flattening ($0 \leq \epsilon \leq0.3$) is observed for $R\geq2$~arcmin, in the region of the stellar envelope, where the SB profiles show a shallower decline. 
Inside 1~arcmin, the image suggests the presence of a small inner bar, characterised by the typical peak in ellipticity profile ($\epsilon\sim0.3$).
In the nuclear regions, for $R\leq5$~arcsec, where the P.A. also shows a twist of about 50 degrees, the color map reveals the presence of a nuclear ring of dust (see Fig.~\ref{col_3}), where $g-i\sim1.4$~mag. A nuclear ring of about 6~arcsec was also found from the near-infrared images in the Ks band by \citet{Laurikainen2006}.
The outer stellar envelope, for $R\geq2$~arcmin, is quite red ($g-i\sim1.6$~mag) compared to the inner and brightest region of the galaxy where the color profile shows a small gradient from $g-i\sim1.4$~mag to $g-i\sim1.2$~mag (see top right panels of Fig.~\ref{col_3}).

\subsection{FCC~190 (NGC~1380B)}
\begin{figure*}
   \centering
   \includegraphics[width=\hsize]{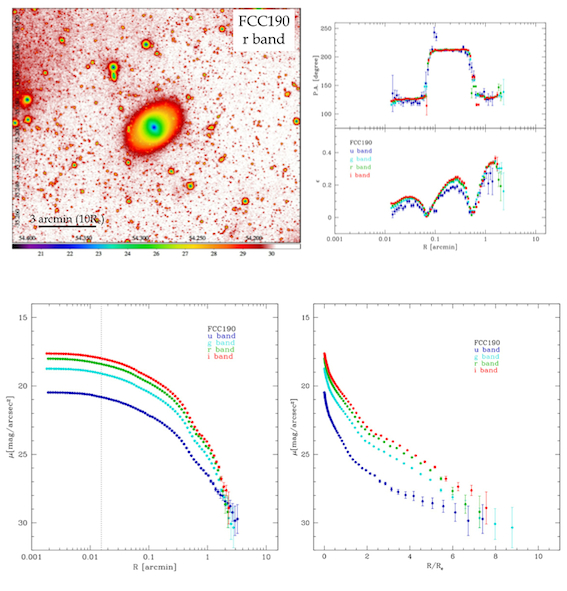}
      \caption{ Surface photometry for FCC~190.}
         \label{FCC190}
   \end{figure*}

This is one of the four barred S0 galaxies close in projection to the cluster center (see Fig.~\ref{type}), on the NW side.
FCC~190 is at a distance of $20.3\pm0.7$~Mpc (see Tab.~\ref{galaxies}). This is consistent with the distance of 
the closest galaxies FCC~182 ($19.6\pm0.8$~Mpc) and FCC~170 ($21.9\pm0.8$~Mpc), 
and  FCC~167 ($21.2\pm0.7$~Mpc) that is located north-west of FCC~190 (see Tab.~\ref{galaxies})
These four galaxies, being close in space and aligned along the north-south direction, could be a primordial sub-group of galaxies in-falling into the cluster. 
In FCC~190 the stellar envelope is at 90 degrees with respect to the 
inner bar,  and extends down to   $\mu_r \sim 29$~mag/arcsec$^2$ and out to $7R_e$ (see Fig.~\ref{FCC190}).
Also the $g-i$ colors are quite different between the two components: the inner bar is red, with $g-i\sim1.13$~mag, 
while the outer isophotes become bluer at $R\sim0.5$~arcmin and $g-i\sim0.9$~mag at the outer radius (see lower left panels of Fig.~\ref{col_3}).

\subsection{FCC~193 (NGC~1389)}
\begin{figure*}
   \centering
   \includegraphics[width=\hsize]{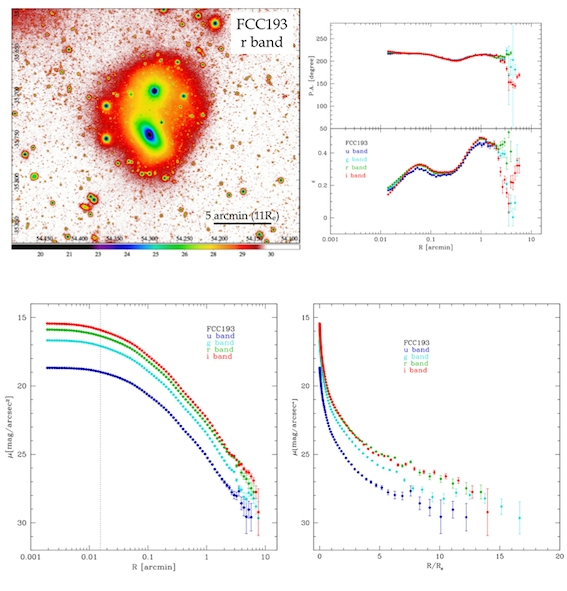}
      \caption{ Surface photometry for FCC~193.}
         \label{FCC193}
   \end{figure*}

This barred S0 galaxy is at a projected distance of 0.39~degrees from NGC~1399 on the SW side (see Fig.~\ref{mosaic_g}). 
South of FCC~193, as well as West and East of it, there are no more ETGs in the high-density region of the cluster (see Sec.~\ref{overview}).
The outer and fainter isophotes, mapped down to $\mu_r \sim 29$~mag/arcsec$^2$ and 
out to $14R_e$ (see Fig.~\ref{FCC193}), show a sharp boundary on the south side, so that they appear more boxy than on the north side.
The position angle of the major axis of the outer envelope differs from the major axis of the bright central bulge by more than 30 degrees (see the top-right panel of Fig.~\ref{FCC193}).
The center of FCC~193 is very red, with $g-i\sim1.2$~mag, while bluer colors are measured with  increasing semi-major axis 
(lower right panels of Fig.~\ref{col_3}).

\subsection{FCC~213 (NGC~1399)}
This is NGC~1399, the brightest cluster galaxy of the Fornax cluster inside the virial radius.
A detailed discussion of FCC~213 based on FDS data is presented by \citet{Iodice2016}. 

\subsection{FCC~219 (NGC~1404)}
\begin{figure*}
   \centering
   \includegraphics[width=\hsize]{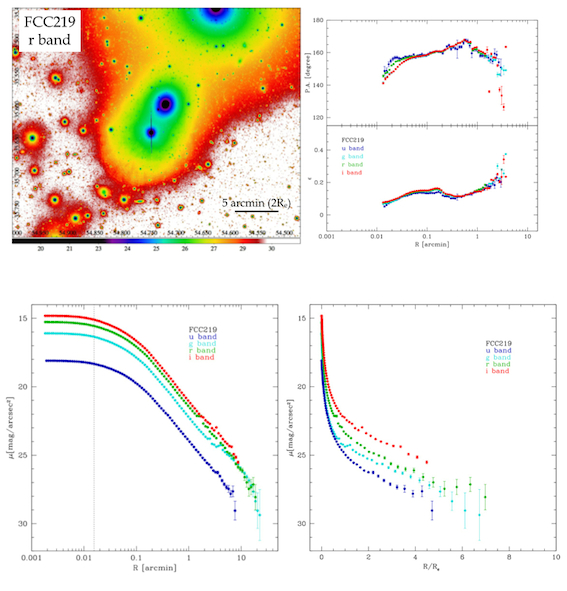}
      \caption{ Surface photometry for FCC~219.}
         \label{FCC219}
   \end{figure*}

\begin{figure*}
   \centering
   \includegraphics[width=16cm]{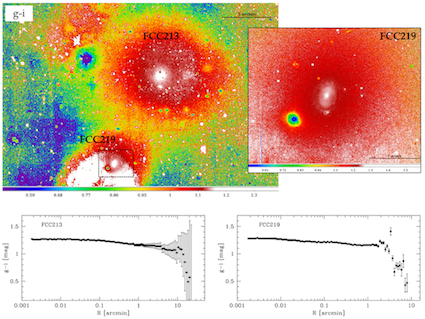}
      \caption{ $g-i$ colomap (top panels) and azimuthally averaged color profiles (bottom panels) for the two bright ETGs in the core of the Fornax cluster, FCC213 and FCC219. The image size of the top-left panel is $0.45 \times 0.38$~degrees. In the top right panel there is an extracted image  ($3.4 \times 2.6$~arcmin) of FCC219 into the central regions (dashed box in the top left panel).}
         \label{col_4}
   \end{figure*}

This is the second brightest ETG of the Fornax cluster, $\sim1$~mag fainter than NGC~1399 (see Tab.~\ref{galaxies}), inside the viral radius.
Even if FCC~219 is superposed on the very luminous outskirts of NGC 1399 and it is in the halo of a bright star south of it, the isophotes have a regular 
shape with semi-major axis, showing an increasing flattening and a twisting of about 10 degrees for $R\geq 1$~arcmin, where $\mu_r \geq 23$~mag/arcsec$^2$ (see Fig.~\ref{FCC219}).
The SB profiles show an extended exponential-like decrease for $R\geq1Re$ (see the lower panels of Fig.~\ref{FCC219}) out to $7R_e$ and down to $
\mu_r \sim 28$~mag/arcsec$^2$. Since FCC~219 is completely superposed on the stellar envelope of NGC~1399, it is reasonable to think that 
the outskirts of both galaxies overlap in projection at the faintest magnitudes, being  indistinguishable by the photometry alone.
The $g-i$ color profile is quite constant $\sim1.3 - 1.2$~mag out to $R\sim 1$~arcmin, very similar to that for NGC~1399 (see lower panels of Fig.~\ref{col_4}). 
At larger radii, the colors are bluer $g-i\sim0.7$, consistent with the $g-i$ colors derived for the stellar envelope of NGC~1399.
The inner regions of FCC219 are  quite red, $g-i\sim1.3$. A small (few arcsecs) red feature  is detected on the south 
of the galaxy center, which seems to be connected to a red tail on its east side. It could be the remnant of a disrupted small galaxy or a background galaxy.

\subsection{FCC~276 (NGC~1427)}
\begin{figure*}
   \centering
   \includegraphics[width=\hsize]{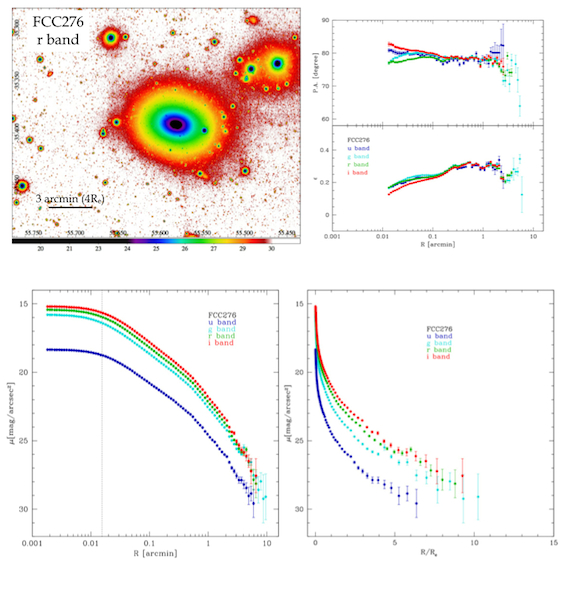}
      \caption{ Surface photometry for FCC~276.}
         \label{FCC276}
   \end{figure*}

This is the brightest ($M_i=-21.3$~mag, see Tab.~\ref{Mtot_tab}) elliptical galaxy located on the east side of the cluster, at a projected distance of 0.79~degrees from 
NGC~1399 (see Fig.~\ref{mosaic_g}), in the transition region of the cluster from  high to low density (see Fig.~\ref{mosaic_g}). 
The SB profiles are mapped out to $\sim9R_e$, except  for the $u$-band SB,  which is less extended ($\sim6R_e$), and 
down to $\mu_r \sim 29$~mag/arcsec$^2$, in the $r$ band (see Fig.~\ref{FCC276}).
The outer and fainter isophotes appear asymmetric: 
they are more extended on the N-NW side, where many diffuse and very faint ($\mu_r \sim 30$~mag/arcsec$^2$) patches 
of light are detected, than the isophotes on the opposite SE side, which show a sharp boundary.
The $g-i$ color profile and color map (see the top-left panels of Fig.~\ref{col_5}) show a very blue nucleus: for $R
\leq2$~arcsec the $g-i$ decreases from 0.85~mag to $\sim0.65$~mag (this is not a seeing effect, since it affects the profile 
only for $R\leq0.01$~arcmin). Similar behaviour was found by \citet{Carollo1997} in the $V-I$ color profile (obtained with HST), where $V-I$ 
decreases by $\sim0.15$ mag inside 0.1 arcsec.
The $g-i$ color map shows a disky and blue ($g-i \sim0.7$~mag) nucleus surrounded by redder regions ($g-i 
\sim0-8-0.9$~mag) with a ring-like feature made of red knots.
The presence of a nuclear disk between 0.3 and 3 arcsec was suggested by \citet{Forbes1995}.
The galaxy becomes bluer with increasing radius and in the outskirts $g-i \sim 0.7-0.6$~mag.

\subsection{FCC~277 (NGC~1428)}
\begin{figure*}
   \centering
   \includegraphics[width=\hsize]{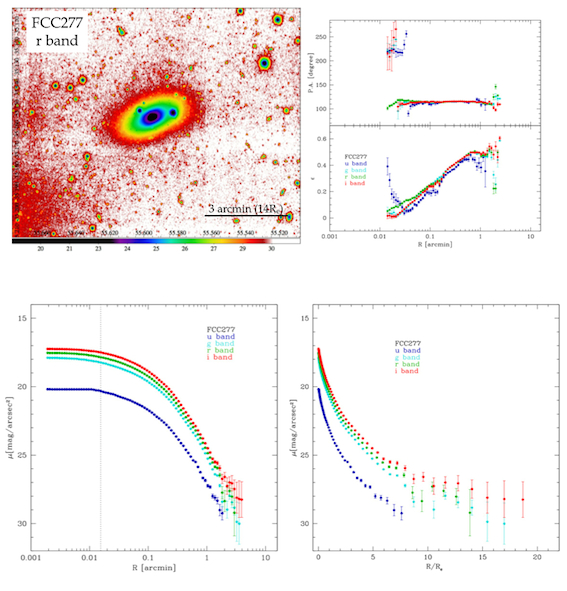}
      \caption{ Surface photometry for FCC~277.}
         \label{FCC277}
   \end{figure*}

This galaxy is located on the east side of the cluster, at a projected distance of 0.86~degrees from 
NGC~1399 (see Fig.~\ref{mosaic_g}),  in the transition region of the cluster from the high to low density (see Fig.~
\ref{mosaic_g}), north of FCC~276, the other elliptical galaxy in this area. 
FCC~277 is about two magnitudes fainter than FCC~276  (see Tab.~\ref{Mtot_tab}) 
and the distance differs by 1.1$\pm1.3$ Mpc \citep{Blakeslee2009}.
Like FCC~276, FCC~277 resembles a "normal" elliptical galaxy without any prominent subcomponents (see Fig.~
\ref{FCC277}), as bars or X-shaped bulge observed for many of the ETGs in the high-density region of the cluster on the west side.
The $gri$ SB profiles extend out to $\sim15R_e$, while the SB in the $u$ band  is less extended ($\sim7R_e$). 
The limiting magnitude is $\mu_r \sim 29$~mag/arcsec$^2$ in the $r$ band (see Fig.~\ref{FCC277}).
The shape of the stellar envelope is quite regular.
As observed in FCC~276, also in FCC~277 the inner regions of the galaxy are bluer: $g-i$ decreases by about 0.1 magnitude 
(from 0.8  to 0.7 mag) for $R\leq 2.4$~arcsec (see the top-right panels of Fig.~\ref{col_5}). The $g-i$ color map shows a 
clear bluer nuclear structure with $g-i\sim0.7$~mag.
The outer regions of the galaxy  tend to get bluer ($0.8 \leq g-i \leq 0.65$~mag) with 
increasing distance from the center $0.3 \leq R \leq 0.9$~arcmin. At larger radii ($R>1$~arcmin), the color profile shows 
redder values ($g-i\sim0.75$~mag): these regions could be affected by the scattered light of the bright and red foreground star on the NE side of the galaxy.

\subsection{FCC~301}
\begin{figure*}
   \centering
   \includegraphics[width=\hsize]{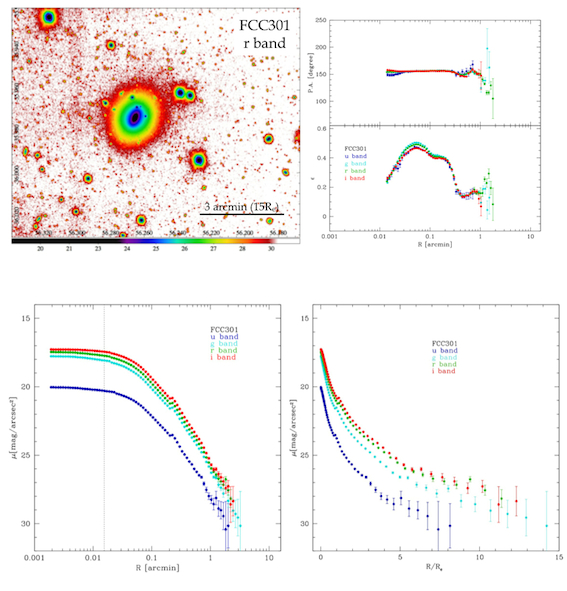}
      \caption{ Surface photometry for FCC~301.}
         \label{FCC301}
   \end{figure*}

This is the faintest ($M_i=-18.82$~mag, see Tab.~\ref{Mtot_tab}) elliptical galaxy in the sample located on the SE side of the cluster, at a projected distance of 1.44~degrees from 
NGC~1399 (see Fig.~\ref{mosaic_g}), in the low density region of the cluster (see Fig.~\ref{mosaic_g}). 
The $gri$ SB profiles extend out to $\sim11-14R_e$, while the SB in the $u$ band  is less extended  and reaches $\sim7R_e$  (see Fig.~\ref{FCC301}).
The inner regions ($R\leq 0.1$~arcmin) are flatter ($\epsilon \sim 0.4-0.5$) than the stellar envelope (see top panels of Fig.~\ref{FCC301}), which has smooth and unperturbed isophotes.
As observed in FCC~276 and in FCC~277, also FCC~301 shows a blue structure in the inner regions  of the galaxy.
The $g-i$ color map and color profile reveal a  blue ($g-i \sim 0.6 $~mag) disk-like structure in the nucleus (see the lower-left 
panels of Fig.~\ref{col_5}),  inside 6~arcsec. 
The flattened isophotes and blue colours are both consistent with an inner disk component in this galaxy.
This is also in agreement with the stellar long-slit kinematics by \citet{Bedregal2006}, where a flat and low velocity 
dispersion is found in the inner 5 arcsec, so that the authors suggested the presence of a counter-rotating stellar disk, consistent with the new FDS data. 
The galaxy is quite red in the regions outside the nuclear disk-like structure, with $g-i\sim0.7-0.8$~mag out to $R\sim0.3$~arcmin. 
At larger radii, the stellar envelope is bluer with $g-i\sim0.65-0.55$~mag out to $R\sim1$~arcmin.

\subsection{FCC~310 (NGC~1460)}
\begin{figure*}
   \centering
   \includegraphics[width=\hsize]{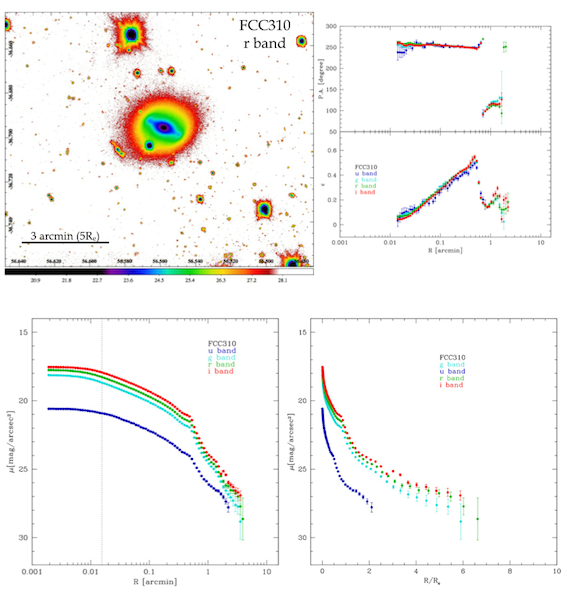}
      \caption{ Surface photometry for FCC~310.}
         \label{FCC310}
   \end{figure*}

\begin{figure*}
   \centering
   \includegraphics[width=16cm]{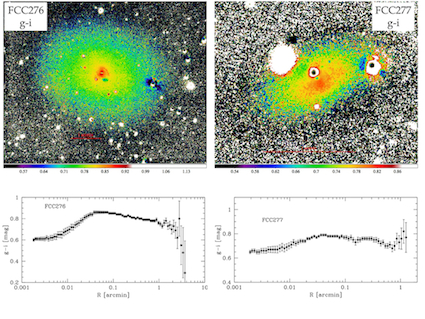}
  \includegraphics[width=16cm]{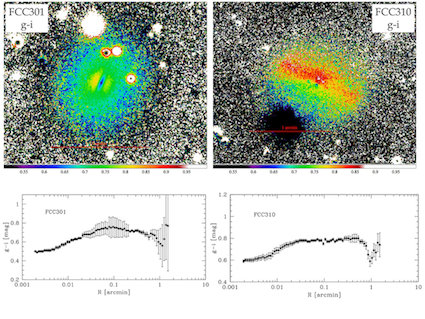}
      \caption{ $g-i$ color maps and color profiles for FCC~276 (top-left panels), FCC~277 (top-right panels),
  FCC~301 (lower-left panels) and FCC~310 (lower-right panels).}
         \label{col_5}
   \end{figure*}

This is the only barred S0 galaxy on the East side of the cluster, south of NGC~1399, at a projected distance of 2 degrees, 
so at the virial radius.
The prominent and bright ($\mu_r \leq 24$~mag/arcsec$^2$) bar is evident in the image as well as in the SB and ellipticity and P.A. profiles (see Fig.~\ref{FCC310}). 
The $gri$ SB profiles extend out to $\sim 6-7 R_e$. They show two breaks, the inner one, due to the bar, at $R
\sim0.5$~arcmin, where the SB profiles become steeper, and the outer one at $R\sim 1$~arcmin and for $\mu_r \geq 
26$~mag/arcsec$^2$, where a shallow and almost exponential-like decrease is observed. The isophotes in these regions are rounder ($\epsilon \sim 0.2$)  and twisted 
by about  100 degrees (see top-right panels of Fig.~\ref{FCC310}).
As observed for all ETGs  on the East side of the cluster (FCC~276, FCC~277 and FCC~301), FCC~310 also shows 
a blue nucleus, with $g-i\sim0.65$ in the center and $g-i\sim0.8$ at $R\sim2$~arcsec (see the lower-right panels of Fig.~\ref{col_5}). Color is almost constant $g-i\sim0.8$~mag along the bar, for $0.03\leq R \leq 0.5$~arcmin. At larger radii, the galaxy outskirts are bluer, with $g-i\sim0.6$~mag at $R\sim 1$~arcmin.


\section{$r$-band VST Images of the ETGs inside the virial radius of the Fornax cluster: surface brightness, ellipticity and P.A. profiles}\label{VST_image}

In each figure of this section we show the results from the fit of the isophote, performed for all ETGs in the Fornax cluster listed in Tab.~\ref{galaxies} in the $ugri$ VST images. See Sec.~\ref{phot} for details.

The top-left panel of each figure shows an extracted region of the VST mosaic in the $r$ band around each galaxy of the sample, plotted in surface brightness levels (shown in the colorbar). {  The black line indicates the scale in arc-minutes and the corresponding value of Re in the $r$ band (listed in Tab.~\ref{mag}).}
In the top-right panels are shown the P.A. (top) and ellipticity (lower) profiles, derived by the isophote fit from $ugri$ VST images.
In the lower panels, we show the azimuthally-averaged surface brightness profiles in $ugri$ bands, plotted in logarithmic scale (left) and in linear scale (right) as function of the effective radius $R_e$ derived in each band (see Tab.~\ref{mag}). The vertical dotted line (left panel) delimits the region affected by the seeing.

 \clearpage

\section{Color distribution: 2D map and profiles}\label{colormap}

In each figure of this section we show the $g-i$ color map and azimuthally averaged color profile for each ETG of the sample.

\clearpage

\clearpage

\clearpage

\clearpage

\end{appendix}

\end{document}